\documentclass[11pt]{article}
\usepackage{jheppub}
\usepackage{physics}
\usepackage{slashed}
\usepackage{bm}
\usepackage{mathrsfs}
\usepackage{amsmath, amsthm}
\usepackage{graphicx}

\usepackage{empheq}
\usepackage[most]{tcolorbox}
\newtcbox{\mymath}[1][]{%
    nobeforeafter, math upper, tcbox raise base,
    enhanced, colframe=blue!30!black,
    colback=blue!30, boxrule=1pt,
    #1}

\newcommand{\pic}[2]{\vcenter{\hbox{\includegraphics[height=#1]{#2}}}}

\graphicspath{ {./images/} }
\newcommand{\pick}[2]{\vcenter{\hbox{\includegraphics[width=#1]{#2}}}}

\usepackage{soul}

\DeclareMathOperator{\sign}{sign}

\newcommand{\calI}{\mathcal{I}}
\newcommand{\bC}{\mathbb{C}}
\newcommand{\bR}{\mathbb{R}}
\renewcommand{\dd}{\mathrm{d}}

\newcommand{\dvol}{\mathrm{dVol}}

\newcommand\cD{\mathcal D}

\newcommand\cI{\mathcal I}\newcommand\cJ{\mathcal J}
\newcommand\cN{\mathcal N}\newcommand\cO{\mathcal O}
\newcommand\cT{\mathcal T}
\newcommand\cV{\mathcal V}

\newcommand\define{\overset{\rm def}{=}}

\def\d{{\rm d}}

\newcommand\calT{\mathcal T}
\newcommand\calO{\mathcal O}
\def\calC{\mathcal{C}}
\def\calD{\mathcal{D}}
\def\calI{\mathcal{I}}

\def\bbR{\mathbb{R}}
\def\bbZ{\mathbb{Z}}
\def\bbC{\mathbb{C}}
\def\bbN{\mathbb{N}}
\def\calL{\mathcal{L}}

\theoremstyle{definition}

\newtheorem{definition}{Definition}

\newcommand\NORD[1]{\ensuremath{:\!\!{#1}\!\!:}}
\newcommand\nord[2]{\ensuremath{:\!\!{#1}\!\!:\!\!({#2})}}

\newcommand{\BRST}{\mathrm{BRST}}

\usepackage{tikz,tikz-3dplot}
\usepackage{pgfplots}
\pgfplotsset{compat=1.17}
\usetikzlibrary{shapes,arrows,cd,chains,decorations.markings,decorations.pathmorphing,calc, patterns}

\tikzset{
    ->-/.style args={#1rotate#2}{
        decoration={
            markings, 
            mark=at position #1 with {\arrow[scale=1.5, rotate=#2]{stealth}}
        }, 
        postaction={decorate}
    }
}

\tikzstyle{GraphNode}=[circle, draw=black, fill=black, inner sep=2pt, minimum size=5pt]
\tikzstyle{GraphNodeH}=[circle, draw=black, fill=none, inner sep=2pt, minimum size=5pt]
\tikzstyle{GraphEdge}=[black]

\tikzstyle{vertex}=[circle, draw, minimum size=0.2cm, inner sep = 1pt, fill=black] 
\tikzstyle{line}=[line width=0.3mm,->- = {0.5rotate0}]
\tikzstyle{lineShimmy}=[line width=0.3mm,->- = {0.53rotate0}]
    
\pgfmathsetmacro{\gS}{1}

\usepackage{tikzit}

\tikzstyle{red dot}=[fill=red, draw=none, shape=circle, minimum size=5pt, inner sep=2pt]
\tikzstyle{black dot}=[fill=black, draw=none, shape=circle, minimum size=5pt, inner sep=2pt]

\tikzstyle{red edge}=[-, draw=red]
\tikzstyle{black edge}=[-]
\tikzstyle{new edge style 0}=[fill={rgb,255: red,191; green,0; blue,64}, draw=black, dots, =]

\author[\pick{1.5ex}{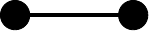}]{Davide Gaiotto,}
\author[\pic{1.2ex}{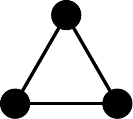},\pic{1ex}{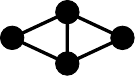}]{Justin Kulp,}
\author[\pic{1ex}{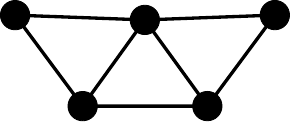}]{and Jingxiang Wu}

\affiliation[\pick{1.5ex}{segment}]{Perimeter Institute for Theoretical Physics, Waterloo, ON N2L 2Y5, Canada}
\affiliation[\pic{1.2ex}{triangle}]{Simons Center for Geometry and Physics, Stony Brook University, Stony Brook, NY 11794, USA}
\affiliation[\pic{1.2ex}{bitriangle}]{C. N. Yang Institute for Theoretical Physics, Stony Brook University, Stony Brook, NY 11794, USA}
\affiliation[\pic{1ex}{tritriangle}]{Mathematical Institute, University of Oxford, Andrew-Wiles Building, Woodstock Road, Oxford, OX2 6GG, UK}

\emailAdd{dgaiotto@perimeterinstitute.ca}
\emailAdd{jkulp@scgp.stonybrook.edu}
\emailAdd{Jingxiang.Wu@maths.ox.ac.uk}

\title{Higher Operations in Perturbation Theory}
\abstract{We discuss the role of formal deformation theory in quantum field theories and present various ``higher operations'' which control their deformations, (generalized) OPEs, and anomalies. Particular attention is paid to holomorphic-topological theories where we systematically describe and regularize the Feynman diagrams which compute these higher operations in free and perturbative scenarios, including examples with defects. We prove geometrically that the resulting higher operations satisfy expected ``quadratic axioms,'' which can be interpreted physically as a form of Wess-Zumino consistency condition for BRST symmetry. We discuss a higher-dimensional analogue of Kontsevich's formality theorem, which proves the absence of perturbative corrections in TQFTs with two or more topological directions. We discuss at some length the relation of our work to the theory of factorization algebras and provide an introduction to the subject for physicists.}

\begin{document}
\maketitle

\section{Introduction}
The main objective of this paper is to give a systematic treatment of perturbative calculations in Holomorphic-Topological Quantum Field Theories (HT QFTs) in flat spacetime. In this paper, the term ``holomorphic-topological'' is used to mean that flat spacetime is given the structure of $\bC^H \times \bR^T$ and both anti-holomorphic translations in $\bC^H$ and translations in $\bR^T$ effectively become gauge symmetries.

Introducing adapted coordinates $(x^\bC, \bar x^\bC, x^\bR)$, the Lagrangian of such HT-theories can be written in a universal form in a BV superspace formalism, with odd superspace coordinates identified with the forms $d x^\bR$, $d \bar x^\bC$:
\begin{equation}\label{eq:HTaction}
    \int \left[ (\Phi, \dd \Phi) + {\cal I}(\Phi) \right] \, d^H\! x^\bC\,.
\end{equation}
Here $\Phi(x, d x^\bR, d \bar x^\bC)$ denotes some collection of superfields, ${\cal I}(\Phi)$ is an interaction term, and we have introduced the mixed Dolbeault-de Rham differential 
\begin{equation}
    \dd = d x^\bR \frac{\partial}{\partial x^\bR}+ d \bar x^\bC \frac{\partial}{\partial \bar x^\bC}\,.
\end{equation}

A common scenario for such theories to arise is in twisted Supersymmetric Quantum Field Theories (SQFTs). Twisted SQFTs are obtained by adding a nilpotent supercharge to the BRST charge of the SQFT \cite{Witten:1988ze} (see also \cite{Johansen:1994aw, Witten:1994ev, nik, LMNS, CostelloHol, Eager:2018oww, Saberi:2019fkq, Elliott:2020ecf}). The resulting quantum field theory captures certain protected properties of the original SQFT and may have other independent applications (see e.g. \cite{Chang:2013fba, Costello:2018zrm, Chang:2022mjp, Budzik:2023vtr, Chang:2023ywj, Choi:2023vdm, Chang:2024zqi} for applications).\footnote{Traditionally, the theory is first modified to make the supercharge a spacetime scalar. This is immaterial in flat spacetime. Even in more general spacetimes, we find it much simpler to first twist the theory and then determine which manifolds it can be placed on. See Appendix F of \cite{Budzik:2023xbr} for a detailed review.} If the original physical SQFT has a Lagrangian description, twisting usually allows for drastic simplifications of the field content.\footnote{Interestingly, in the twisted scenario, the kinetic term $(\Phi,\dd\Phi)$ may not arise from the kinetic term of the original untwisted theory, see \cite{Saberi:2019ghy} for some concrete examples.}

The defining data of an HT theory, and main output of a twisted theory in flat space, will be a holomorphic-topological {\it factorization algebra} \cite{CG1,CG2}, i.e. a space of local operators equipped with a product on observables which is compatible with the holomorphic-topological structure of the theory (see \cite{Costello:2023knl} for a recent simplified review or Appendix \ref{app:FactAlg}). We will discuss the general form of the HT factorization algebra data momentarily, and in greater detail in the main body of the paper, but simple examples of data in a factorization algebra include the ring structure controlling the OPE of a topological field theory, or the chiral algebra structure in a 2d holomorphic theory. The OPE structure can be enriched by adding defects of various codimension, which may include the twist of supersymmetric defects in the original theory. The defects will support their own spaces of defect local operators. We will discuss defects which can be described by a standard action built from defect superfields, generalizing (\ref{eq:HTaction}), in Section \ref{sec:DefectIntegrals}.

The OPE data of an interacting theory may receive perturbative and non-perturbative contributions. In this work we focus on perturbative contributions, which can be systematically formulated in terms of the factorization algebra data of the free theory.\footnote{Note that the definition of a factorization algebra does not require a Lagrangian description. We focus on a specific class of Lagrangian theories in order to give an explicit computation of an important part of the OPE data.} In light of this, the main objective of this paper can be rephrased as follows: \textit{formulate a perturbative computational scheme for the OPE data which is uniform in $H$ and $T$ and makes manifest all the axioms that the OPE data is expected to satisfy}. We will accomplish this by manipulating the relevant Feynman integrals to a manifestly finite form, with a simple function integrated over an intricate region called the ``operatope'' \cite{Budzik:2022mpd}. All OPE coherence axioms will follow directly from the combinatorial geometry of the operatope. 

We will illustrate our results with various examples. In particular, we will recover Kontsevich deformation quantization from boundary conditions in 2d TFTs \cite{Kontsevich:1997vb}, and gauged $\beta \gamma$ systems from 2d holomorphic theories \cite{Li:2016gcb}. We also present computational details for recent work on 4d holomorphic theories \cite{Budzik:2023xbr}. We will formulate a combinatorial conjecture which implies a non-renormalization theorem for systems with at least two topological directions. 

The generality of our formalism and concerns of space prevented us from a systematic exploration of the zoo of HT theories and defects in various dimensions. Doing so is a natural future direction of inquiry. We should highlight in particular the case of 3d Holomorphic-Topological theories \cite{Oh:2019mcg} and their boundary conditions \cite{Costello:2020ndc}, which we expect is related to the problem of deformation quantization of Poisson vertex algebras. The lack of a non-renormalization theorem for $T=1$ suggests that the perturbative calculations in this paper will be relevant to describe obstructions to the deformation problem.

The results of the paper will be organized in the form of multilinear operations (``brackets'') on the local operators of the free theory. The brackets systematically organize a lot of information: pieces of the operator product expansion, anomalies, deformations of the space of local operators due to interactions and more. Most generally, the brackets are a device for tracking the effect of and obstructions to deformations of a QFT, even beyond the HT scenario. Related operations have appeared earlier in the context of descent operations in topological \cite{Witten:1992yj, Lian:1992mn, Getzler:1994yd} and holomorphic \cite{Beem:2018fng, Oh:2019mcg, Garner:2022its} twists. Mathematically, the brackets are homotopical objects which naturally arise in a BV-BRST formulation of QFT (see Section 4 of \cite{Douglas:2012tca} for a great discussion anticipating these structures). We speculate that the brackets we define and compute capture a large part of the perturbative factorization algebra data of twisted theories.

\subsection{The Many Faces of Factorization Algebras}\label{sec:ManyFaces}
The OPE data in HT factorization algebras is often presented in the language of descent relations. For simplicity, we focus first on the topological case, where derivatives of local operators are BRST exact. Identifying each differential in a topological direction $dx^{\bbR}$ as a Grassmann odd superspace coordinate $\theta$ allows one to organize local operators into ``superfields'' (see e.g. \cite{Axelrod:1991vq}), which are formal sums of operator-valued forms
\begin{equation}
    \calO[\theta] := e^{\theta \cdot Q} \calO^{(0)} = \calO^{(0)} + \calO^{(1)} + \calO^{(2)} + \dots \,
\end{equation}
satisfying the \textit{descent relations}
\begin{equation}\label{eq:descentrelations}
    Q_\BRST \cO^{(k)} + \dd \cO^{(k-1)} = (Q_\BRST \calO)^{(k)}\,,
\end{equation}
with $\dd$ the de Rham differential in this purely topological case. 

The descent relation allows one to define integrated observables which are BRST closed up to integration by parts. In particular, the OPE/factorization algebra data contains expressions of the form
\begin{equation}
    \{ \cO_1, \dots, \cO_n \}_{\gamma_n} = \int_{\gamma_n \subset \mathrm{Conf}_n} \cO_1(x_1) \cdots \cO_n(x_n)\,,
\end{equation}
where $\gamma_n$ is a closed integration contour in the configuration space $\mathrm{Conf}_n$ of $n$ distinct points in $\bR^T$. Although this expression is non-local, deformations of $\gamma_n$ only change the answer by a BRST-exact quantity, hence we can identify it with a local operator up to BRST-exact corrections.\footnote{The identification of such ``non-local'' expressions with local operators can be improved by a careful use of scale transformations and renormalization. In Section \ref{sec:PointSplittingRevisted} we accordingly enlarge the superspace formalism to include superpartners for both translations and scale transformations. The definition of $\mathrm{Conf}_n$ can be extended accordingly. The discussion in this section should be understood as an intuitive presentation of a more careful procedure described in Section \ref{sec:PointSplittingRevisted}.} At the level of BRST cohomology, these operations are labelled by integration contours, which are homology classes of cycles in $\mathrm{Conf}_n$. With some effort, one can compute that homology and present a full collection of possible operations. Unless $\gamma$ is a point, the integral expression above employs descendants $\cO^{(k>0)}$ and is sometimes denoted as a ``secondary operation'' in the literature \cite{Beem:2018fng, Oh:2019mcg, Garner:2022its}.

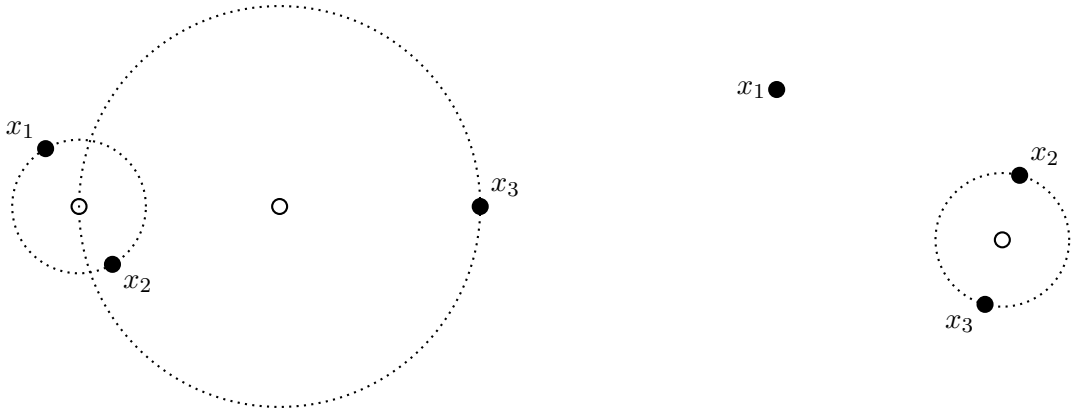
\begin{figure}[t]
\centering
\begin{minipage}{0.45\textwidth}
\centering
\begin{tikzpicture}[thick, baseline={(current bounding box.center)}]
    \def\gS{1/3};
    \def\gR{8};
    \def\th{60};
    \def\phi{45};

    \coordinate (c1) at (0,0);
    \coordinate (c2) at (-\gS*\gR,0);
    \draw[dotted] (c1) circle (\gS*\gR);
    \draw[dotted] (c2) circle (\gS*\gS*\gR);
    \draw (c1) node[GraphNodeH] {};
    \draw (c2) node[GraphNodeH] {};

    \coordinate (x1) at ({-\gS*\gR+\gS*\gS*\gR*cos(120)},{\gS*\gS*\gR*sin(120)});
    \coordinate (x2) at ({-\gS*\gR+\gS*\gS*\gR*cos(120+180)},{\gS*\gS*\gR*sin(120+180)});
    \coordinate (x3) at (\gS*\gR,0);

    \draw (x1) node[GraphNode] {};
    \node[above left] at (x1) {$x_1$};
    \draw (x2) node[GraphNode] {};
    \node[below right] at (x2) {$x_2$};
    \draw (x3) node[GraphNode] {};
    \node[above right] at (x3) {$x_3$};
\end{tikzpicture}
\end{minipage}
\hspace{1.25cm}
\begin{minipage}{0.45\textwidth}
\centering
\begin{tikzpicture}[thick, baseline={(current bounding box.center)}]
    \def\gS{1/3};
    \def\gR{8};

    \coordinate (c1) at (0,0);
    \draw[dotted] (c1) circle (\gS*\gS*\gR);
    \draw (c1) node[GraphNodeH] {};

    \coordinate (x1) at ({-3},{2});
    \coordinate (x2) at ({\gS*\gS*\gR*cos(75)},{\gS*\gS*\gR*sin(75)});
    \coordinate (x3) at ({\gS*\gS*\gR*cos(75+180)},{\gS*\gS*\gR*sin(75+180)});

    \draw (x1) node[GraphNode] {};
    \node[left] at (x1) {$x_1$};
    \draw (x2) node[GraphNode] {};
    \node[above right] at (x2) {$x_2$};
    \draw (x3) node[GraphNode] {};
    \node[below left] at (x3) {$x_3$};
\end{tikzpicture}
\end{minipage}
\caption{Left, a cycle in the configuration space $\mathrm{Conf}_3(\bbR^2)$ corresponding to a 2-bracket of $\calO_1$ and $\calO_2$ concatenated with a 2-bracket with $\calO_3$. Right, a cycle in the same configuration space of 3-points corresponding to a 2-bracket of $\calO_2$ and $\calO_3$; $\calO_1$ is fixed ``at infinity.''}
\label{fig:Brackets}
\end{figure}

These operations can be naturally composed. Up to some contour deformations, we can focus on configurations of $n+m-1$ points 
where $m$ points are close to each other. This allows one to combine a path $\gamma$ in the configuration space of $n$ points with a 
small copy of a path $\gamma'$ in the configuration space of $m$ points, identified with the $k$'th point out of $n$, to give a path $\gamma \circ_k \gamma'$ in $\mathrm{Conf}_{n+m-1}$. An example is depicted in Figure \ref{fig:Brackets}. Accordingly, 
\begin{equation}
    \{ \cO_1, \cdots,\{ \cO'_1, \cdots, \cO'_m \}_{\gamma'},\cdots \cO_n \}_{\gamma} = \{ \cO_1, \cdots, \cO'_1, \cdots, \cO'_m ,\cdots \cO_n \}_{\gamma \circ_k \gamma'}\,.
\end{equation}
If we have enough control over the homology of $\mathrm{Conf}_{n+m-1}$, we can find linear relations between compositions $\gamma_i \circ_k \gamma_j$ and derive quadratic identities satisfied by the associated operations. In Appendix \ref{sec:operadsAndQFT} we recount some of the mathematical underpinnings of this structure.

With some extra work, one can also learn how to express the operations for the product of two theories in terms of operations for the individual theories.\footnote{This notion includes coupling to background fields.} This requires decomposing a diagonal copy of $\gamma$ in $\mathrm{Conf}_n \times \mathrm{Conf}_n$ into our favourite basis. This type of definition can be readily extended to situations with defects, etc. 

In practice, identifying and describing the correct integration cycles in $\mathrm{Conf}_n$ is cumbersome. Furthermore, all the BRST-exact terms one naively disregards in contour deformations must be carefully accounted for in order to define the OPE data at the level of the BRST complex, rather than on the cohomology only. Finally, the physical meaning of the OPE data associated with some arbitrary contour $\gamma$ is a bit obscure. The complexity increases further for HT theories, where the integrand can include a non-trivial $\dd$-closed form defined in a neighbourhood of $\gamma$.\footnote{For topological theories, the insertion of a closed form can be traded for a change of integration contour. In HT theories, this is not the case. For example, consider the OPE data of a 2d holomorphic theory 
    \begin{equation*}
        \oint dz\, z^n \cO_1(z) \cO_2(0)\,,
    \end{equation*}
Here $dz\,z^n$ is the ``extra closed form''. We cannot change $n$ by changing the integration contour.
}

A way out of this thicket is to focus on OPE data which has a precise and transparent physical meaning. As we review at length in the main body of the paper, some specific OPE data controls BRST anomalies which may occur as we deform the theory or couple it to auxiliary degrees of freedom. This data satisfies simple quadratic relations which are a form of Wess-Zumino consistency conditions. We suspect that this ``BRST anomaly data'' may well capture the full information of an HT factorization algebra, though we have not analyzed this question in depth.\footnote{A skeptical reader with a string theory background may recall how the beta functions of the 2d QFT describing the motion of a string are encoded into the worldsheet BRST anomaly. In general, a great amount of information about a QFT can be extracted from the anomalies which appear when the QFT is coupled perturbatively to some exotic gauge fields. Even classically, the BV formalism encodes the equations of motion in the BRST variation of anti-fields.}

Practically, we can restate the main point of the paper as follows: \textit{formulate fully regularized Feynman diagrams which compute the perturbative BRST anomaly data of a theory of the form (\ref{eq:HTaction}) and verify that they satisfy the expected quadratic consistency conditions}. As we will see, the geometric OPE data is related to BRST anomaly data computed in a point-splitting regularization scheme, which is available for abstract theories. We will instead work with a regularization scheme suitable for perturbation around a free theory. 

\subsection{Outline of the Paper}
In Section \ref{sec:QFT} we outline the structure of the BRST anomaly brackets in general QFTs. We argue that such homotopical structures naturally arise in quantum field theory, and help to explain the deformation theory of QFTs, and connect to familiar ideas of anomalies.

In Section \ref{sec:fey} we specialize to the HT theories of interest and discuss the $n$-Laman conditions that define all relevant Feynman diagrams. We provide a regularization scheme for these Feynman diagrams, and show that they satisfy ``quadratic relations,'' which hold diagram-by-diagram. After the general analysis, we turn to topological quantum mechanics and 2d chiral algebras as concrete examples (especially in preparation for Section \ref{sec:brackets}). We end the section with a different perspective on our Feynman integrands and provide evidence of a non-renormalization theorem for HT theories with $T \geq 2$ topological directions. 

In Section \ref{sec:brackets} we assemble the Feynman diagrams from the previous section into BRST brackets and describe how computations are actually performed (especially in perturbation theory around free theories). We include examples in topological quantum mechanics, where we make contact with techniques from deformation quantization and the Moyal product; chiral algebras, where we encounter the $\lambda$-bracket; as well as 2-complex dimensional scenarios, which connects to earlier works on twists of 4d SQFTs.

In Section \ref{sec:DefectIntegrals} we repeat our previous analysis for anomalies supported at defects. We include a number of examples, and connect BRST anomaly brackets to the familiar $A_\infty$ and $L_\infty$-algebra structures (and their generalizations) that show up in QFTs with topological defects. We briefly discuss the generalization to boundaries and conclude the section with a general argument that HT theories with $T \geq 2$ topological directions receive no perturbative loop corrections.

Finally, in Section \ref{sec:PointSplittingRevisted}, we analyze BRST anomalies in a point-splitting perspective and derive a geometric form for the BRST anomaly brackets, with focus on the topological scenario for simplicity.

We also include a number of appendices, in Appendix \ref{app:VFCalcs} we include some details for Section \ref{sec:QFT}. In Appendix \ref{app:SchwingerParametrization} we provide our conventions/details for the Schwinger parametrization of propagators and regularized propagators that underpins Section \ref{sec:fey}. In Appendix \ref{app:Associativity} we provide some more detailed derivations of associativity relations for brackets from the quadratic identities on graphs. And, lastly, in Appendix \ref{sec:operadsAndQFT} we provide a review for physicists of the operad and factorization algebra mathematics that underlie this work.

\section{Brackets and QFTs} \label{sec:QFT}
As we describe the OPE data captured by holomorphic-topological factorization algebras, we will encounter rather forbidding-looking algebraic structures such as $L_\infty$-algebras and various generalizations. Although this fact is poorly advertised, these structures are actually an integral part of \textit{any QFT} which is defined in a BV-BRST formalism \cite{Hohm:2017pnh, CG1, CG2, Jurco:2018sby}. Accordingly, we will begin our discussion in great generality and then specialize to the case of theories which are Holomorphic/Topological in Section \ref{sec:fey}.

One of the most important properties of QFTs is that they can be deformed. Indeed, essentially every observable in a quantum field theory can be understood as the result of turning on some coupling between the QFT and an external system. Most quantum field theories belong to families parameterized by the values of some couplings and can thus be deformed along the family. Even in the absence of finite deformations, we can always consider an infinite number of formal deformations, essentially by adding some local operators to the Lagrangian with infinitesimal couplings and working perturbatively in the couplings. This is a crucial tool, e.g. in effective field theory. If the couplings are position-dependent, such formal deformations give generating functions for the correlation functions of local operators in the theory. Many important objects in QFT, such as the stress tensor or conserved currents, are precisely defined as the response of the theory to the deformation of some (position-dependent) coupling, e.g. the background metric of spacetime or the background connection for a global symmetry. 

Accordingly, it is useful to define a \textit{formal neighbourhood} of a given QFT ${\cal T}$ as the space of all effective QFTs which can be defined as a perturbative deformation of ${\cal T}$. We will parameterize the deformation by a collection of couplings $g^i$, with each quantity in the deformed theory being a formal power series in $g^i$. The precise definition of the $g^i$ is scheme-dependent, but different definitions will be related by redefinitions $g^i \to \widetilde g^i(g)$ which are also formal power series.\footnote{The term ``formal neighbourhood'' refers to the fact that the deformations only parametrize an infinitesimal neighbourhood of the point $\calT$ as opposed to a larger ``open neighbourhood'' of $\calT$ in theory space. This is analogous to how a formal Taylor series (generically) has 0 radius of convergence, as we expect in perturbation theory. Together, the point $\calT$ and its formal neighbourhood of deformations define a ``formal pointed manifold.'' ``Pointed'' because $\calT$ is a distinguished base point where the coordinates $g^i$ are all 0. Indeed, the simplest examples of formal pointed manifolds are dual spaces of formal Taylor series on finite dimensional vector spaces \cite{Kontsevich:1997vb}.} We will find it useful to introduce formal couplings for fermionic operators as well, so that some $g^i$ will be Grassmann odd. Mathematically, the 
formal neighbourhood of a QFT ${\cal T}$ is a {\it formal pointed supermanifold} $\cD[\cT]$.

The formal neighbourhood $\cD[\cT]$ inherits all the symmetries of $\cT$ as actions on the space of couplings. A notable example of this action comes from the RG flow: an infinitesimal scale transformation can be traded for a change of the coupling, described by the $\bm\beta$ function:
\begin{equation}
   \bm{\beta} = \sum_i \beta^i(g) \frac{\partial}{\partial g^i}\,.
\end{equation} 
The $\bm\beta$ function is a vector field describing the RG flow within the space of formal deformations of a (typically free) scale-invariant theory. Deformations of $\cT$ which preserve scale invariance are recognizable as zeroes of $\bm{\beta}$. This property is not unique to scale transformations. The existence of vector-fields associated to other transformations often goes unremarked. Another notable example is the action of anomalous axial transformations on the $\theta$ angles of a gauge theory.

Next, we focus on a theory $\cT$ which is described in a BRST formalism in terms of an ambient theory $\cT_{\BRST}$ equipped with a Grassmann odd nilpotent symmetry $Q_{\BRST}$, such that observables in $\cT$ are given by the $Q_{\BRST}$ cohomology of observables in $\cT_{\BRST}$. 
In order to deform $\cT$, we need to deform the ambient theory $\cT_{\BRST}$ without breaking the BRST symmetry. A convenient way to express that is to consider the formal neighbourhood $\cD[\cT_{\BRST}]$ and look at the action of the BRST symmetry on the couplings. This will be given by an odd nilpotent vector field on $\cD[\cT_{\BRST}]$:
\begin{equation}\label{eq:Eta}
    \bm{\eta} = \sum_i \eta^i(g) \frac{\partial}{\partial g^i}\,.
\end{equation}
such that consistent deformations of $\cT$ are zeroes of $\bm{\eta}$ (i.e. simultaneous zeroes of the $\eta^i(g)$ or ``eta functions''). Mathematically, this describes the 
formal neighbourhood of ${\cal T}$ as a {\it formal pointed dg-supermanifold} $\cD[\cT]$.

The vector field $\bm{\eta}$ thus quantifies the BRST anomalies which may arise as we deform $\cT$. The observation that $\bm{\eta}$ is nilpotent, i.e. that
\begin{equation}
    \bm{\eta}^2 = \sum_j \left[\sum_i \eta^i(g) \frac{\partial}{\partial g^i} \eta^j(g) \right]\frac{\partial}{\partial g^j}=0 \label{eq:WZConsistency}
\end{equation}
will be key in the following. It can be thought of as a Wess-Zumino consistency condition for the BRST symmetry \cite{WessZumino}, see Appendix \ref{app:eta2} for details.

The analogy with the $\bm\beta$ function is instructive. Many deformations are either relevant or irrelevant and explicitly break scale symmetry at the leading order, i.e.
\begin{equation}
    \beta^i(g) = (d-\Delta_{i}) g^i+ \calO(g^2)\,.
\end{equation}
For this reason, we tune all relevant terms to $0$, then study the $\bm\beta$-function as a measure of what happens to classically marginal deformations due to ``quantum corrections.'' Analogously, BRST anomalies can be generated at the linear order by operators which explicitly are not BRST closed. Just like the case of scale invariance, deformations of the theory which do preserve the BRST symmetry at the linear order in perturbation theory may fail to do so at higher order. We will see this explicitly in examples in Sections \ref{sec:PointSplitting} and \ref{sec:BRSTLocal}. 

If we expand 
\begin{equation}\label{eq:expansionG}
\eta^i(g) = \sum_{n>0} \frac{1}{n!}\sum_{j_1\cdots j_n} \eta^i_{j_1 \cdots  j_n} g^{j_1} \cdots g^{j_n}\,,
\end{equation}
then nilpotency becomes a collection of quadratic constraints on the $\eta^i_{j_1 \cdots  j_n}$ coefficients. Essentially by definition, these are the same relations which must be satisfied for the $\eta^i_{j_1 \cdots  j_n}$ to define the structure constants of an $L_\infty$-algebra. More precisely, we can associate to each coupling $g^i$ a Lagrangian density $\cI_i$, defined up to total derivatives, so that the theory is deformed by\footnote{The Lagrangian density will always come equipped with a volume form so that integrals are always performed over all of $\bbR^d$ or many copies of $\bbR^d$ at higher orders in perturbation theory.}
\begin{equation}
    \int_{\bbR^d} \cI\,,
\end{equation}
with
\begin{equation}
    {\cal I} := \sum_i g^i {\cal I}_{i}\,.
\end{equation}

Denote the space of gauge invariant local operators as $\mathrm{Op}$, and the deformations parameterized by the $g^i$ by $\mathrm{In}$. Schematically, $\mathrm{In} \sim \mathrm{Op}/\dd\mathrm{Op}$, since the interactions are only defined up to total derivatives.\footnote{More precisely, $\mathrm{In}$ is a top cohomology of $Q + \dd$ acting on form-valued local operators. Lower dimensional forms control defect couplings. As we will mention momentarily, ghost zeromodes are also treated differently in $\mathrm{In}$ and in $\mathrm{Op}$.} We can write the BRST variation as
\begin{empheq}[box={\mymath[colback=gray!10, sharp corners]}]{equation}
   \bm{\eta}\calI = \sum_i \eta^i(g) {\cal I}_{i} = \{{\cal I}\} + \frac12 \{{\cal I},{\cal I}\}+ \frac16 \{{\cal I},{\cal I},{\cal I}\} + \cdots \,, 
   \label{eq:MCE}
\end{empheq}
where the brackets $\mathrm{In}^{\otimes n} \to \mathrm{In}$ are defined so that 
\begin{equation}
    \{g^{j_1} {\cal I}_{j_1}, \dots, g^{j_n} {\cal I}_{j_n} \} = \eta^i_{j_1 \cdots  j_n} g^{j_1} \cdots g^{j_n} {\cal I}_{i}\,.
\end{equation}
By this identification (see Section 4.3 of \cite{Kontsevich:1997vb}),  
\begin{empheq}[box={\mymath[colback=gray!10, sharp corners]}]{equation}
    \bm{\eta}^2=0 \quad\Leftrightarrow\quad \substack{\text{\normalsize{The coefficients $\eta^{i}_{j_1\cdots j_n}$ and brackets $\{\cdot,\dots,\cdot\}$}}\\\text{\normalsize define an $L_\infty[1]$ algebra structure on $\mathrm{In}$.}} 
\end{empheq}
The condition that $\eta^i(g)=0$ is called the Maurer-Cartan equation for the $L_\infty[1]$ algebra $\mathrm{In}$. In other words, the condition that a deformed theory remain BRST-invariant is equivalent to the Maurer-Cartan equation.

We can briefly discuss the scheme-dependence of this $L_\infty$-algebra. Just as with $\bm\beta$ function coefficients, the specific functional form of $\bm{\eta}$ will depend very explicitly on the renormalization scheme we employ in the perturbative calculations. If we fix a specific collection of infinitesimal deformations, then different schemes will be related by a perturbative redefinition of the couplings which are invertible near the origin. This coordinate redefinition must intertwine the corresponding $\bm{\eta}$'s. 

More generally, different schemes may involve different choices of unphysical deformations which lead to the same space of physical deformations. In such a situation we would expect to be able to find a formal map between the space of couplings which intertwine the corresponding $\bm{\eta}$'s and reduces to a quasi-isomorphism of $L_\infty$-algebras at the linear level. \cite{Kontsevich:1997vb}. It is only a quasi-isomorphism, as opposed to an isomorphism, because the two only need to agree in $Q_{\mathrm{BRST}}$ cohomology.

Dually, the power series coefficients of such morphisms $g'(g)$ give collections of multilinear graded-symmetric maps $\mathrm{In}^{\otimes n} \to \mathrm{In}'$ which define (degree-shifted) $L_\infty$-morphisms between the $L_\infty$-algebras of interactions $\mathrm{In}$ and $\mathrm{In}'$ in the two schemes.

\subsection{Point-Splitting Example}\label{sec:PointSplitting}
From now on, we denote $Q_\BRST = Q$. In order to set up perturbation theory, it is useful to pick specific local operator representatives for the $\cI_i$, i.e. choose a scheme for dealing with total derivatives.

At the linear order in perturbation theory, the BRST anomaly generated by an interaction of the form
\begin{equation}
    \int_{\bbR^d} {\cal I}_i \,,
\end{equation} is simply\footnote{In this section we use the traditional physics convention of denoting the action of $Q$ on local operators ${\cal O}$ as a commutator $[Q,{\cal O}]$. In later sections we will just write this as $Q {\cal O}$.}
\begin{equation}
    \int_{\bbR^d} [Q,{\cal I}_i] \,.
\end{equation}
We can represent it as a deformation of the theory by some $\eta^i$ linear in $g$ by expanding $[Q,{\cal I}_i]$ in our basis:
\begin{equation}
    [Q,{\cal I}_i] = \sum_j \eta^j_i {\cal I}_j + d {\cal J}_i\,.
\end{equation}
Here we included a possible total derivative on the right hand side.

If we could find a collection of ${\cal I}_i$ representatives such that the ${\cal J}_i$ vanish, a point-splitting regularization would have no quantum BRST anomalies. For example, at the second order in perturbation theory we could regularize the deformation to 
\begin{equation}\label{eq:noJ}
    \int_{\bbR^{2d}} f_\epsilon^{(2)}(x_1,x_2) \cI(x_1) \cI(x_2)
\end{equation}
for some cutoff function $f_\epsilon^{(2)}$ that sufficiently regulates the divergence, e.g. a hard cutoff function if $\abs{x_1 - x_2} < \epsilon$. This deformation and possible counterterms would be annihilated by the action of $Q$.

Instead, when the ${\cal J}$ do not vanish, we get additional quantum corrections to the BRST anomaly. For example, the correction to the second-order deformation in \eqref{eq:noJ} due to total derivative terms is:
\begin{align}
    \left.\big[Q, \int_{\bbR^{2d}} f_\epsilon^{(2)}(x_1,x_2) \, \calI(x_1)\calI(x_2)\big]\, \right\vert_{d\cal{J}}
        &=\int_{\bbR^{2d}} f_\epsilon^{(2)}(x_1,x_2) (\cI(x_1) d\cJ(x_2)+ d\cJ(x_1) \cI(x_2))\nonumber \\
        &=-\!\int_{\bbR^{2d}} \!\!d f_\epsilon^{(2)}(x_1,x_2) (\cI(x_1) \cJ(x_2)+ \cJ(x_1) \cI(x_2))
\end{align}
where we have used the fact that $\calI$ is a top form and thus $d\calI(x) = 0$ in passing from the first line to the second line. Because $f^{(2)}_\epsilon$ is constant at large distances, this integral is localized near $x_1 = x_2$. As we remove the cutoff $\epsilon \to 0$, the BRST anomaly is thus given by some ``local operator'' $\{ {\cal I}, {\cal I}\}(x)$: 
\begin{equation}
\int_{\bbR^d} \{ {\cal I}, {\cal I}\}(x) = \int_{\bbR^{2d}} d f_\epsilon^{(2)}(x_1,x_2) \left(\cI(x_1) \cJ(x_2)+ \cJ(x_1) \cI(x_2)\right) + d \cdots 
\end{equation}

In particular, if we use a sharp cutoff regularization scheme, with $f_\epsilon(x_1,x_2) = 1$ if $\abs{x_1-x_2} > \epsilon$ and $0$ otherwise, and arbitrarily declare the ``local operator'' to be at $x_2$, then the BRST anomaly is given by an integral on a sphere of radius $\epsilon$ around $x_2$:
\begin{equation}
    \{\calI,\calI\}(x_2)
    \stackrel{\substack{\text{\tiny Sharp}\\\text{\tiny Cutoff}}}{=}
    \int_{\abs{x_{12}} = \epsilon} 
    \!\!
    \cI(x_1) \cJ(x_2)+ \cJ(x_1) \cI(x_2)\,.
\end{equation}
As previously mentioned, the bracket of two operators is (generically) scheme dependent, just like beta-function coefficients and composite operators are (generically) scheme dependent.
 
Pushing the point-splitting regularization to higher orders is cumbersome. Intuitively, we expect that it should be possible to set it up so that the resulting higher brackets would be computed as integrals of
\begin{equation}
    \sum_i {\cal J}_i(x_i) \prod_{j \neq i} {\cal I}_j(x_j) 
\end{equation}
over appropriate compact codimension 1 contours in the configuration space of distinct points in $\mathbb{R}^d$. We will return to the technical details of this construction in Section \ref{sec:PointSplittingRevisted}.

\subsubsection{Example: Anomaly in 2d Gauge Theory}\label{sec:2dGaugeTheory}
A simple example of this calculation occurs in 2d gauge theory. Consider an action
\begin{equation}
    S_{\calT} = -\frac{1}{4}\int d^2x\, F_{\mu\nu} F^{\mu\nu} + S_{\mathrm{Matter}}
\end{equation}
where $S_{\mathrm{Matter}}$ is some theory with $G$ global symmetry and symmetry current $J^\mu_a$. For example, free fermions with vector current $J^\mu_a = \bar\psi \gamma^\mu t_a \psi$. The gauge interaction term $\cI = A_\mu J^\mu$ is added to ``gauge the $G$ symmetry'' of $S_{\mathrm{Matter}}$.

Adding ghost fields in the standard way embeds $\calT$ in the ambient theory $\calT_{\BRST}$, with observables of $\calT$ identified with $Q_{\BRST}$ cohomology of observables of $\calT_{\BRST}$. Under BRST transformations, the gauge interaction $\calI$ transforms as:
\begin{equation}
    \delta(A_\mu J^\mu) = (\epsilon D_\mu c) J^\mu + A_\mu (i \epsilon g c J^\mu) = \epsilon \partial_\mu c J^\mu\,,
\end{equation}
where $c$ is the ghost field. In particular, we find that $\calI$ is BRST-closed up to the total derivative term $\cJ = c J$.

The two-bracket receives a contribution from the 2d $JJ$ OPE:
\begin{align}
    \{\calI,\calI\}(x_2)
        &=
    \int_{\abs{x_{12}} = \epsilon} 
    \nord{A J}{x_1}\nord{c J}{x_2}\, + \nord{c J}{x_1}\nord{A J}{x_2}\\
        &= \oint_{S^1_{x_2}} (\,\NORD{A(x_1) c(x_2)} + \NORD{c(x_1) A(x_2)}\,) \expval{J(x_1)J(x_2)}\\
        &= \# \,\NORD{c\,d\!A}(x_2)\,.
\end{align}
Here we have used the fact that only the $JJ \sim \abs{x_{12}}^{-2}$ contraction/OPE is non-trivial in the first line, and have Taylor expanded and integrated by parts in passing to the last line (and judiciously normalized all of our operators). The $\#$ denotes the various combinatorial and representation theoretic factors (possibly 0 overall) that weight the anomaly. Ultimately we recover the well-known 1-loop BRST anomaly proportional to $c F_{12}$. We can thus write
\begin{equation}
    \{ A_\mu J^\mu, A_\nu J^\nu \} =\# \,c F_{12}\,.
\end{equation}
In $2k$-dimensions, we expect to recover the analogous 1-loop $F^k$ anomaly from the $k+1$ bracket of $\calI$s. See \cite{Bomans:2023mkd} for an analogous computation of central charge and (higher-) Kac-Moody anomalies in 4d.

\subsection{BRST Symmetry and Local Operators}\label{sec:BRSTLocal}
Going back to the RG flow analogy, recall that the concrete manifestation of the RG flow on correlation functions is encoded in the Callan–Symanzik equation, which has two ingredients: the beta functions and the anomalous dimensions of local operators. The beta function ${\bm \beta}$ is the vector field describing the flow in ${\cal D}[\calT]$ induced by scale transformations, and the anomalous dimensions are linear transformations $\bm\gamma$ on the space of local operators of the theory, comparing the spaces of local operators along the flow. The Callan-Symanzik equation is just the statement that renormalized correlation functions are independent of the arbitrarily chosen renormalization scale $\mu$:
\begin{equation}
    \mu \frac{d}{d\mu} G^{(n)} = \left( \mu \frac{\partial}{\partial \mu} + \bm{\beta} + \bm{\gamma}\right) G^{(n)} = 0\,.
\end{equation}
Here, $\mu$ is the parameter for the flow generated by $\bm{\beta}$, and the combination $\bm{\beta} + \bm{\gamma}$ is just a decomposition of the Lie derivative $\calL_{\bm{\beta}}$ on the correlation function $G^{(n)}$ viewed as a section of a rank-$n$ bundle over $\mathcal{D}[\mathcal{T}]$ \cite{PolchinskiRG, Lassig:1989tc, Osborn:1991mk, Dolan:1994pq}.

More precisely, gauge-invariant local operators $\mathrm{Op}$ in the theory are not exactly the same as interactions $\mathrm{In}$ in the theory. For one, the latter are only defined up to total derivatives. A further difference in gauge theory is that the $c$ ghosts can only appear in local operators through their derivatives, but can appear directly in interactions as seen in the previous example.\footnote{The standard BRST quantization of a gauge theory involves a ghost field $c$ valued in the Lie algebra. If the gauge group is compact, local operators are defined by a {\it relative} BRST cohomology, where one only allows derivatives of the $c$ ghost and projects by hand onto $G$-invariant operators. Conversely, Lagrangian interactions can directly contain the $c$ ghost. These simple differences lead to some important distinctions.} 

Geometrically, $\mathrm{In}$ is the tangent bundle to $\cD[\cT]$ and no connection is required to compare vectors at different points along a flow. As operators are not quite the same as interactions, the space $\mathrm{Op}$ of local operators is a vector bundle on $\cD[\cT]$ distinct from the tangent bundle, and requires additional data to compare local operators at different points along the deformation. A symmetry of $\cT$ must be able to act simultaneously on the space of couplings and on local operators. 

In the BRST context, the action of the BRST symmetry on local operators (analogous to $\bm{\gamma}$) in the deformed theory is captured by an odd linear map ${\bf Q}$ on $\mathrm{Op}$. Nilpotency of the BRST transformations gives a relation  
\begin{equation}
    \calL_{\bm{\eta}}^2 = \{ \bm{\eta}, {\bf Q} \} + {\bf Q}^2 =0\,.
\end{equation}

Again, the coefficients $Q^i(g)$ of the power series expansion of ${\bf Q}$ can be identified as the coefficients for a tower of multilinear operations $\mathrm{In}^{\otimes n} \otimes \mathrm{Op} \to \mathrm{Op}$.
In the higher algebra language, this equips the space of local operators with the structure of a (right) $L_\infty$-module for the algebra of interactions. We thus write
\begin{equation}
    \bm{Q} \cO = \{\cO\}+ \{\cI, \cO\} + \frac12 \{\cI, \cI, \cO\} +\cdots\,.
\end{equation}
For example, the second term can be computed in a hard-cutoff point-splitting regularization as before:
\begin{equation}
\{ {\cal I}, {\cal O}\}(x_2) = \int_{|x_{12}|=\epsilon} {\cal J}(x_1) {\cal O}(x_2)\,.
\end{equation}
We find that ${\cal J}$ can be interpreted as the first order deformation of the BRST current. 

\subsection{\texorpdfstring{$p$}{p}-Brackets}\label{sec:pBracket}
The discussion up to now assumed implicitly that both the reference theory and the deformations were translation invariant. This assumption is not necessary: one could just as well consider a reference theory and interactions which break translation symmetry.

There is a particularly interesting situation when the reference theory ${\cal T}$ is translation invariant, but we allow interactions which break translation symmetry. Because of translation invariance, we can consider infinitesimal deformations by interactions with definite momentum $p$. We are interested in situations where the position dependence is infinitely slow, i.e. the couplings are supported on an infinitesimal neighbourhood of $p=0$. 

More concretely, we are considering formal deformations of the form 
\begin{equation}\label{eq:positionDependentCouplings}
    g^i_{\mu_1 \cdots \mu_k} \int_{\bbR^d} x^{\mu_1} \cdots x^{\mu_k} {\cal I}_i\,.
\end{equation}
All vector fields describing transformations are now lifted to this bigger space of couplings. In particular, the $L_\infty$-brackets corresponding to $\bm{\eta}$ now act on interactions of the form $x^{\mu_1} \cdots x^{\mu_k} {\cal I}_i$ modulo total derivatives, i.e.
\begin{equation}\label{eq:bigEta}
    \bm{\eta} = \sum_i \sum_{k} \sum_{\mu_1 \cdots \mu_k} \eta^i_{\mu_1\cdots\mu_k}(g) \frac{\partial}{\partial g^i_{\mu_1\cdots\mu_k}}\,,
\end{equation}
where the $\eta^i_{\mu_1\cdots\mu_k}(g)$ can be expanded into $L_\infty$ structure coefficients analogous to \eqref{eq:expansionG}.

As a concrete example, any theory with continuous $G$-global symmetry can be coupled to a \textit{background} gauge field $A_\mu$ through terms like $A_\mu j^\mu$, (with possible higher order terms in $A$ required). When $A_\mu$ is a background gauge field, it can be viewed as a position dependent coupling. The support on infinitesimal neighbourhoods of $p = 0$ means that the background field $A_\mu$ is essentially constant. The couplings $g_{\mu_1\cdots\mu_k}$ are just the Taylor series coefficients of $A(x)$.

The breaking of translation symmetry by position dependent interactions is described by even vector fields $\bm{P}_\mu$ which commute with $\bm{\eta}$. We expect all $\bm{P}_\mu$ to receive no quantum corrections,\footnote{Pedantically, we expect translation-invariant renormalization schemes to exist and be useful since we are deforming a translation invariant theory. This is in contrast to scale invariance, since regularization and renormalization intrinsically breaks scale symmetry.} i.e. the $g$-expansion of coefficients in $\bm{P}_{\mu}$ analogous to \eqref{eq:expansionG} contains only the linear term
\begin{equation}\label{eq:bigP}
    \bm{P}_\mu = \sum_i\sum_k \sum_{\mu_1\cdots \mu_k} g^i_{\mu \mu_1 \cdots \mu_k} \frac{\partial}{\partial g^i_{\mu_1 \cdots \mu_k}}\,,
\end{equation}
so that
\begin{equation}
    \bm{P}_\mu g^i_{\mu_1 \cdots \mu_k} = g^i_{\mu \mu_1 \cdots \mu_k}\,.
\end{equation}
Translation invariance $[\bm{P}_\mu,\bm{\eta}]=0$ fixes
\begin{equation}\label{eq:PfromEta}
    \eta^i_{\mu \mu_1 \cdots \mu_k} = \sum_\ell \sum_{\nu_1\cdots\nu_\ell} g^i_{\mu \nu_1 \cdots \nu_\ell} \frac{\partial}{\partial g^i_{\nu_1 \cdots \nu_\ell}} \eta^i_{\mu_1 \cdots \mu_k} \,,
\end{equation}
see Appendix \ref{app:posDep} for details. In the end, we find that the $\eta^i_{\mu_1\cdots \mu_k}$ for couplings with many $x$'s are determined recursively in terms of those with fewer $x$'s, hence the full information on the extended space of couplings is captured by just the $\eta^i(g)$.

It is convenient to form generating series of the form $e^{p \cdot x} {\cal I}_i(x)$, this gives
\begin{equation}
    \partial_\mu(e^{p\cdot x}\calI_i(x)) = e^{p\cdot x} (\partial_\mu+p_\mu)\calI_i(x)\,.
\end{equation}
We can view the interactions ${\cal I}_i$ as elements of deformed spaces of interactions $\mathrm{In}_p$, which are defined up to $\partial_\mu + p_\mu$ derivatives of other operators. Because of translation invariance, the BRST anomaly due to the regularization of a collection of interactions $\cI_i$ of momenta $p^{(i)}$ has momentum $\sum_i p^{(i)}$. The $L_\infty$-brackets are thus extended to multilinear brackets $\otimes_i \mathrm{In}_{p^{(i)}} \to \mathrm{In}_{\sum_i p^{(i)}}$
\begin{equation}
    \{{\cal I}_{i_1}\,  {}_{p^{(1)}} \, {\cal I}_{i_2}\,  {}_{p^{(2)}}\,  \dots \, {}_{p^{(n-1)}} \, {\cal I}_{i_n} \}\,,
\end{equation}
where the final $p^{(n)}$ is implied by momentum conservation. This momentum conservation of the multilinear operation reflects the underlying (momentum-coloured) operad algebra structure carried by the brackets, see Appendix \ref{sec:operadsAndQFT}.

We can also consider the space of local operators, placed at the origin, and compute ${\bf Q}$ for this bigger space of deformations. This will be encoded in multilinear brackets $\otimes_i \mathrm{In}_{p^{(i)}} \otimes \mathrm{Op} \to \mathrm{Op}$.

Holomorphic analogues of these ``$p$-brackets'' will play an important role in the rest of the paper. Specifically, $\lambda$-brackets are just $p$-brackets where holomorphic momentum can flow in at an interaction vertex.

\subsection{Free Fields and Brackets}\label{sec:FreeFieldsAndBrackets}
A natural setting to apply perturbation theory is in deformations of a free theory. Local operators in the free theory are normal-ordered collections of letters $\phi_a$, which includes the field and their derivatives
(either up to the equations of motion, or including anti-fields in a BV formalism).\footnote{Normal-ordering is the associative commutative prescription defining a composite operator in a free theory by subtracting all possible Wick contractions, i.e. removing UV divergences from self-contractions. This is in contrast to the ``regularized product'' which is the non-associative non-commutative prescription defining composite operators as the first regular term in an OPE. See Section 3.2 of \cite{Chang:2023ywj} for further discussion and application.} A perturbative deformation of the free theory involves the exponential of a normal-ordered interaction, which is formally rewritten as a sum of normal-ordered terms by Wick's theorem. The associated Wick contractions insert propagators
\begin{equation}
    P_{ab}(x,y) := \expval{\phi_a(x) \phi_b(y)}
\end{equation}
and give rise to Feynman diagrams which need to be regulated. 

A standard regularization procedure is to replace the propagators used in the Wick contractions with some regularized version $P_{ab}(x,y;\epsilon)$ and add counterterms, if needed, to cancel divergences as $\epsilon \to 0$. Concretely, the bare coefficients $g_0^i$ of some given normal-ordered interactions ${\cal I}_i$ will be some $\epsilon$-dependent formal power series in the actual couplings $g^i$.

Now we turn to the example of BRST symmetry. We take the BRST symmetry of the free theory to act linearly on the free letters $\phi_a$, and by the Leibniz rule on more complicated local operators built from (normal-ordered) products of free fields.\footnote{The reader is cautioned that the BRST symmetry in an interacting theory does {\it not} necessarily satisfy the Leibniz rule on the regularized products of local operators.} By assumption, the BRST-symmetry annihilates the propagator $P_{ab}(x,y)$ at separate points
\begin{equation}
    \mathscr{D}_Q P_{ab}(x,y) = 0\,,\quad x\neq y\,.
\end{equation}
Here, $\mathscr{D}_Q$ is the differential operator implementing the free BRST transformation on $\phi_a(x) \phi_b(y)$. In such a scheme, BRST anomalies will arise from a failure of the regularized propagators $P_{ab}(x,y;\epsilon)$ to be BRST invariant. We can denote the BRST variation of the regularized propagators as
\begin{equation}
    \mathscr{D}_Q P_{ab}(x,y;\epsilon) = - K_{ab}(x,y;\epsilon) \,.
\end{equation}

At zeroth-order the local BRST anomaly arises from the undeformed, i.e. free, BRST variation of the interactions ${\cal I}_i$. We write again 
\begin{equation}
    [Q,{\cal I}_i] = \sum_j \eta^j_i {\cal I}_j + d {\cal J}_i \,.
\end{equation}
The BRST variation of a given Feynman diagram in the expansion of the exponentiated interaction is a sum over the action on the $\phi_a$ that survive the Wick contractions. Up to total derivatives that do not contribute local terms, a sum over all the ways to replace an interaction vertex with $[Q,{\cal I}]$ cancels these terms and replaces them with a sum over all possible ways to act with the BRST current on the Wick-contracted letters, i.e. a sum over all possible way to replace a single propagator $P_{ab}(x,y;\epsilon)$ with $K_{ab}(x,y;\epsilon)$. For example, the tree-level contribution to $\{{\cal I}_i, {\cal I}_j\}$ comes from a single Wick contraction with propagator $K_{ab}(x,y;\epsilon)$. 

\subsection{Further Structures and Defect Brackets} \label{sec:furtherStructures}
The ``BRST flow'' data comes with some useful extra structure. Given two theories ${\cal T}_1$ and ${\cal T}_2$, we can build a composite theory ${\cal T}_1 \times {\cal T}_2$. The space of couplings of the combined theory is, to first approximation, the tensor product of the space of couplings of the two original theories (this is not exactly true because of subtleties concerning total derivatives). The flow vector ${\bm \eta}_{12}$ in this larger space contains much more information than the ${\bm \eta}_{1}$ and ${\bm \eta}_{2}$ vector fields for the individual theories, for the simple reason that one function in two variables has more information than two functions in one variable.

We can use this fact: given a theory ${\cal T}_1$, we can extract information about it by coupling it with an auxiliary theory ${\cal T}_2$ and computing the BRST anomaly brackets. The extra information may be non-trivial even if the auxiliary theory 
is free. An 't Hooft anomaly for ${\cal T}_1$ is a natural example of such extra information: the anomaly may be captured by tensoring $\calT_1$ with a free gauge theory ${\cal T}_2$, and turning on a gauge interaction, leading to a BRST anomaly in the combined theory. 

We saw an example of this in the 2d gauge theory example of Section \ref{sec:2dGaugeTheory}. The reference theory $\calT_1$ that we want to study is the ``matter theory'' with action $S_{\mathrm{Matter}}$ and $G$ global symmetry. In order to detect the 't Hooft anomaly of $\calT_1$, we couple it to a free $G$ gauge theory $\calT_2$ by the usual $\calI = J\wedge A$ interaction. The 't Hooft anomaly for the $G$ global symmetry of $\calT_1$ (an intrinsic property of $\calT_1$ alone), is manifest from the non-trivial BRST anomaly bracket $\{\calI,\calI\} = c F_{12}$ of $\calT_1\times \calT_2$. The only way the BRST anomaly bracket will vanish is if the combinatorial/representation theoretic factors weighting $\{\calI,\calI\}$ are judiciously chosen to vanish, in which case, the theory $\calT_1$ has no $G$ 't Hooft anomaly.

The discussion in this section can apply as well to the deformation of theories with defects. Defects are simply probes or external systems of higher codimension. By locality, the presence of a defect does not affect the space of bulk couplings and the bulk BRST anomalies. It introduces new defect couplings and potential BRST anomalies localized at the defect. This is analogous to the situation in DCFT where bulk 3-point functions remain unchanged, but defects introduce new defect and mixed bulk-defect OPEs/structure constants.

A simple way (but not the only way!) to produce a defect is to couple some lower-dimensional theory $\cT_\partial$ to the bulk theory $\cT$. As in the previous (codimension-0) case, coupling theories of different dimensions and computing associated mixed BRST anomalies can be used as way to probe one of the two theories. 

Geometrically, we can think of the formal space of defect deformations as being fibered over the formal space of deformations of the underlying theory. The BRST vector field $\bf{\eta}$ of the bulk is lifted to an odd nilpotent vector field on the total space of couplings. Dually, one gets a variety of multilinear operations which take as inputs various numbers of bulk and defect interactions/local operators and output interactions/local operators on the defect. The structure extends in a hierarchical manner if we consider defects-for-defects, etc. We compute examples of such defect brackets in Section \ref{sec:DefectIntegrals}.

\section{Feynman Diagrams in Holomorphic-Topological twists}\label{sec:fey}
In our companion paper \cite{Budzik:2022mpd}, we formulate and study Feynman integrals which we will now employ in the calculation of the factorization algebra structure on local operators in free holomorphic-topological theories with a first order action. 

\subsection{Superspace Notation}
We consider theories on flat spacetime with the structure of $\bC^H \times \bR^T$, with coordinates $(x^\bC, \bar x^\bC, x^\bR)$, corresponding to holomorphic $x^\bC$, anti-holomorphic $\bar x^\bC$, and topological $x^\bR$ directions respectively. We further identify the forms $d \bar x^\bC$ and $d x^\bR$ with Grassmann odd superspace coordinates. We denote the total set of coordinates on this superspace by:
\begin{equation}
    x := (x^\bC, \bar x^\bC, x^\bR, d \bar x^\bC, d x^\bR)\,.
\end{equation}
The number of holomorphic and topological directions will be given by $H$ and $T$ respectively. In this notation, a function $f(x)$ is actually a sum of forms in spacetime, built from $d \bar x^\bC$ and $d x^\bR$ differentials only. We will refer to $f(x)$ as a $(0,*)$-form, and may also consider $(0,*)$-forms valued in holomorphic bundles. 

We will often employ formal holomorphic shifts of the coordinates, typically denoted by $z$ with some subscripts, so that 
\begin{equation}
    f(x+z) \define \sum_{n_*=0}^\infty \prod_{a=1}^H \left[\frac{(z^a)^{n_a}}{n_a!} \partial_{(x^\bC)^a}^{n_a} \right] f(x)
\end{equation}
is a generating function for holomorphic derivatives of $f(x)$.

In order to integrate a top form on spacetime, i.e. a $(0,H+T)$-form, we need to include a holomorphic volume form:
\begin{equation}
    \dvol = \prod_{a=1}^H \frac{d (x^\bC)^a}{2 \pi i} \, .
\end{equation}
We orient spacetime ${\cal M} = \bR^T \times \bC^H$ so that the following normalization holds:
\begin{equation}
   \int_{\cal M} 
        \pi^{-\frac{T}{2}} 
        e^{- x^\bR \cdot x^\bR - x^\bC \cdot \bar x^\bC} 
        \,\dvol 
        \,\prod_{a=1}^H d (\bar x^\bC)^a 
        \,\prod_{b=1}^T d ( x^\bR)^b = 1 \, .
\end{equation}

We further introduce a holomorphic-topological differential
\begin{equation}
    \dd = d x^\bR \frac{\partial}{\partial x^\bR}+ d \bar x^\bC \frac{\partial}{\partial \bar x^\bC}\,,
\end{equation}
which enters the kinetic terms of the holomorphic-topological theory. A choice of Green's function for $\dd$ will be denoted as $P(x)$, and a UV-regulated version of the Green's function will be denoted by $P_{\epsilon}(x)$. We discuss our choice in Appendix \ref{app:SchwingerParametrization}. The Feynman diagrams which arise in BRST anomaly calculations will involve the regulated propagator $P_{\epsilon}(x)$ on all edges except one, which involves 

\begin{equation}
    K_\epsilon(x) := \dd P_{\epsilon}(x)\,,
\end{equation}
as seen previously in Section \ref{sec:FreeFieldsAndBrackets}. In Section \ref{sec:FeynmanIntegrals} we will define universal generating functions for such ``cut'' Feynman diagrams.

\subsection{Feynman Diagrams in Holomorphic-Topological Field Theories}
Our objective is to study theories of the general form
\begin{equation}
    \int \left[ (\Phi, \dd \Phi) + {\cal I}(\Phi) \right] \dvol
\end{equation}
in perturbation theory, with $\Phi$ denoting some collection of superfields. In particular, we want to systematically compute all the possible brackets we discussed in the previous section, but specialized to interactions and operators which are polynomials in superfields and are thus superfields themselves. 

In the free theory, every BRST closed superfield satisfies a descent relation:
\begin{equation}
    Q\cO + \dd \cO = 0\,.
\end{equation}
The actual interaction density is the $(0,H+T)$ part of the interaction ${\cal I}(\Phi)$. It is obviously BRST closed up to the total derivative $\dd$ of the $(0,H+T-1)$ part of the superfield. 

We express inputs and outputs of the operations we compute in terms of the full superfields, leading to a degree shift in the conventions for the brackets. So, for example, we may write 
\begin{equation}
    \{\cI_1, \cI_2\} 
\end{equation}
to denote the superfield whose $(0,H+T)$ component appears in the BRST anomaly associated to the $(0,H+T)$ components of $\cI_1$ and $\cI_2$. If $\cI_i$ has ghost number $n_i$, the $(0,H+T)$ part of $\cI_i$ will have ghost number $n_i - H - T$. As a result, the associated BRST anomaly of $\{\cI_1, \cI_2\}$ will have ghost number $n_1 + n_2 - 2 H - 2 T + 1$, but the interaction $\{\cI_1, \cI_2\}$ has ghost number $n_1 + n_2 - H - T + 1$.

\subsection{Graph Combinatorics}\label{sec:GraphCombinatorics}
As we will see, the relevant Feynman diagrams will be labelled by graphs $\Gamma$ which we will call \textit{$n$-Laman graphs}.\footnote{In this paper, graphs will not have edges joining a vertex to itself, as each vertex will represent a normal-ordered operator and Wick contractions are done between distinct operators.} The vertices and edges of $\Gamma$ will be denoted $\Gamma_0$ and $\Gamma_1$ respectively. Such $n$-Laman graph are defined by two constraints:\footnote{Note: In this notation, the ``standard'' Laman graphs that appear in the study of minimally rigid graphs in the plane \cite{henneberg1911graphische, pollaczek1927gliederung, laman1970graphs} are actually 2-Laman graphs; satisfying $2\abs{\Gamma_0} = \abs{\Gamma_1} + 3$ globally, and $2\abs{\Gamma[S]_0} \geq \abs{\Gamma[S]_1} + 3$ locally (see also \cite{Gaiotto:2015zna, Budzik:2022mpd}).} 
\begin{enumerate}
    \item \textbf{Global Constraint.} An $n$-Laman graph must have at least two vertices and satisfy the global constraint
        \begin{equation}\label{eq:globalconstraint}
            n |\Gamma_0| = (n-1) |\Gamma_1| +n+1\,,
        \end{equation}
    \item \textbf{Local Constraint.} For subgraphs $\Gamma[S]$ induced by subsets of vertices $S \subset \Gamma_0$ containing at least two vertices
        \begin{equation}
            n \abs{\Gamma[S]_0} \geq (n-1) \abs{\Gamma[S]_1}+n+1\,.
        \end{equation}
\end{enumerate}

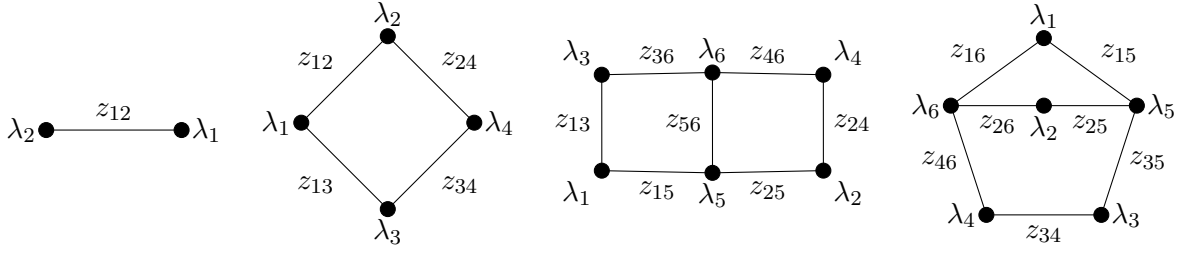
\begin{figure}
\centering
\setlength\tabcolsep{4pt} 
\renewcommand{\arraystretch}{3} 
\begin{tabular}{cccc}
   \begin{tikzpicture}
	[
	baseline={(current bounding box.center)},
	line join=round
	]
        \def\gS{0.9};
	\coordinate (pd1) at (1.*\gS,0.*\gS);
	\coordinate (pd2) at (-1.*\gS,0.*\gS);
	\draw (pd1) node[GraphNode] {} node[right] {$\lambda_{1}$};
	\draw (pd2) node[GraphNode] {} node[left] {$\lambda_{2}$};
	\draw[GraphEdge] (pd1) -- (pd2) node[midway, above] {$z_{12}$};
\end{tikzpicture}
&
    \begin{tikzpicture}
	[
	baseline={(current bounding box.center)},
	line join=round
	]
        \def\gS{1.15};
	\coordinate (pd1) at (-1.*\gS,0.*\gS);
	\coordinate (pd2) at (0.*\gS,1.*\gS);
	\coordinate (pd3) at (0.*\gS,-1.*\gS);
	\coordinate (pd4) at (1.*\gS,0.*\gS);
	\draw (pd1) node[GraphNode] {} node[left] {$\lambda_{1}$};
	\draw (pd2) node[GraphNode] {} node[above] {$\lambda_{2}$};
	\draw (pd3) node[GraphNode] {} node[below] {$\lambda_{3}$};
	\draw (pd4) node[GraphNode] {} node[right] {$\lambda_{4}$};
	\draw[GraphEdge] (pd1) -- (pd2) node[midway, above left] {$z_{12}$};
	\draw[GraphEdge] (pd1) -- (pd3) node[midway, below left] {$z_{13}$};
	\draw[GraphEdge] (pd2) -- (pd4) node[midway, above right] {$z_{24}$};
	\draw[GraphEdge] (pd3) -- (pd4) node[midway, below right] {$z_{34}$};
\end{tikzpicture}
&
    \begin{tikzpicture}
	[
	baseline={(current bounding box.center)},
	line join=round
	]
        \def\gS{1.40};
	\coordinate (pd1) at (0.0005*\gS,0.0207*\gS);
	\coordinate (pd2) at (2.1045*\gS,0.0211*\gS);
	\coordinate (pd3) at (0.*\gS,0.9324*\gS);
	\coordinate (pd4) at (2.104*\gS,0.9327*\gS);
	\coordinate (pd5) at (1.0516*\gS,0.*\gS);
	\coordinate (pd6) at (1.0511*\gS,0.9535*\gS);
	\draw (pd1) node[GraphNode] {} node[below left] {$\lambda_{1}$};
	\draw (pd2) node[GraphNode] {} node[below right] {$\lambda_{2}$};
	\draw (pd3) node[GraphNode] {} node[above left] {$\lambda_{3}$};
	\draw (pd4) node[GraphNode] {} node[above right] {$\lambda_{4}$};
	\draw (pd5) node[GraphNode] {} node[below] {$\lambda_{5}$};
	\draw (pd6) node[GraphNode] {} node[above] {$\lambda_{6}$};
	\draw[GraphEdge] (pd1) -- (pd3) node[midway, left] {$z_{13}$};
	\draw[GraphEdge] (pd1) -- (pd5) node[midway, below] {$z_{15}$};
	\draw[GraphEdge] (pd2) -- (pd4) node[midway, right] {$z_{24}$};
	\draw[GraphEdge] (pd2) -- (pd5) node[midway, below] {$z_{25}$};
	\draw[GraphEdge] (pd3) -- (pd6) node[midway, above] {$z_{36}$};
	\draw[GraphEdge] (pd4) -- (pd6) node[midway, above] {$z_{46}$};
	\draw[GraphEdge] (pd5) -- (pd6) node[midway, left] {$z_{56}$};
\end{tikzpicture}
&
\begin{tikzpicture}
	[
	baseline={(current bounding box.center)},
	line join=round
	]
        \def\gS{1.30};
	\coordinate (pd1) at (0.*\gS,1.*\gS);
	\coordinate (pd2) at (0.*\gS,0.309*\gS);
	\coordinate (pd3) at (0.5878*\gS,-0.809*\gS);
	\coordinate (pd4) at (-0.5878*\gS,-0.809*\gS);
	\coordinate (pd5) at (0.9511*\gS,0.309*\gS);
	\coordinate (pd6) at (-0.9511*\gS,0.309*\gS);
	\draw (pd1) node[GraphNode] {} node[above] {$\lambda_{1}$};
	\draw (pd2) node[GraphNode] {} node[below] {$\lambda_{2}$};
	\draw (pd3) node[GraphNode] {} node[right] {$\lambda_{3}$};
	\draw (pd4) node[GraphNode] {} node[left] {$\lambda_{4}$};
	\draw (pd5) node[GraphNode] {} node[right] {$\lambda_{5}$};
	\draw (pd6) node[GraphNode] {} node[left] {$\lambda_{6}$};
	\draw[GraphEdge] (pd1) -- (pd5) node[midway, above right] {$z_{15}$};
	\draw[GraphEdge] (pd1) -- (pd6) node[midway, above left] {$z_{16}$};
	\draw[GraphEdge] (pd2) -- (pd5) node[midway, below] {$z_{25}$};
	\draw[GraphEdge] (pd2) -- (pd6) node[midway, below] {$z_{26}$};
	\draw[GraphEdge] (pd3) -- (pd4) node[midway, below] {$z_{34}$};
	\draw[GraphEdge] (pd3) -- (pd5) node[midway, right] {$z_{35}$};
	\draw[GraphEdge] (pd4) -- (pd6) node[midway, left] {$z_{46}$};
\end{tikzpicture}
\end{tabular}
\caption{We give the first four 3-Laman graphs (up to 2 loops). In general, for $n$-Laman graphs, there will only be one $0$-loop graph given by a single edge, one $1$-loop graph given by an $(n+1)$ gon, and $(n-1)$ $2$-loop graphs obtained by sewing an $(n+1)$-gon to an $(n+k)$-gon along $k$ consecutive edges, for $k=1,\dots, n-1$.}\label{fig:canLabLaman}
\end{figure}

Our interest will be in $n$-Laman graphs with $n = H+T$. These defining constraints for the $n$-Laman graphs guarantee that the integrand of the corresponding Feynman integrals have appropriate form degrees to contribute to the integrals over the anti-holomorphic/topological coordinates. In particular, our Feynman integrals will involve: 
\begin{itemize}
    \item $(H+T) \abs{\Gamma_0}$ integration variables: one for each vertex of the graph. However, $(H+T)$ integrals can be removed by overall translation symmetry (i.e. by fixing one vertex), leaving just $(H+T) (\abs{\Gamma_0}-1)$ integrals.
    \item $\abs{\Gamma_1}-1$ (regulated) propagators $P_\epsilon$, which are $(0,H+T-1)$-forms, as well as a single cut-propagator $K_\epsilon = \dd P_{\epsilon}$ which is a $(0,H+T)$ form (see Section \ref{sec:FeynmanIntegrals} for an explanation of this point).
\end{itemize}
See Figure \ref{fig:canLabLaman} for some examples of 3-Laman graphs.

We will call a graph ``almost $n$-Laman'' if the global constraint in \eqref{eq:globalconstraint} is replaced by a $\geq$ inequality. In this case, we will define the the \textit{degree} $\tau(\Gamma)$ of the graph as
\begin{equation}
    \tau(\Gamma) := n |\Gamma_0| - (n-1)|\Gamma_1| - n-1
\end{equation}
These graphs will give useful integrands which are not top forms. 

\subsection{Feynman Integrals}\label{sec:FeynmanIntegrals}
The Feynman integrals we will employ are formally defined by the integral:
\begin{equation}
    I_\Gamma(\lambda;z) \define \int_{{\cal M}^{|\Gamma_0|-1}} \left[\prod^{v\neq v_*}_{v \in \Gamma_0} \dvol_v e^{\lambda_v \cdot x^\bC_v}\right] \dd \left[\prod_{e \in \Gamma_1} P_\epsilon(x_{e(0)}-x_{e(1)}+z_e) \right]\,.\label{eq:FeynIntegralP}
\end{equation}
$I_\Gamma(\lambda;z)$ should be understood as a formal power series in the holomorphic shifts $z_e$, and in the holomorphic external momenta $\lambda_v$. Later on, in concrete calculations, we will make use of a judicious analytic continuation to an analytic function of real $\lambda_v$ and imaginary $z_e$. 

In the definition of the Feynman integral, each vertex $v$ of $\Gamma$ is associated to some point $x_v$ in spacetime. We also specified a vertex $v_*$ and set $x_{v_*} = 0$. The integral $I_\Gamma(\lambda;z)$ is thus a function of the $\lambda_v$ for $v \neq v_*$. The integral also depends on a choice of ordering of the edges of the graph, which is reflected in the second product above, containing the Green's functions. The ordering of the vertices employed in the first product cancels in part against the same choice of ordering in the integration contour ${\cal M}^{|\Gamma_0|-1}$, unless $H+T$ is odd.  

We can include the choice of $v_*$ and edge orderings in the definition of $\Gamma$ and keep track of how the answer changes if these choices are modified. If we define $\lambda_{v_*}$ so that $\sum_v \lambda_v=0$, the choice of special vertex is immaterial (up to the possible change in ordering of the vertices). Permutations in the order of vertices or edges changes $I_{\Gamma}(\lambda;z)$ by the sign of the permutation.

In our companion paper \cite{Budzik:2022mpd}, we express the propagator $P_\epsilon$ as an integral over an auxiliary Schwinger time $t\in[\epsilon,\infty)$ and combine it with $K_\epsilon$ to form a propagator ${\cal P}_\epsilon$. The $t$-dependent propagator ${\cal P}_\epsilon$ is an $H+T$ form in space and the Schwinger time $t$, which evaluates to $P_{\epsilon}$ or $K_{\epsilon}$ when integrated on cycles of dimension $1$ or $0$ in the Schwinger time direction (see also Appendix \ref{app:SchwingerParametrization}). All-in-all, this reformulation gives a greatly simplified expression:
\begin{equation}
    I_\Gamma(\lambda;z) = \int_{{\cal M}^{|\Gamma_0|-1} \times \bR \mathbb{P}_>^{|\Gamma_1|-1}} \left[\prod^{v\neq v_*}_{v \in \Gamma_0} \dvol_v e^{\lambda_v \cdot x^\bC_v}\right]\left[\prod_{e \in \Gamma_1} {\cal P}(x_{e(0)}-x_{e(1)}+z_e;t_e) \right]\,. \label{eq:FeynIntegralcalP}
\end{equation}
Notice that we integrate over both spacetime and the positive projective space of Schwinger times $t_{e} \in \bR \mathbb{P}_>^{|\Gamma_1|-1}$.

The propagators ${\cal P}(x_{e(0)}-x_{e(1)}+z_e;t_e)$ are best expressed in a set of auxiliary coordinates:
\begin{equation}
    {\cal P}(x;t) 
        =  \pi^{-\frac{T}{2}} e^{- s^2 - x^\bC \cdot y} \prod_{a=1}^H d y^a \prod_{b=1}^T d s^b \,,
    \qquad 
    y 
        = \frac{\bar x^\bC}{t} \,, 
    \qquad
    s 
        = \frac{x^\bR}{t^{\frac12}} \,,
        \label{eq:super-propagator}
\end{equation}
see \cite{Budzik:2022mpd} for details. The $(H+T)$-Laman condition precisely ensures that the product of propagators gives a non-trivial top $(0,*)$-form. 

We have removed the UV regulator in the above expressions, as we will now show that the integral is finite in the limit $\epsilon \to 0$. It is useful to analytically continue the Gaussian integrals to a contour where $\bar x^\bC$ is real and  $x^\bC$ pure imaginary, so that the integral is recast as:
\begin{equation}\label{eq:RecastInt}
    I_\Gamma(\lambda;z) = \int_{\prod^{v\neq v_*}_{v \in \Gamma_0} (i\bR)^{|\Gamma_0|-1} \times \Delta_\Gamma} \left[\prod^{v\neq v_*}_{v \in \Gamma_0} \dvol_v \,e^{\lambda_v \cdot x^\bC_v}\right]\left[\prod_{e \in \Gamma_1} {\cal P}(x^\bC_{e(0)}-x^\bC_{e(1)}+z_e;y_e,s_e) \right]\,.
\end{equation}
Here $\Delta_\Gamma$ is the \textit{image} of the analytically continued integration contour for the anti-holomorphic $\bar x_v^\bC$ and topological $x_v^\bR$ coordinates in the space of $(y_e, s_e)$. Notice that depending on the parity of $H$, $T$ and $H+T$, the ordering of the edges and vertices may affect the overall sign convention for the integration cycle $\Delta_\Gamma$.

The integral in \eqref{eq:RecastInt} is best understood by doing the $x^{\mathbb{C}}_v$ integral first, which imposes holomorphic momentum conservation relations:
\begin{equation}\label{eq:holoMomCons}
    \sum_{e|e(0)=v} y_e - \sum_{e|e(1)=v} y_e = \lambda_v\,.
\end{equation}
These relations cut out a slice of $\Delta_\Gamma$ which is bounded in the $y$ direction, so that the remaining integrals are finite (some related integral finiteness results were obtained in \cite{wang2024feynman}). We end up with the generalized Fourier transform of a bounded region in $\Delta_{\Gamma}$ called the \emph{operatope}:

\begin{empheq}[box={\mymath[colback=gray!10, sharp corners,boxsep=-3pt,top=13pt, bottom=13pt]}]{equation}
\resizebox{0.85\textwidth}{!}{$
    \displaystyle
    I_\Gamma(\lambda;z) = \int_{\Delta_\Gamma} 
    \left[ 
        \prod_{v \in \Gamma_0}^{v \neq v_*} \delta\!\left(\lambda_v -\!\, \sum_{e|e(0)=v} \!y_e + \,\,\sum_{e|e(1)=v} \!y_e \right) 
    \right] 
    \left[
        \prod_{e \in \Gamma_1} \pi^{-T/2} e^{-s_e^2 - y_e \cdot z_e} d^H y_e \, d^Ts_e 
    \right]\,.
    $
    }
   \label{eq:operatope}
\end{empheq}

The greatest benefit of this reformulation is that it makes many properties of the $I_\Gamma$ integrals transparent. As we will see, the expected properties of BRST anomalies will follow from explicit geometric identities satisfied by $\Delta_\Gamma$ (see also \cite{Budzik:2022mpd}).

We should mention an alternative way to use \eqref{eq:FeynIntegralcalP}, which we discuss further in Section \ref{sec:projective}. One can simply do the Gaussian integrals over the spacetime coordinates $(x^\bC, \bar x^\bC, x^\bR)$ to produce a top form $\omega_{\Gamma}$ on the positive projective space $ \bR \mathbb{P}_>^{|\Gamma_1|-1}$ of Schwinger times (aka Feynman parameters). It is actually a convenient and familiar approach for practical calculations. The form is singular along some locus which touches the integration region at locations on the boundary, so finiteness of the integral is not immediately manifest in that presentation, but can be argued for via some careful blowup of the dangerous regions.

We should also mention the opposite possibility of working directly in position space. We will come back to this point momentarily.

\subsection{Identities}\label{sec:GraphIdentities}
The integral $I_{\Gamma}$ has a number of symmetries and useful identities. Firstly, the integral $I_{\Gamma}$ has the same symmetries as the graph $\Gamma$, up to reordering signs (recall $\Gamma$ includes a choice of edge orderings). Moreover, it is also invariant under the shifts $z_e \to z_e + \delta_{e(0)} - \delta_{e(1)}$, up to an overall factor of $e^{-\sum_v \delta_v \lambda_v}$, i.e. 
\begin{equation}
    \left(\lambda_v + \sum_{e|v = e(0)} \partial_{z_e}-\sum_{e|v = e(1)} \partial_{z_e}\right)I_\Gamma(\lambda;z) = 0\,.\label{eq:identitylambda+partial}
\end{equation}

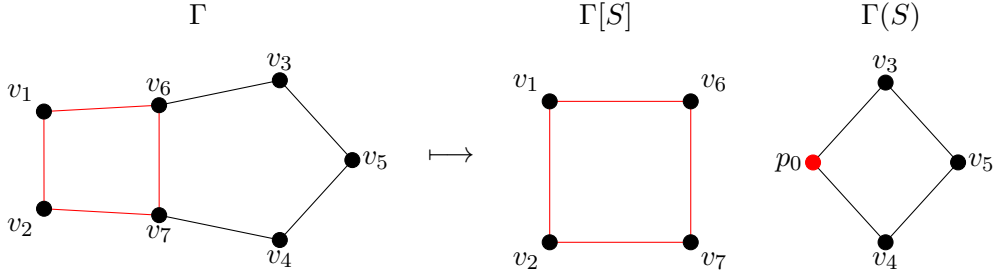
\begin{figure}
    \centering
    \begin{equation*}
    \begin{tikzpicture}
	[
	baseline={(current bounding box.center)},
	line join=round
	]
    \def\gS{1.5};
	\coordinate (pd1) at (0.*\gS,1.1395*\gS);
	\coordinate (pd2) at (0.0011*\gS,0.2772*\gS);
	\coordinate (pd3) at (2.0911*\gS,1.4167*\gS);
	\coordinate (pd4) at (2.0908*\gS,0.*\gS);
	\coordinate (pd5) at (2.7335*\gS,0.7086*\gS);
	\coordinate (pd6) at (1.0209*\gS,1.1956*\gS);
	\coordinate (pd7) at (1.0215*\gS,0.2203*\gS);
    \coordinate (label) at (1.36675*\gS,2*\gS);
    \draw (label) node {$\Gamma$};
	\draw[GraphEdge,red] (pd1) -- (pd2) node[midway, above] {};
	\draw[GraphEdge,red] (pd1) -- (pd6) node[midway, above] {};
	\draw[GraphEdge,red] (pd2) -- (pd7) node[midway, above] {};
	\draw[GraphEdge] (pd3) -- (pd5) node[midway, above] {};
	\draw[GraphEdge] (pd3) -- (pd6) node[midway, above] {};
	\draw[GraphEdge] (pd4) -- (pd5) node[midway, above] {};
	\draw[GraphEdge] (pd4) -- (pd7) node[midway, above] {};
	\draw[GraphEdge,red] (pd6) -- (pd7) node[midway, above] {};
	\draw (pd1) node[GraphNode] {} node[above left] {$v_{1}$};
	\draw (pd2) node[GraphNode] {} node[below left] {$v_{2}$};
	\draw (pd3) node[GraphNode] {} node[above] {$v_{3}$};
	\draw (pd4) node[GraphNode] {} node[below] {$v_{4}$};
	\draw (pd5) node[GraphNode] {} node[right] {$v_{5}$};
	\draw (pd6) node[GraphNode] {} node[above] {$v_{6}$};
	\draw (pd7) node[GraphNode] {} node[below] {$v_{7}$};
\end{tikzpicture}
	\quad\raisebox{-10pt}{$\longmapsto$}\quad
 \begin{tikzpicture}
	[
	baseline={(current bounding box.center)},
	line join=round
	]
    \def\gS{1.5};
	\coordinate (pd1) at (0*\gS,1.25*\gS);
	\coordinate (pd2) at (0*\gS,0.0*\gS);
	\coordinate (pd6) at (1.25*\gS,1.25*\gS);
	\coordinate (pd7) at (1.25*\gS,0.0*\gS);
        \coordinate (label) at (1/2*\gS,2*\gS);
        \draw (label) node {$\Gamma[S]$};
	\draw[GraphEdge,red] (pd1) -- (pd2) node[midway, above] {};
	\draw[GraphEdge,red] (pd1) -- (pd6) node[midway, above] {};
	\draw[GraphEdge,red] (pd2) -- (pd7) node[midway, above] {};
	\draw[GraphEdge,red] (pd6) -- (pd7) node[midway, above] {};
	\draw (pd1) node[GraphNode] {} node[above left] {$v_{1}$};
	\draw (pd2) node[GraphNode] {} node[below left] {$v_{2}$};
	\draw (pd6) node[GraphNode] {} node[above right] {$v_{6}$};
	\draw (pd7) node[GraphNode] {} node[below right] {$v_{7}$};
\end{tikzpicture}
	\quad
 \begin{tikzpicture}
	[
	baseline={(current bounding box.center)},
	line join=round
	]
    \def\gS{1.5};
	\coordinate (pd3) at (2.0911*\gS,1.4167*\gS);
	\coordinate (pd4) at (2.0908*\gS,0.*\gS);
	\coordinate (pd5) at (2.7335*\gS,0.7086*\gS);
	\coordinate (pd6) at (1.4481*\gS,0.70795*\gS);
    \coordinate (label) at (2.1408*\gS,2*\gS);
    \draw (label) node {$\Gamma(S)$};
	\draw (pd3) node[GraphNode] {} node[above] {$v_{3}$};
	\draw (pd4) node[GraphNode] {} node[below] {$v_{4}$};
	\draw (pd5) node[GraphNode] {} node[right] {$v_{5}$};
	\draw[GraphEdge] (pd3) -- (pd5) node[midway, above] {};
	\draw[GraphEdge] (pd3) -- (pd6) node[midway, above] {};
	\draw[GraphEdge] (pd4) -- (pd5) node[midway, above] {};
	\draw[GraphEdge] (pd4) -- (pd6) node[midway, above] {};
 	\draw (pd6) node[GraphNode, red] {} node[left] {$p_0$};
\end{tikzpicture}
	\end{equation*}
    \caption{Left, an almost $3$-Laman graph $\Gamma$ of degree $1$. A square $3$-Laman subgraph $\Gamma[S]$ is marked in red. If the red square is shrunk down to a point $p_0$, the remaining graph $\Gamma(S)$ is also $3$-Laman.}
    \label{fig:shrinkingGraph}
\end{figure}

Finally, and most importantly, the $I_{\Gamma}$ integrals satisfy a set of \textit{quadratic identities} labelled by almost $(H+T)$-Laman graphs of degree $1$. These quadratic identities satisfied by the integration regions $\Delta_{\Gamma}$ and, consequently, $I_{\Gamma}$, give a diagram-by-diagram explanation for the associativity of the deformed factorization algebra and nilpotency of the BRST differential. The quadratic identities are obtained as follows:
\begin{enumerate}
    \item Start with an almost $(H+T)$-Laman graph $\Gamma$ of degree $\tau(\Gamma) = 1$.
    \item Consider a subset of vertices $S \subset \Gamma$ such that the induced graph $\Gamma[S]$ is $(H+T)$-Laman.
    \item Shrink the components of the induced subgraph $\Gamma[S] \subset \Gamma$ to a new vertex, and call the resulting graph $\Gamma(S)$. $\Gamma(S)$ will satisfy the \textit{global} $(H+T)$-Laman condition. We will only consider $S$ such that the local condition is also satisfied, so that $\Gamma(S)$ is actually $(H+T)$-Laman.
\end{enumerate}
See Figure \ref{fig:shrinkingGraph} for an example of this shrinking procedure.
The resulting quadratic identity, labelled by $\Gamma$, on integration regions takes the form:
\begin{empheq}[box={\mymath[colback=gray!10, sharp corners]}]{equation}
\label{eq:DD}
    \sum_{\mathrm{Laman}\, S}^{\hphantom{\mathrm{Laman} S}}\sigma(\Gamma,S) \Delta_{\Gamma[S]} \times \Delta_{\Gamma(S)} =0 \,.
\end{empheq}
Here we identify the summands as chains in the space of $(y_e, s_e)$ labelled by edges of $\Gamma$. The signs $\sigma(\Gamma,S) $ account for the reordering of edges and vertices induced by the identification of edges and vertices of $\Gamma$ with edges and vertices of $\Gamma[S]$ and $\Gamma(S)$. 

Let $\calC = (\bR^{H+T})^{|\Gamma_0|-1} \times \bR \mathbb{P}_>^{|\Gamma_1|-1}$ be the configuration space of a graph $\Gamma$ in an $(H+T)$-twist. For an almost $(H+T)$-Laman graph $\Gamma$ of degree $k$, the map $\calC \to \Delta_\Gamma$ has a kernel of dimension $k$. When $k=1$, the kernel is generically either a closed loop or a segment in $\calC$. Near each endpoint of the segment in $\calC$, a set of vertices $S$ is shrinking down, and the edge variables give a point in $\Delta_{\Gamma[S]} \times \Delta_{\Gamma(S)}$. The converse is also true. Keeping track of relative orientations allows one to show that the images of the two endpoints cancel out in the sum \eqref{eq:DD}, and thus the sum vanishes. This is a variant of the Morse-theoretic ideas employed in \cite{Gaiotto:2015aoa}.

The identity \eqref{eq:DD} for the integration contours leads to an analogous identity for the $I_{\Gamma}$ integrals, labelled by $\lambda_v$ and $z_e$ variables attached to vertices and edges of $\Gamma$. The only subtlety is to keep track of the holomorphic momenta on the vertices of $\Gamma$ which end up as the same vertex in $\Gamma(S)$. The result is a quadratic identity for the integrals of the form:
\begin{equation}\label{eq:II}
    \sum_{\mathrm{Laman}\, S}\sigma'(\Gamma,S) I_{\Gamma[S]}\left(\lambda+ \partial;z\right) \cdot I_{\Gamma(S)}\left(\lambda;z\right) = 0\,.
\end{equation}
Note that while the total $(H+T)$ value determines the Feynman \textit{diagrams} that appear, the corresponding Feynman \textit{integrals} are different.

\begin{figure}
    \centering
\begin{gather*}
0 = 
    \begin{tikzpicture}
    [
	baseline={(current bounding box.center)},
	line join=round
	]
    \def\gS{0.8};
	\coordinate (pd1) at (1.*\gS,0.*\gS);
	\coordinate (pd2) at (-1.*\gS,0.*\gS);
	\draw (pd1) node[GraphNode] {} node[above] {$\lambda_{4}$};
	\draw (pd2) node[GraphNode] {} node[above] {$\lambda_{7}$};
	\draw[GraphEdge] (pd1) -- (pd2) node[midway, above] {$z_{47}$};
      \end{tikzpicture}
      \begin{tikzpicture}
	[
	baseline={(current bounding box.center)},
	line join=round
	]
	\coordinate (pd1) at (0.0005*\gS,0.0207*\gS);
	\coordinate (pd2) at (2.1045*\gS,0.0211*\gS);
	\coordinate (pd3) at (0.*\gS,0.9324*\gS);
	\coordinate (pd4) at (2.104*\gS,0.9327*\gS);
	\coordinate (pd5) at (1.0516*\gS,0.*\gS);
	\coordinate (pd6) at (1.0511*\gS,0.9535*\gS);
	\draw (pd1) node[GraphNode] {} node[below left] {$\lambda_{3}$};
	\draw (pd2) node[GraphNode] {} node[below right] {$\lambda_{1}$};
	\draw (pd3) node[GraphNode] {} node[above left] {$\lambda_{4 + 7}$};
	\draw (pd4) node[GraphNode] {} node[above right] {$\lambda_{5}$};
	\draw (pd5) node[GraphNode] {} node[below] {$\lambda_{2}$};
	\draw (pd6) node[GraphNode] {} node[above] {$\lambda_{6}$};
	\draw[GraphEdge] (pd1) -- (pd3) node[midway, left] {$z_{37}$};
	\draw[GraphEdge] (pd1) -- (pd5) node[midway, below] {$z_{23}$};
	\draw[GraphEdge] (pd2) -- (pd4) node[midway, right] {$z_{15}$};
	\draw[GraphEdge] (pd2) -- (pd5) node[midway, below] {$z_{12}$};
	\draw[GraphEdge] (pd3) -- (pd6) node[midway, above] {$z_{46}$};
	\draw[GraphEdge] (pd4) -- (pd6) node[midway, above] {$z_{56}$};
	\draw[GraphEdge] (pd5) -- (pd6) node[midway, left] {$z_{26}$};
\end{tikzpicture}
-
    \begin{tikzpicture}
    [
	baseline={(current bounding box.center)},
	line join=round
	]
    \def\gS{0.8};
	\coordinate (pd1) at (1.*\gS,0.*\gS);
	\coordinate (pd2) at (-1.*\gS,0.*\gS);
	\draw (pd1) node[GraphNode] {} node[above] {$\lambda_{4}$};
	\draw (pd2) node[GraphNode] {} node[above] {$\lambda_{5}$};
	\draw[GraphEdge] (pd1) -- (pd2) node[midway, above] {$z_{45}$};
      \end{tikzpicture}
      \begin{tikzpicture}
	[
	baseline={(current bounding box.center)},
	line join=round
	]
	\coordinate (pd1) at (0.0005*\gS,0.0207*\gS);
	\coordinate (pd2) at (2.1045*\gS,0.0211*\gS);
	\coordinate (pd3) at (0.*\gS,0.9324*\gS);
	\coordinate (pd4) at (2.104*\gS,0.9327*\gS);
	\coordinate (pd5) at (1.0516*\gS,0.*\gS);
	\coordinate (pd6) at (1.0511*\gS,0.9535*\gS);
	\draw (pd1) node[GraphNode] {} node[below left] {$\lambda_{1}$};
	\draw (pd2) node[GraphNode] {} node[below right] {$\lambda_{3}$};
	\draw (pd3) node[GraphNode] {} node[above left] {$\lambda_{2}$};
	\draw (pd4) node[GraphNode] {} node[above right] {$\lambda_{4+5}$};
	\draw (pd5) node[GraphNode] {} node[below] {$\lambda_{6}$};
	\draw (pd6) node[GraphNode] {} node[above] {$\lambda_{7}$};
	\draw[GraphEdge] (pd1) -- (pd3) node[midway, left] {$z_{12}$};
	\draw[GraphEdge] (pd1) -- (pd5) node[midway, below] {$z_{16}$};
	\draw[GraphEdge] (pd2) -- (pd4) node[midway, right] {$z_{35}$};
	\draw[GraphEdge] (pd2) -- (pd5) node[midway, below] {$z_{36}$};
	\draw[GraphEdge] (pd3) -- (pd6) node[midway, above] {$z_{27}$};
	\draw[GraphEdge] (pd4) -- (pd6) node[midway, above] {$z_{47}$};
	\draw[GraphEdge] (pd5) -- (pd6) node[midway, left] {$z_{67}$};
\end{tikzpicture}
\nonumber\\
\hphantom{0}+
    \begin{tikzpicture}
    [
	baseline={(current bounding box.center)},
	line join=round
	]
    \def\gS{0.8};
	\coordinate (pd1) at (1.*\gS,0.*\gS);
	\coordinate (pd2) at (-1.*\gS,0.*\gS);
	\draw (pd1) node[GraphNode] {} node[above] {$\lambda_{3}$};
	\draw (pd2) node[GraphNode] {} node[above] {$\lambda_{6}$};
	\draw[GraphEdge] (pd1) -- (pd2) node[midway, above] {$z_{36}$};
      \end{tikzpicture}
      \begin{tikzpicture}
	[
	baseline={(current bounding box.center)},
	line join=round
	]
	\coordinate (pd1) at (0.0005*\gS,0.0207*\gS);
	\coordinate (pd2) at (2.1045*\gS,0.0211*\gS);
	\coordinate (pd3) at (0.*\gS,0.9324*\gS);
	\coordinate (pd4) at (2.104*\gS,0.9327*\gS);
	\coordinate (pd5) at (1.0516*\gS,0.*\gS);
	\coordinate (pd6) at (1.0511*\gS,0.9535*\gS);
	\draw (pd1) node[GraphNode] {} node[below left] {$\lambda_{1}$};
	\draw (pd2) node[GraphNode] {} node[below right] {$\lambda_{3+6}$};
	\draw (pd3) node[GraphNode] {} node[above left] {$\lambda_{5}$};
	\draw (pd4) node[GraphNode] {} node[above right] {$\lambda_{4}$};
	\draw (pd5) node[GraphNode] {} node[below] {$\lambda_{2}$};
	\draw (pd6) node[GraphNode] {} node[above] {$\lambda_{7}$};
	\draw[GraphEdge] (pd1) -- (pd3) node[midway, left] {$z_{15}$};
	\draw[GraphEdge] (pd1) -- (pd5) node[midway, below] {$z_{12}$};
	\draw[GraphEdge] (pd2) -- (pd4) node[midway, right] {$z_{34}$};
	\draw[GraphEdge] (pd2) -- (pd5) node[midway, below] {$z_{23}$};
	\draw[GraphEdge] (pd3) -- (pd6) node[midway, above] {$z_{57}$};
	\draw[GraphEdge] (pd4) -- (pd6) node[midway, above] {$z_{47}$};
	\draw[GraphEdge] (pd5) -- (pd6) node[midway, left] {$z_{27}$};
\end{tikzpicture}
-
\begin{tikzpicture}
    [
	baseline={(current bounding box.center)},
	line join=round
	]
    \def\gS{0.8};
	\coordinate (pd1) at (1.*\gS,0.*\gS);
	\coordinate (pd2) at (-1.*\gS,0.*\gS);
	\draw (pd1) node[GraphNode] {} node[above] {$\lambda_{3}$};
	\draw (pd2) node[GraphNode] {} node[above] {$\lambda_{5}$};
	\draw[GraphEdge] (pd1) -- (pd2) node[midway, above] {$z_{35}$};
      \end{tikzpicture}
      \begin{tikzpicture}
	[
	baseline={(current bounding box.center)},
	line join=round
	]
	\coordinate (pd1) at (0.0005*\gS,0.0207*\gS);
	\coordinate (pd2) at (2.1045*\gS,0.0211*\gS);
	\coordinate (pd3) at (0.*\gS,0.9324*\gS);
	\coordinate (pd4) at (2.104*\gS,0.9327*\gS);
	\coordinate (pd5) at (1.0516*\gS,0.*\gS);
	\coordinate (pd6) at (1.0511*\gS,0.9535*\gS);
	\draw (pd1) node[GraphNode] {} node[below left] {$\lambda_{1}$};
	\draw (pd2) node[GraphNode] {} node[below right] {$\lambda_{35}$};
	\draw (pd3) node[GraphNode] {} node[above left] {$\lambda_{2}$};
	\draw (pd4) node[GraphNode] {} node[above right] {$\lambda_{4}$};
	\draw (pd5) node[GraphNode] {} node[below] {$\lambda_{6}$};
	\draw (pd6) node[GraphNode] {} node[above] {$\lambda_{7}$};
	\draw[GraphEdge] (pd1) -- (pd3) node[midway, left] {$z_{12}$};
	\draw[GraphEdge] (pd1) -- (pd5) node[midway, below] {$z_{16}$};
	\draw[GraphEdge] (pd2) -- (pd4) node[midway, right] {$z_{34}$};
	\draw[GraphEdge] (pd2) -- (pd5) node[midway, below] {$z_{36}$};
	\draw[GraphEdge] (pd3) -- (pd6) node[midway, above] {$z_{27}$};
	\draw[GraphEdge] (pd4) -- (pd6) node[midway, above] {$z_{47}$};
	\draw[GraphEdge] (pd5) -- (pd6) node[midway, left] {$z_{67}$};
\end{tikzpicture}\nonumber\\
+
    \begin{tikzpicture}
	[
	baseline={(current bounding box.center)},
	line join=round
	]
	\coordinate (pd1) at (-1.*\gS,0.*\gS);
	\coordinate (pd2) at (0.*\gS,1.*\gS);
	\coordinate (pd3) at (0.*\gS,-1.*\gS);
	\coordinate (pd4) at (1.*\gS,0.*\gS);
	\draw (pd1) node[GraphNode] {} node[left] {$\lambda_{1}$};
	\draw (pd2) node[GraphNode] {} node[above] {$\lambda_{2}$};
	\draw (pd3) node[GraphNode] {} node[below] {$\lambda_{6}$};
	\draw (pd4) node[GraphNode] {} node[right] {$\lambda_{7}$};
	\draw[GraphEdge] (pd1) -- (pd2) node[midway, above left] {$z_{12}$};
	\draw[GraphEdge] (pd1) -- (pd3) node[midway, below left] {$z_{16}$};
	\draw[GraphEdge] (pd2) -- (pd4) node[midway, above right] {$z_{27}$};
	\draw[GraphEdge] (pd3) -- (pd4) node[midway, below right] {$z_{67}$};
\end{tikzpicture}
\begin{tikzpicture}
	[
	baseline={(current bounding box.center)},
	line join=round
	]
	\coordinate (pd1) at (-1.*\gS,0.*\gS);
	\coordinate (pd2) at (0.*\gS,1.*\gS);
	\coordinate (pd3) at (0.*\gS,-1.*\gS);
	\coordinate (pd4) at (1.*\gS,0.*\gS);
	\draw (pd1) node[GraphNode] {} node[left] {$\lambda_{1+2+6+7}$};
	\draw (pd2) node[GraphNode] {} node[above] {$\lambda_{3}$};
	\draw (pd3) node[GraphNode] {} node[below] {$\lambda_{4}$};
	\draw (pd4) node[GraphNode] {} node[right] {$\lambda_{5}$};
	\draw[GraphEdge] (pd1) -- (pd2) node[midway, above left] {$z_{36}$};
	\draw[GraphEdge] (pd1) -- (pd3) node[midway, below left] {$z_{47}$};
	\draw[GraphEdge] (pd2) -- (pd4) node[midway, above right] {$z_{35}$};
	\draw[GraphEdge] (pd3) -- (pd4) node[midway, below right] {$z_{45}$};
\end{tikzpicture}
\end{gather*}
    \caption{The quadratic identity for $3$-Laman graphs implied by the almost $3$-Laman square-pentagon graph in Figure \ref{fig:shrinkingGraph}. We label all of the vertices and edges following the ordering conventions implied by Figures \ref{fig:canLabLaman} and \ref{fig:shrinkingGraph}. We employ here the shorthand $\lambda_{i_1 + \dots + i_n} := \sum_{j=1}^n \lambda_{i_j}$.}
    \label{fig:quadraticDoubleBox}
\end{figure}
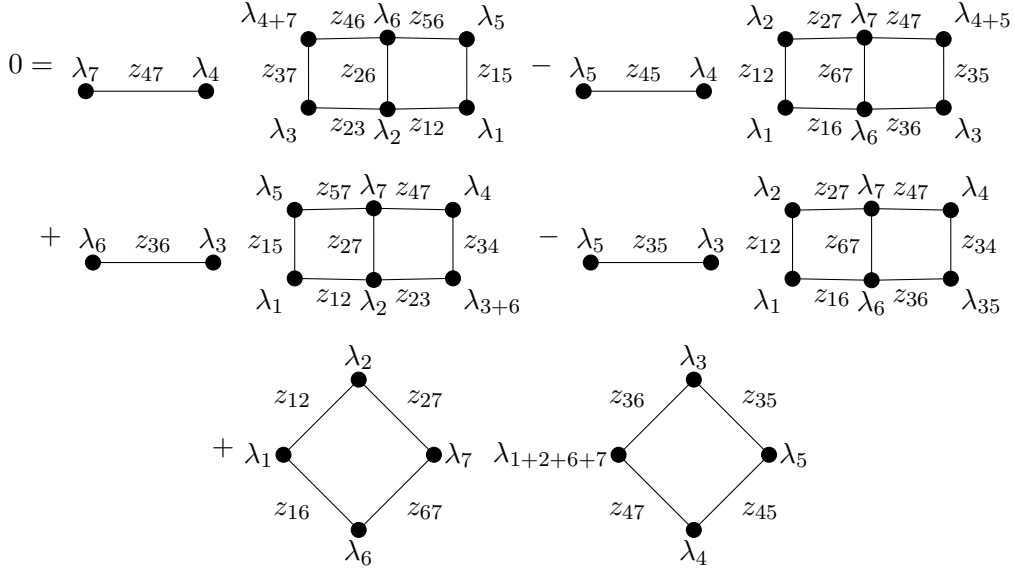

In $I_{\Gamma[S]}$, the $\lambda$ arguments are the holomorphic momenta inherited from $\Gamma$, and the $\partial$ shifts denote
\begin{equation}
    \partial_v = \sum_{\substack{e|e(0) = v \\ e(1)\notin S}} \partial_{z_e} - \sum_{\substack{e|e(1) = v \\ e(0)\notin S}} \partial_{z_e} \, .
\end{equation}
The $\lambda + \partial$ operators add up to $0$ when acting on $I_{\Gamma(S)}$, so they are valid arguments for $I_{\Gamma[S]}$. Likewise, the $z$ arguments in $I_{\Gamma[S]}$ are inherited from $\Gamma$. For $I_{\Gamma(S)}$, the $\lambda$ arguments are again the holomorphic momenta inherited from $\Gamma$, and the ``new vertex'' (where the $I_{\Gamma[S]}$ was shrunken to a point) has a sum over all $\lambda_v$ for $v \in S$. The $z$ arguments in $I_{\Gamma(S)}$ are inherited from $\Gamma$. See Figure \ref{fig:quadraticDoubleBox} for an example of a quadratic identity on integrals.

These quadratic identities satisfied by $I_{\Gamma}$ give a large variety of quadratic identities between the brackets we compute through the corresponding Feynman diagrams. This includes the $L_\infty$ axioms for the basic BRST anomaly brackets as well as their generalizations which capture various parts of the OPE/factorization algebra structure of the theory.

\subsection{Configuration Space Perspective}\label{sec:twistedScaling}
In order to make contact with more geometric perspectives on the factorization algebra operations, e.g. point-splitting regularizations of perturbation theory, it is useful to do yet another change of variables. 

Define a ``twisted scaling'' action on analytically continued spacetime:
\begin{equation}
x^\bC \to x^\bC\,, \qquad \bar x^\bC \to R^2 \bar x^\bC\,, \qquad x^\bR \to R x^\bR\,.
\end{equation}
Now consider the configuration space $\widetilde{\mathrm{Conf}}_{\abs{\Gamma_0}}$ of $\abs{\Gamma_0}$ points in this analytically continued spacetime, modulo translations and twisted scale transformations. We can trade the position of a point $x$ in spacetime for a point $(R,u) \in \bbR^+ \times \widetilde{\mathrm{Conf}}_{\abs{\Gamma_0}}$.

The radial coordinate $R$ can be combined with the Schwinger time coordinates into a new rescaled Schwinger time $T_e = R^{-2} t_e \in \bR^+$. We have thus traded the overall scale $R$ in spacetime for a scale in the space of Schwinger times. The integral over the $T_e$ variables can be performed first, mapping each propagator factor to the unregularized propagator $P(x_{e(0)}(u) -x_{e(1)}(u) + z_e)$. The integral in \eqref{eq:FeynIntegralcalP} then becomes 
\begin{equation}\label{eq:unregulated}
    I_\Gamma(\lambda;z) = \int_{\widetilde{\mathrm{Conf}}_{\abs{\Gamma_0}}} \left[\prod^{v\neq v_*}_{v \in \Gamma_0} \dvol_v e^{\lambda_v \cdot x^\bC_v}\right] \left[\prod_{e \in \Gamma_1} P(x_{e(0)}-x_{e(1)}+z_e)\right]\,.
\end{equation}
We are thus tasked with the problem of identifying an integration cycle in the (complexified) configuration space of points in spacetime which represents $\widetilde{\mathrm{Conf}}_{\abs{\Gamma_0}}$ and such that the integral converges.

A possible route would be to define a careful compactification of the configuration space of points, such that the Feynman integrands extend to non-singular forms on the compactification, and consider a representative integration contour in the compactified space. This is a standard approach in the literature on 2d and 3d topological systems \cite{kontsevich1994feynman, Kontsevich:1997vb} (see also \cite{Axelrod:1991vq, Axelrod:1993wr}).

In Section \ref{sec:PointSplittingRevisted} we discuss at length a point-splitting regularization of Feynman integrals in topological examples and relations to topological factorization algebra perspectives. It should be possible to use similar ideas here to extend the configuration space of spacetime points by including rescaling factors $R_v$ and present $\widetilde{\mathrm{Conf}}_{\abs{\Gamma_0}}$ as a safe (divergence free), compact contour in the extended configuration space. Accordingly, we expect the BRST anomalies computed in this chapter to fit well in an HT-factorization algebra perspective as well. 

\subsection{Examples}
We will now go over a few simple examples with a possibly excessive level of detail.

\subsubsection{Example: \texorpdfstring{$H=0$}{H=0}, \texorpdfstring{$T=1$}{T=1} Diagrams in Topological Quantum Mechanics}\label{sec:exampleH0T1}
Free topological quantum mechanics is already an interesting example. The 1d topological action (with bosonic or fermionic parts) is
\begin{equation}\label{eq:BVQM}
    S =\int_{\mathbb{R}} \left[p \,\dd q+ \psi \,\dd\psi \right] = \int_{\mathbb{R}} \left[p^{(0)} \dot q^{(0)}+ \psi^{(0)} \dot \psi^{(0)} \right]dx\, ,
\end{equation}
where $dx= dx^{\bR}$. In the BV formalism employed in this paper, the variables are superfields $p = p^{(0)} + p^{(1)}dx$, $q = q^{(0)} + q^{(1)}dx$ and $\psi = \psi^{(0)} + \psi^{(1)}dx$ and we interpret it as a BV action. The 0-form components of the superfields are the usual physical fields, with the usual action. In particular, $p^{(0)}$, $q^{(0)}$ and $\psi^{(0)}$ are coordinates on the phase space. The 1-form components of the superfields do not enter the action and are anti-fields in a BV formalism: their BRST variation are the equations of motion $\dot p=0$, $\dot q = 0$ and $\dot \psi=0$.

The classical observables of topological quantum mechanics are obviously polynomials $\cI(p,q,\psi)$ in positions, momenta and fermions. Adding an ``interaction'' $\cI(p,q, \psi)$ to the \textit{BV action} does not actually change the action for the physical fields, but rather introduces (or further deforms if another interaction was already present) a differential $\{\cI, \,\cdot\,\}_{\mathrm{PB}}$ given by the classical Poisson bracket with $\cI$, making the phase space into a dg-manifold. This works because a fermionic $\cI$ will classically square to $0$, so that $\{\cI, \{\cI, \,\cdot\,\}_{\mathrm{PB}}\}_{\mathrm{PB}}=0$ as well.\footnote{More explicitly, adding an interaction term $\calI(p,q,\psi)$ to the BV action in \eqref{eq:BVQM} and expanding in the ``superspace variable'' $dx$, we find that we just add terms linear in the anti-fields $q^{(1)}$, $p^{(1)}$, and $\psi^{(1)}$, and so simply introduce/modify the BV-BRST differential.}

Quantum mechanically, we need a scheme to deal with ordering ambiguities in $\cI$, which becomes a local operator, and ensure the ``Maurer-Cartan equation'' $\cI^2=0$ holds operatorially as well, so that the (anti-)commutator $[\cI, \,\cdot\,]$ in the operator sense still defines a differential. 

Working in perturbation theory, i.e. with Wick contractions and deformation quantization, a natural resolution of ordering ambiguities is the Moyal product and the MC equation involves the Moyal (anti-)commutator. Here we will see how our standard scheme and universal Feynman rules reproduce coefficients in the standard Moyal (anti-)commutator. In a later section we will upgrade the analysis to combined systems to recover the complete Moyal product.

The 1-Laman constraints only allow graphs with two vertices and at least one edge, i.e. ``banana'' or ``melon'' graphs. Let $N = |\Gamma_1|$ denote the number of edges. We aim to compute the corresponding integrals $I_N$. We will then illustrate how to derive these relations from a careful analysis of the integration contours $\Delta_{\Gamma}$. 

The propagator $P$ is just the solution to $\dd P(x) = \delta(x)$. In 1 dimension this is just the step function:
\begin{equation}
    P(x-y) = \frac12\mathrm{sign}(x-y)\,,
\end{equation}
with source 
\begin{equation}
    K(x-y) = \delta(x-y) \dd(x-y)\,.
\end{equation}
The integral $\int P(x)^{N-1} K(x)$ is naturally regularized to:
\begin{equation}
    I_N := \begin{cases}
        0   & \text{if $N$ is even}\,,\\
        2^{1-N} & \text{if $N$ is odd}\,.
    \end{cases}
\end{equation}
It is also intuitive that this controls the discontinuity of $P(x-y)^N$, which would enter in the expansion of the Moyal (anti-)commutator for an operator at $x$ and one at $y$. We will see geometrically that $I_N = 2^{1-N} = 2^{-N}(1-(-1)^N)$ is computing the volume of $\Delta_N$ relative to $\mathbb{R}^N$, weighted by signs. And we will see yet another proof that $I_N=0$ for even $N$ in Section \ref{sec:projective}.

The identity
\begin{equation}\label{eq:moyalJacobiGraph}
   (-1)^{N_1 + N_2} I_{N_{3}}I_{N_1 + N_2} - (-1)^{N_3} I_{N_2} I_{N_{1}+N_3}+ I_{N_1}I_{N_2+N_3} =0
\end{equation}
can be shown by writing $I_N =(1-(-1)^N)2^{-N}$. It will lead to the Jacobi identity for the Moyal commutator \eqref{eq:moyalCom}. 

We now reproduce these results from our general formalism. First of all, the region $\Delta_N$ in the space of $s_e$ is the union of the two sectors where all $s_e$ have the same sign. To compute the local orientation, we observe
\begin{equation}\label{eq:measure}
\prod_e ds_e = \frac{t_1 dt_2 \dots dt_N - dt_1 t_2 \dots dt_N + \cdots + (-1)^{N-1} dt_1 \dots dt_{N-1}t_N }{\prod_e t_e^{\frac32}} \left(\frac{x^\bR}{2} \right)^{N-1} d x^\bR \, .
\end{equation}
We see that the local orientation of $\Delta_N$ differs from the one given by the order of the edges by $\sign(x^\bR)^{N-1}$.

It is also useful to describe $\Delta_\Gamma$ when the edge orientations are mixed. Call the vertices of the graph $v_1$ and $v_2$, and suppose there are $N_{12}$ edges with orientation from $v_1$ to $v_2$ and $N_{21}$ with orientation $v_2$ to $v_1$. Order the edges in the graph so that the $12$ edges all come before the $21$ edges. Then the region $\Delta_{N_{12},N_{21}}$ in the space of $s_e$'s, consists of the orthants where all the $s_{12}$'s have the same fixed sign, and all the $s_{21}$'s have the same fixed sign opposite to the $s_{12}$'s. If we set the position of the second vertex $x^\bR_2 = 0$, the local orientation of $\Delta_N$ differs from the local orientation given by ordering the edges by
\begin{equation}
    \sign(s_{12})^{N-1}(-1)^{N_{21}}\,.
\end{equation} 
This can also be written as $\sign(s_{21})^{N-1}(-1)^{N_{12}+1}$. This is consistent with the region $\Delta_\Gamma$ being odd under permutation of vertices when $H+T=1$. We can thus write
\begin{equation}
\Delta_{N_{12},N_{21}} = \Delta^+_{N_{12}}\times \Delta^-_{N_{21}}-  \Delta^-_{N_{12}}\times \Delta^+_{N_{21}}\,,
\end{equation}
where $\Delta^\pm_N$ are the orthants of $s_e$ space consisting of all positive or all negative $s_e$, with weight $(-1)^N$ for $\Delta^-_N$.

\begin{figure}
    \centering
    \begin{tikzpicture}[
	baseline={(current bounding box.center)},
	line join=round
	]
    \def\gS{4};
	\coordinate (pd3) at (0.5*\gS,0.866*\gS);
	\coordinate (pd1) at (0.*\gS,0.*\gS);
	\coordinate (pd2) at (1.*\gS,0.*\gS);
 
    \node[vertex, label={west:$v_1$}] at (pd1) {$\cdot$};
    \node[vertex, label={east:$v_2$}] at (pd2) {$\cdot$};
    \node[vertex, label={west:$v_3$}] at (pd3) {$\cdot$};

    \draw[lineShimmy] (pd1) to[bend left=20] (pd2);
    \draw[lineShimmy] (pd1) to[bend right=20] (pd2);
    \node[rotate=90] at ($(pd1)!0.5!(pd2)$) {$\cdots$};
    
    \draw[lineShimmy] (pd2) to[bend left=20] (pd3);
    \draw[lineShimmy] (pd2) to[bend right=20] (pd3);
    \node[rotate=30] at ($(pd2)!0.5!(pd3)$) {$\cdots$};

    \draw[lineShimmy] (pd1) to[bend left=20] (pd3);
    \draw[lineShimmy] (pd1) to[bend right=20] (pd3);
    \node[rotate= 150] at ($(pd3)!0.5!(pd1)$) {$\cdots$};
\end{tikzpicture}
    \caption{Degree $1$ almost 1-Laman graphs have three vertices, labeled as $v_1$, $v_2$, $v_3$, with $N_{12}$, $N_{13}$ and $N_{23}$ edges. Note, when there is only one spacetime dimension, i.e. $(H,T)=(0,1)$, these graphs should be viewed as one dimensional, not stretching into a second direction.}
    \label{fig:almost-1Laman}
\end{figure}
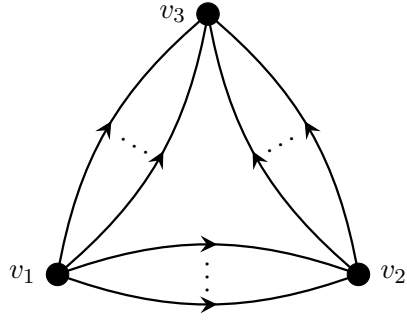

We can now observe the quadratic identities graphically. As shown in Figure \ref{fig:almost-1Laman}, the almost 1-Laman graphs of degree $1$ have three vertices, labelled $v_1$, $v_2$, $v_3$, with $N_{12}$, $N_{13}$ and $N_{23}$ edges between them, as above, oriented according to the subscripts. We order the edge set of the graph first within each ``grouping,'' and then ``globally'' with respect to a global lexicographic ordering: $12$, $13$, then $23$.  
The corresponding region $\Delta_{N_{12},N_{13},N_{23}}$ in $s_e$ space consists of sectors where the $s_e$ in each grouping have the same sign. Naively, there are $2^3 = 8$ choices of independent signs for the groupings. However, the edges are also constrained by the ordering of the vertices in space: the sign of $s_e$ must coincide with the sign given by the difference of the corresponding vertices in space. So there are actually only $6$ choices, corresponding to the $6$ different ways to place the points $\{x_1^\bR,x_2^\bR,x_3^\bR\}$ in space.
For example, if $x^\bR_2$ lies between $x^\bR_1$ and $x^\bR_3$, then $s_{12}$, $s_{13}$ and $s_{23}$ will all have the same sign. When we derive a quadratic identity, such a region will contribute a term where vertices $v_1$ and $v_2$ shrink, and a term where the $v_2$ and $v_3$ vertices shrink, etc. 

We pick conventions so that the ``new vertex'' in $\Gamma(S)$ is last in the ordering.  If we shrink the vertices $v_1$ and $v_2$, we see that $\Gamma[S]$ has $N_{12}$ properly oriented edges and $\Gamma(S)$ has $N_{13} + N_{23}$ edges oriented backwards. We can repeat this shrinking all of the other combinations of vertices. We thus expect an identity:
\begin{equation}
    \Delta_{N_{12},0} \times \Delta_{0,N_{13}+ N_{23}} -(-1)^{N_{13} N_{12}} \Delta_{N_{13},0} \times \Delta_{N_{23},N_{12}} + (-1)^{N_{23}N_{12}+ N_{23}N_{13}} \Delta_{N_{23},0} \times \Delta_{N_{12}+ N_{13},0}
\end{equation}
We have included signs from reordering the edges, as well as a sign from picking two shrinking vertices out of three ordered vertices. 

Expanding into $\Delta^\pm$ regions, the sum becomes 
\begin{align}
    & \left[ \Delta^+_{N_{12}}- \Delta^-_{N_{12}} \right] \times \left[ \Delta^-_{N_{13}+ N_{23}}- \Delta^+_{N_{13}+ N_{23}} \right] + \cr &-(-1)^{N_{13} N_{12}}\left[  \Delta^+_{N_{13}}-   \Delta^-_{N_{13}}\right] \times \left[  \Delta^+_{N_{23}}\times \Delta^-_{N_{12}}-  \Delta^-_{N_{23}}\times \Delta^+_{N_{12}} \right] + \cr &+(-1)^{N_{23}N_{12}+ N_{23}N_{13}} \left[ \Delta^+_{N_{23}}- \Delta^-_{N_{23}} \right] \times \left[ \Delta^+_{N_{12}+N_{13}}-\Delta^-_{N_{12}+N_{13}}\right]
\end{align}
which indeed vanishes identically. 

It is useful to also define integrals $I_{N_{12},N_{21}}$ with $N_{12}$ edges oriented in one direction and $N_{21}$ in the other direction. This differs from $I_{N_{12} + N_{21}}$ by a sign:
\begin{equation}
    I_{N_{12},N_{21}} = \left[(-1)^{N_{21}}-(-1)^{N_{12}} \right] 2^{-N_{12}-N_{21}}\,,
\end{equation}
which accordingly satisfies
\begin{equation}
    I_{N_{12},0}I_{0,N_{13}+ N_{23}} - I_{N_{13},0} I_{N_{23},N_{12}}+ I_{N_{23},0}I_{N_{12}+ N_{13},0} =0\,,
\end{equation}
which is the appropriate generalization of \eqref{eq:moyalJacobiGraph} with arbitrary edge orientations.

\subsubsection{Example: \texorpdfstring{$H=1$}{H=1}, \texorpdfstring{$T=0$}{T=0} Diagrams in 2d Chiral Algebras}\label{sec:H1T0ExampleDiagrams}
Next, we can consider 1-dimensional holomorphic theories, aka chiral algebras. The free BV action
\begin{equation}
    \int_{\bbC} \beta \bar \partial \gamma \, \dd z
\end{equation}
simply describes a standard $\beta \gamma$ system. We can also include free fermions with action $\int_{\bbC} \psi \bar \partial \psi\, \dd z$.

As in the quantum mechanics case, an interaction $\cI$ does not change the physical action, but only introduces a new BRST differential. BRST anomalies will arise from short distance singularities captured by the standard OPE; we will come back to this point in greater detail in the Section \ref{sec:brackets}. Here we present the relevant Feynman diagram computations from our general formalism.

We use the same 1-Laman graphs as for the 1d topological case, which have two vertices and any number of propagators between them. The propagator is simply $P(w_1,w_2) = (w_1-w_2)^{-1}$, and the Feynman integral with $N$ propagators is
\begin{equation}
    I_{N}(\lambda;z) = \oint_{|w|> |z_e|} \frac{dw}{2 \pi i} e^{\lambda  w}\prod_{e=1}^N \frac{1}{w+ z_e}\,.\label{eq:INH1T0}
\end{equation}
The $\lambda$ parameter allows one to catch higher terms in the OPE and the $z_e$ parameters allow one to deal with derivatives by the Taylor series $(w+z)^{-1} = \sum_n (-1)^n z^n w^{-n-1}$, which we note converges since $\abs{w} > \abs{z_e}$.

These integrals satisfy a quadratic identity which ensures associativity of the OPE. In order to write down the quadratic recursions, it is instructive to consider a two-variable contour integral appearing in the Feynman integral defined by the almost 1-Laman graph in Figure \ref{fig:almost-1Laman}:
\begin{equation}\label{eq:IntChiral1}
   \oint_{|w_2|=1} \frac{dw_2}{2 \pi i} e^{\lambda_2  w_2} \oint_{|w_1-w_2|=\epsilon} \frac{dw_1}{2 \pi i} e^{\lambda_1  w_1} \frac{1}{
   \prod_{e=1}^{N} (w_1+ z_{e}) \prod_{e'=1}^{N'} (w_2+ z'_{e'}) 
   \prod_{e''=1}^{N''} (w_1- w_2+ z''_{e''})}\,,
\end{equation}
where all shifts $z, z', z'' \ll \epsilon \ll 1$. The integral in \eqref{eq:IntChiral1} involves two ``coupled'' integrals in the variables $w_1$ and $w_2$. By using derivative (shift) operators as arguments, we can decouple these integrals into an integral over $w_2$ and and integral over $w_1-w_2$ by rewriting it as follows: 
\begin{equation}\label{eq:IntChiral2}
\begin{aligned}
   &\oint_{|w_2|=1} \frac{dw_2}{2 \pi i} e^{(\lambda_1 +\lambda_2)  w_2}\frac{1}{\prod_{e=1}^{N} (w_2+ z'_{e})\prod_{e'=1}^{N'} (w_2 + \partial_{\lambda_1}+ z_{e})}\\
    &\quad\quad\times\oint_{|w_1-w_2|=\epsilon} \frac{dw_1}{2 \pi i} e^{\lambda_1 ( w_1-w_2) } \frac{1}{\prod_{e''=1}^{N''} (w_1 - w_2 + z''_{e''})}\,.
\end{aligned}
\end{equation}
This is a useful trick. We can mimic the derivation of \eqref{eq:IntChiral2} in the opposite order to obtain the non-trivial identity:
\begin{equation}
   I_{N+N'}(\lambda_1 + \lambda_2; z', z+\partial_{\lambda_1}) \cdot I_{N''}(\lambda_1; z'') = I_{N''}(\lambda_1+ \partial_{z}; z'')\cdot  I_{N+N'}(\lambda_1 + \lambda_2; z, z')\,.
\end{equation}
We have introduced the short hand notation  $\partial_{z'} := \sum_{e=1}^{N'}\partial_{z'_e}$  and 
\begin{equation}
    \begin{aligned}
    I_{N+N'}(\lambda;z,z') \define \oint_{|w_1|=1} \frac{dw_1}{2 \pi i} e^{\lambda_1  w_1}\prod_{e=1}^N \frac{1}{w_1+ z_e}\prod_{e'=1}^{N'} \frac{1}{w_1+ z_{e'}'}\,.
    \end{aligned}
\end{equation}

On the other hand, we can return to the integral \eqref{eq:IntChiral1} and reorganize it as the difference of two contours in the standard way familiar to 2d chiral algebras:
\begin{equation}
   \oint_{|w_2|=1} \frac{dw_2}{2 \pi i} e^{\lambda_2  w_2} \left[\oint_{|w_1|>1}-\oint_{|w_1|<1} \right]\frac{dw_1}{2 \pi i} e^{\lambda_1  w_1} \frac{1}{\prod_{e=1}^N (w_1+ z_e)\prod_{e'=1}^{N'} (w_2+ z'_{e'}) \prod_{e''=1}^{N''} (w_1- w_2+ z''_{e''})}\,.
\end{equation}
Evaluating the first contour gives 
\begin{equation}
    I_{N'+N''}(\lambda_1; z',z''-\partial_{\lambda_2}) I_N(\lambda_2;z')\,,
\end{equation}
which by rearranging the order of integration has the same functional form as 
\begin{equation}
    I_N(\lambda_2- \partial_{z''};z')I_{N'+N''}(\lambda_1; z,z'')\,\,.
\end{equation}
By evaluating the second contour we have 
\begin{equation}
   -(-1)^{N''} I_{N'}(\lambda_1+ \partial_{z''};z) I_{N+N''}(\lambda_2; z', -z'') \,.
\end{equation}

Putting the results together, and converting to the naming conventions in Figure \ref{fig:almost-1Laman}, we have $z \mapsto z_{13}$, $z' \mapsto z_{23}$, and $z'' \mapsto z_{12}$. The quadratic identity associated with the almost 1-Laman graph becomes:
\begin{align}
&I_{N_{13}}(\lambda_2- \partial_{z_{12}};z_{23})I_{N_{23}+N_{12}}(\lambda_1; z_{13},z_{12}) \cr
&- (-1)^{N_{12}} I_{N_{23}}(\lambda_1+ \partial_{z_{12}};z_{13}) I_{N_{13}+N_{12}}(\lambda_2; z_{23}, -z_{12})  \cr
&-I_{N_{12}}(\lambda_1+ \partial_{z_{13}}; z_{12}) I_{N_{13}+N_{23}}(\lambda_1 + \lambda_2; z_{13}, z_{23}) =0\,.\label{eq:IdentityH1T0}
\end{align}

We will now reproduce this result by the general strategy. The regions $\Delta_\Gamma$ are defined as in the previous Section \ref{sec:exampleH0T1}, but utilizing only the $y_e$ variables instead of the $s_e$. The integral over the $x_1^\bC$ coordinate forces the $y_e$ to add up to $\lambda_1$, i.e. the holomorphic momentum conservation \eqref{eq:holoMomCons}, so we only get contributions from the positive or negative $y_e$ sub-regions $\Delta^\pm_N$ depending on the sign of $\lambda$. 

The integral for positive $\lambda_1$ is most easily computed by going back to the original expression
\begin{equation}
    I_N(\lambda_1;z) = \int_{ (i\bR)\times \Delta^+_N} \left[ \dvol_1 e^{\lambda_1 x^\bC_1}\right]\left[\prod_{e=1}^N e^{- y_e (z_e+x^\bC_1)} dy_e  \right] \,.
\end{equation}
Giving $x_1^\bC$ a small positive real part $\epsilon$ and taking $z_e<0$ gives 
\begin{equation}
    I_N(\lambda_1;z) = \int_{ i\bR + \epsilon} \dvol_1 e^{\lambda_1  x^\bC_1}\prod_{e=1}^N \frac{1}{x^\bC_1+ z_e}\,.
\end{equation}
Closing the contour towards the negative reals (left-side of the plane) picks up the simple poles: 
\begin{equation}
    I_N(\lambda_1;z) = \sum_{e=1}^N e^{-\lambda_1 z_e}\prod_{e'=1|e' \neq e}^N \frac{1}{z_{e'}-z_e}\,.
\end{equation}
Of course, the integral for negative $\lambda_1$ gives the same answer. They both agree with the answer from the standard contour-integral \eqref{eq:INH1T0} and have the expected symmetries of the $1$-Laman graphs:
\begin{align}
    I_N(\lambda_1;z) &= (-1)^{N-1} I_N(\lambda_2 = -\lambda_1; -z)\,, \cr
    I_N(\lambda_1;z) &= e^{\lambda_1 \delta_1}I_N(\lambda_1;z+ \delta_1)\,.
\end{align}
The three-term identity \eqref{eq:IdentityH1T0} can be recognized as arising from the quadratic identity for $\Delta_N$ regions associated to a triangle graph with $N_{ij}$ edges. The complicated patterns of shifts of the $\lambda_v$ arguments by $\partial_{z_e}$ derivatives ensure that the arguments of propagators in the second $I_\bullet$ function in each product are shifted by $x^\bC_v$ coordinates from the first $I_\bullet$ function, representing the insertion of the first graph at a vertex of the second.

\subsection{A Projective Perspective and a Non-Renormalization Theorem}\label{sec:projective}
Both for direct calculations, and for a uniform treatment across dimensions, it is useful to take a different route to the $I_\Gamma$ integral: we can perform the Gaussian integral over positions $x_v$ first, leading to a top form $\omega_\Gamma$
on the positive projective space $\bR \mathbb{P}_>^{|\Gamma_1|-1}$ of Schwinger times.

A remarkable fact is that the ${\cal P}$ propagator in \eqref{eq:super-propagator} factors over contributions from individual dimensions of spacetime. Correspondingly, $\omega_\Gamma$ factors into the topological contributions and the holomorphic ones
\begin{empheq}[box={\mymath[colback=gray!10, sharp corners]}]{equation}
   \omega_\Gamma = \alpha_\Gamma^T \prod_{i=1}^H \rho_\Gamma[\lambda^i;z^i]\,. \label{eq:omegafactorize}
\end{empheq}
Here, $\lambda_v^i$ and $z_e^i$ are the components in the $i$-th holomorphic direction. Each of the $\alpha_\Gamma$ and $\rho_\Gamma$ factors is a closed form of degree $|\Gamma_1|- |\Gamma_0|+1$, i.e. the number of loops, and can be defined for general connected graphs. 

We can start with the topological directions. For each topological direction, the corresponding form is:
\begin{equation}
    \alpha_\Gamma = \int_{\bR^{|\Gamma_0|-1}} \left[\prod_{e \in \Gamma_1}  \pi^{-\frac{1}{2}} e^{- s_e^2}  d s_e \right]\,, 
\end{equation}
where we replace 
\begin{equation}
    s_e = t_e^{-\frac12}(x_{e(0)}-x_{e(1)})
\end{equation}
in the integrand. 

The graphs with non-trivial $\alpha_{\Gamma}$ are highly constrainted. For example, if $\Gamma$ is a segment with two nodes (propagator), we get
\begin{equation}
    \alpha_{\,\pick{1.5ex}{segment}} = \int_{\bR}\pi^{-\frac{1}{2}} e^{-t_e^{-1}  x_2^2}  t_e^{-\frac12} d x_2 =1\,.
\end{equation}
More generally, $\alpha_\Gamma = \pm 1$ if $\Gamma$ is any tree diagram since we can change variables in the integral by writing $x_v$ as a linear combination of $t_e^{\frac12} s_e$. We can also collapse any tree branches attached to a more complicated graph by the same strategy.

We can also consider loop diagrams. The form $\alpha_\Gamma$ vanishes for any $\Gamma$ with an odd number of loops. This is because a reflection $x \mapsto -x$ will reflect the odd number of factors $\frac{ds_e}{dt_e}dt_e$ which appear in the form and the integrand is thus odd under the reflection. 

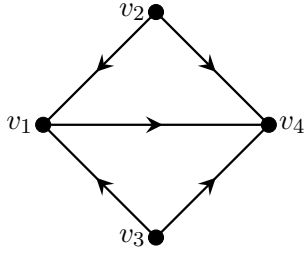
\begin{figure}
\centering
\begin{tikzpicture}
	[
	baseline={(current bounding box.center)},
	line join=round
	]
    \def\gS{3};
	\coordinate (pd1) at (0.5*\gS,0.5*\gS);
	\coordinate (pd2) at (0.5*\gS,-0.5*\gS);
	\coordinate (pd3) at (0.*\gS,0.*\gS);
	\coordinate (pd4) at (1.*\gS,0.*\gS);

	\draw (pd1) node[vertex] {} node[left] {$v_2$};
	\draw (pd2) node[vertex] {} node[left] {$v_3$};
	\draw (pd3) node[vertex] {} node[left] {$v_1$};
	\draw (pd4) node[vertex] {} node[right] {$v_4$};

	\draw[lineShimmy] (pd1) -- (pd3) {};
	\draw[lineShimmy] (pd1) -- (pd4) {};
	\draw[lineShimmy] (pd2) -- (pd3) {};
	\draw[lineShimmy] (pd2) -- (pd4) {};
	\draw[lineShimmy] (pd3) -- (pd4) {};
\end{tikzpicture}
\caption{The ``bitriangle'' graph, the only $2$-loop 2-Laman graph.}
\label{fig:bitriangle}
\end{figure}

An interesting non-vanishing  two-loop example is the bitriangle graph shown in Figure \ref{fig:bitriangle}.
We find
\begin{equation}
    \begin{aligned}
    \alpha_{\pic{1ex}{bitriangle}} = 
    \frac{1}{8\pi} &
\Big[t_{31}\left(t_{24}+t_{43}\right)+t_{12}\left(t_{23}+t_{24}+t_{43}\right)+t_{23}\left(t_{24}+t_{31}+t_{43}\right)\Big]^{-3 / 2}\\
    \Big[ & (t_{24}+t_{43})(dt_{12}dt_{23} + dt_{31}dt_{23}) + (t_{12}+t_{31})\left(dt_{23}dt_{24} + dt_{23}dt_{43} \right) \\
    &-t_{23}\left(dt_{12}dt_{24}+dt_{12}dt_{43}+dt_{31}dt_{24}+dt_{31}dt_{43}\right) \Big]\,,
    \end{aligned}
\end{equation}
where $t_{ij}$ denotes the Schwinger time associated to the edge from vertex $i$ to vertex $j$. We observe a surprising fact: $\alpha_{\pic{1ex}{bitriangle}}^2 = 0$. 

More generally, we find a remarkable fact:
\begin{equation}
    \alpha_\Gamma^2=0
\end{equation}
for any graph $\Gamma$ that we have checked with at least one loop! This combinatorial identity appears magical. It would imply a general non-renormalization theorem: \textit{loop corrections vanish in systems with $T\geq 2$.} This combinatorial statement is compatible with (and generalizes) Kontsevich's formality theorem, which claims specifically the absence of loop corrections for $H=0$, $T=2$. We will give a more detailed account of this non-renormalization theorem in Section \ref{sec:MismatchingTwists}.

Next, we turn to the holomorphic directions. For each holomorphic direction, the corresponding form is: 
\begin{equation}
    \rho_\Gamma[\lambda;z] = \int_{\bC^{|\Gamma_0|-1}} \left[\prod^{v\neq v_*}_{v \in \Gamma_0} \dvol_v e^{\lambda_v x^\bC_v}\right]\left[\prod_{e \in \Gamma_1}  e^{-(z_e + x^\bC_{e(0)}-x^\bC_{e(1)}) y_e} d y_e\right]\,,
\end{equation}
with $y_e = t_e^{-1} \left(\bar x^\bC_{e(0)}-\bar x^\bC_{e(1)}\right)$.

As a concrete example, we can study the bitriangle $2$-Laman graph in Figure \ref{fig:bitriangle} again. For $H=T=1$ the Schwinger integrand $\omega_{\pic{1ex}{bitriangle}} = \alpha_{\pic{1ex}{bitriangle}} \rho_{\pic{1ex}{bitriangle}} [\lambda,z]$ is found to take a very simple form
\begin{equation}
   \begin{aligned}
       \omega_{\pic{1ex}{bitriangle}} = & -\frac{\pi^2}{4}\left(t_{12}t_{23}+t_{12}t_{24}+t_{23}t_{24}+t_{23}t_{31}+t_{24}t_{31}+t_{12}t_{43} + t_{23} t_{43}+t_{31}t_{43} \right)^{-5/2}\\
  & (t_{12}t_{23}t_{31}t_{24}t_{43})  e^{-\lambda A^{-1} B z} \lambda_1(\lambda_1+\lambda\,,_2+\lambda_3)\ d\mathrm{Vol}_t
   \end{aligned}
\end{equation}
where $d\mathrm{Vol}_t$ is the volume form on $\mathbb{RP}_{>}^{4}$:
\begin{equation}
    \begin{aligned}
    d\mathrm{Vol}_t =
    -\frac{dt_{12}}{t_{12}}\frac{dt_{23}}{t_{23}}\frac{dt_{24}}{t_{24}}\frac{dt_{31}}{t_{31}}
    +\frac{dt_{12}}{t_{12}}\frac{dt_{23}}{t_{23}}\frac{dt_{24}}{t_{24}}\frac{dt_{43}}{t_{43}}
    -\frac{dt_{12}}{t_{12}}\frac{dt_{23}}{t_{23}}\frac{dt_{31}}{t_{31}}\frac{dt_{43}}{t_{43}}\\
    +\frac{dt_{12}}{t_{12}}\frac{dt_{24}}{t_{24}}\frac{dt_{31}}{t_{31}}\frac{dt_{43}}{t_{43}}
    -\frac{dt_{23}}{t_{23}}\frac{dt_{24}}{t_{24}}\frac{dt_{31}}{t_{31}}\frac{dt_{43}}{t_{43}}\,,
    \end{aligned}
\end{equation}
and the exponent is $\sum_{v,v'= 1}^3 \lambda_v A^{-1}_{vv'} (Bz)_{v'}$. $A$ is the weighted adjacency matrix
\begin{equation}
A= \left(\begin{array}{ccc}
t_{12}^{-1}+t_{31}^{-1} & -t_{12}^{-1} & -t_{31}^{-1} \\
-t_{12}^{-1} & t_{12}^{-1}+t_{23}^{-1}+t_{24}^{-1} & -t_{23}^{-1} \\
-t_{31}^{-1} & -t_{23}^{-1} & t_{23}^{-1}+t_{31}^{-1}+t_{43}^{-1}
\end{array}\right)
\end{equation}
and the vector $(Bz)_{v}$, $v= 1,2,3$ is
\begin{equation}
\left\{-\frac{z_{12}}{t_{12}}+\frac{z_{31}}{t_{31}},\ \frac{z_{12}}{t_{12}}-\frac{z_{23}}{t_{23}}-\frac{z_{24}}{t_{24}},\ \frac{z_{23}}{t_{23}}-\frac{z_{31}}{t_{31}}+\frac{z_{43}}{t_{43}}\right\}\,.
\end{equation}

As mentioned in the general discussion, finiteness of the $I_\Gamma$ integrals is not quite obvious in this presentation, due to potential divergences as we approach the boundary of $\bR \mathbb{P}_>^{|\Gamma_1|-1}$.\footnote{In order to discuss the boundary behaviour, we believe it is best to do a real blowup of the naive $\bR \mathbb{P}_\geq^{|\Gamma_1|-1}$ compactification of $\bR \mathbb{P}_>^{|\Gamma_1|-1}$: for each subset $S$ of vertices in $\Gamma$, we can define a co-dimension $1$ boundary component where we send the Schwinger times $t_e$ corresponding to edges in $\Gamma[S]$ to $0$, but keep track of their ratios. We expect $\alpha_\Gamma$ to remain finite in such a compactification, approaching $\alpha_{\Gamma[S]} \times \alpha_{\Gamma(S)}$ at the boundary component labelled by $S$, up to an appropriate sign. We leave a careful investigation of this to future work.}

\section{From Integrals to Brackets}\label{sec:brackets}

\newcommand{\Sym}{\mathrm{Sym}}
\newcommand{\del}{\partial}
We now explain how to employ the $I_\Gamma$ integrals in the computation of BRST anomalies/factorization algebra brackets in HT QFTs.

The free field action has the same first order form for all the (super)fields, so we collect them into a single object 
\begin{equation}
    \Phi \define \sum_a \phi^a v_a\,, \qquad v_a \in V\,,
\end{equation}
valued in some auxiliary super vector space $V$. Given the quantum numbers of an individual superfield $\phi^a$, we can adjust the quantum numbers (e.g. fermionic parity, ghost number, etc.) of the auxiliary vector $v_a$ so that the total field $\Phi$ has some well-defined convenient quantum numbers. We are not averse to working directly with the $\phi^a$, but a judicious choice of $V$ can tame a lot of sign annoyances. For example, taking $\Phi$ to be fermionic is surprisingly convenient, preventing annoying signs when $v_a$ is brought across $\phi_a$. We will take $\Phi$ of ghost number $0$, though other choices may be more convenient for specific $H$ and $T$.

We write the kinetic term using an appropriate pairing on $V$:
\begin{equation}
    \int \left( \Phi, \dd\Phi \right) \dvol\,.
\end{equation}
More specifically, the pairing on $V$ must violate ghost number and fermion number precisely by an amount $(H+T-1)$ so that the top form component of the integrand is a boson of ghost number $0$. In perturbation theory, we only really care about the inverse pairing
\begin{equation}
    \eta = \eta^{ab} v_a \otimes v_b    
\end{equation}
which enters the propagator.

In order to tame the contribution of derivatives to our calculations, we introduce generating functions of the form
\begin{equation}
    \phi_a(z) \define \sum_{n_*=0}^\infty \prod_{i=1}^H \left[\frac{(z^i)^{n_i}}{n_i!} \partial_{(x^\bC)^i}^{n_i} \right] \phi_a\,.
\end{equation}
Likewise, the total generating function $\Phi(z)$ is valued in the space of formal power series $V[[z]]$, so that the local operator $\partial_{1}^{n_1} \cdots \partial_{H}^{n_H} \phi_a$ is just the coefficient of the $(z^1)^{n_1} \cdots (z^H)^{n_H} \otimes v_a$ term in $\Phi(z)$. We can just as well think of the local operator above as an element $\calO \in V[[z]]^\vee$, defined as the dual to $(z^1)^{n_1} \cdots (z^H)^{n_H} \otimes v_a$.

More generally, a local operator can be be made from (a linear combination of) products of $m$ letters with arbitrarily many derivatives, i.e. (linear combinations) of terms of the form:
\begin{equation}
    \prod_{j=1}^m \partial_1^{n_1^{\!(j)}} \cdots \partial_H^{n_H^{\!(j)}} \phi_{a_j}\,.
\end{equation}
Such a term is of course just the coefficient of the monomial $\prod_{j=1}^m (z_j^1)^{n_1^{\!(j)}} \cdots (z_j^H)^{n_H^{\!(j)}}$ in the formal power series defined by
\begin{equation}
    \prod_{j=1}^m \phi_{a_j}(z_j)\,.
\end{equation}
If we collect our fields into the total field $\Phi$ (and include arbitrary linear combinations of the products above), the most general operator is just the coefficient of a power series in $S^\bullet V[[z]]$. Thus a general local operator $\calO$ is identified with a multilinear symmetric map in
\begin{equation}
     S^\bullet V[[z]]^\vee\,.
\end{equation}

At the end of a calculation, we can expand the answer in powers of $z$'s to get the contribution of individual derivatives. Whenever we Wick-contract shifted fields $\phi_a(z_a)$ and $\phi_b(z_b)$, the result is a propagator with holomorphic arguments shifted by the difference $z_a - z_b$:
\begin{equation}
    \langle \Phi(x_1^\bC+z_1, \bar x_1^\bC, x_1^\bR) \otimes \Phi(x_2^\bC+z_2, \bar x_2^\bC, x_2^\bR) \rangle = P(x_1-x_2+z_1-z_2) \eta
\end{equation}
as an element in $V[[z_1]]\otimes V[[z_2]]$. This is the origin of the $z_e$ variables in $I_\Gamma$, they are holomorphic shifts which appear as generating function variables to capture Feynman integrals for interactions with derivative operators.

\subsection{The BRST Anomaly}\label{sec:BRSTAnomaly}
We are considering collections of interactions of the form\footnote{Recall the general discussion in Section \ref{sec:pBracket}, where position-dependent interactions live in deformed spaces of interactions $\mathrm{In}_p$ and had accompanying $p$-dependent multilinear operations. Here we are in the same situation, but specifically with holomorphic position-dependent interactions.}
\begin{equation}
    e^{\lambda_v x^\bC_v}\cI_v = e^{\lambda_v x^\bC_v} \prod_{i} \phi_{a_{i,v}}(x^\bC_v+z_{i,v}, \bar x^\bC_v, x^\bR_v) \,.
\end{equation} 
In order to compute the perturbative BRST anomaly, we simply normal order the products of such interactions by performing Wick contractions with regularized propagators $P_{\epsilon}(x)$ and act with the free BRST differential on the result. Since the interactions are superfields, they satisfy the free $Q+\dd$ descent relations, and the BRST differential hitting a surviving superfield can be traded for the differential $-\dd$ acting on the same superfield. Integration by parts at each interaction vertex removes $\dd$ from superfields and move it to the propagators $\dd P_{\epsilon}(x) = K_{\epsilon}(x)$.

In order to get a universal answer, we now drop terms which are BRST exact in the free theory by bringing the surviving normal-ordered superfields back to the origin:
\begin{equation}
     \prod'_{i} e^{(\lambda_v+ \partial_{z_{i,v}}) x^\bC_v} \phi_{a_{i,v}}(z_{i,v}, 0, 0) \,.
\end{equation}  
We use the $'$ in the product to indicate that we only include superfields $\phi_{a_i,v}$ which have not been contracted into propagators. We also remove the overall translation from the integration region (by setting $x_{v^*}=0$ for a chosen vertex $v^*$) as the anomaly is an integrated interaction and we want to compute the integrand.

At $n$-th order in perturbation theory, there will be a collection of Feynman diagrams consisting of $n$ partially contracted interaction vertices which contribute to a BRST anomaly bracket. This can be written schematically as
\begin{equation}
    \{ \cI_1 \, {}_{\lambda_1} \,\dots\, {}_{\lambda_{n-1}}\, \cI_{v^*} \} = \sum_\Gamma \pm I_\Gamma(\lambda_v + \partial_v;z_e) \prod_v \prod'_{i} \phi_{a_{i,v}}(z_{i,v})\,,
\end{equation}
where $\partial_v := \sum_j \partial_{z_{j,v}}$. The Grassmann signs accrue from reordering superfields to act with Wick contractions, reordering vertices, etc. Working with a generating field $\Phi$ can help systematize them. The derivative in the $\lambda_v$ argument of $I_\Gamma$ acts on the $z_{i,v}$ for all the surviving superfields associated to $v$, while the $z_e$ shifts are differences $z_{i,e(0)} - z_{j,e(1)}$ for the Wick-contracted fields. 

In conclusion, we have defined multilinear $\lambda$-brackets:
\begin{equation}
    \{ {\cal O}_1 \, {}_{\lambda_1} \, \dots \, {\cal O}_{n-1} \, {}_{\lambda_{n-1}} \, {\cal O}_{n} \}
\end{equation}
as a sum over $(H+T)$-Laman graphs with $n$ vertices. Each term involves a sum over all in-equivalent ways to associate the $n$ arguments to the vertices of $\Gamma$. The order of the arguments in the $\lambda$-bracket gives an order to the vertices. The last vertex will be $v_*$ and the $\lambda_i$ are mapped to $\lambda_v$ for the remaining vertices. The $v_*$ vertex will have $\lambda_{v_*} = -\sum_{i=1}^{n-1} {\lambda_{v_i}}$, which is consistent with (holomorphic) momentum conservation (see Section \ref{sec:pBracket}).

Suppose the entries of an $n$-ary bracket are of the special form ${\cal O}_v = \prod_{i=1}^{n_v} \phi_{a_{v,i}}(z_{v,i})$. To compute the contribution from $\Gamma$, we must sum over all inequivalent ways to associate the arguments to vertices of $\Gamma$. Focus on an edge $e$, it has two endpoint vertices $e(0)$ and $e(1)$, which are associated with some local operators $\calO_{e(0)}$ and $\calO_{e(1)}$ built as normal ordered collection of fields. Suppose that the edge $e$ connects the factors $i_e$ and $j_e$ in $\calO_{e(0)}$ and $\calO_{e(1)}$ respectively, and define $z_e = z_{e(0), i_e} - z_{e(1), j_e}$. Then we pair up these fields $\phi_{a_{e(0),i_e}}$ and $\phi_{a_{e(1),j_e}}$ with $\eta$ in the order dictated by the ordering of edges of $\Gamma$, removing them from the overall product ${\cal O}_1 \cdots {\cal O}_{n}$, and act with $I_\Gamma(\lambda+ \partial;z)$ on the surviving $\phi_a(z_a)$. Here, $\partial_v$ acts on the $z_a$ arguments of the symbols coming from ${\cal O}_v$.

Alternatively, if we represent local operators as functions in $S^\bullet V[[z]]^\vee$, we can represent the output of the bracket as the action of the tensor product of the functions on an element in $V[[z]]^{\otimes \bullet}$ built from a product of $\eta$'s, $I_\Gamma$, and $\Phi$'s. In such a presentation, the brackets could then be represented, perhaps more economically, as elements in $V[[z]]^{\otimes \bullet}$. 

\subsection{Properties of \texorpdfstring{$\lambda$}{λ}-Brackets} \label{sec:PropertiesofLambdaBracket}
Thanks to the basic properties of the $I_\Gamma$ integrals (see Section \ref{sec:GraphIdentities}), the brackets satisfy a number of properties which give it the structure of a (homotopical generalization of a) Lie conformal (super)algebra.\footnote{We thank Ahsan Khan for educating one of us on this topic at an early stage of the project.} Translation invariance implies that
\begin{equation}
   \partial \{ {\cal O}_1 \, {}_{\lambda_1} \, \cdots \, {\cal O}_{n-1} \, {}_{\lambda_{n-1}} \, {\cal O}_{n} \} = \sum_{i=1}^n \{ {\cal O}_1 \, {}_{\lambda_1} \, \cdots \, \partial {\cal O}_{i} \, {}_{\lambda_{i}} \, \cdots \, {\cal O}_{n} \}\,,
\end{equation}
while the shift property of \eqref{eq:identitylambda+partial} implies:
\begin{equation}
   \{ {\cal O}_1 \, {}_{\lambda_1} \, \cdots \, (\partial+ \lambda_i) {\cal O}_{i} \, {}_{\lambda_{i}} \, \dots \, {\cal O}_{n} \} = 0\,, \qquad \forall i<n\,.
\end{equation}
Combining the two identities, we can re-write this as
\begin{equation}
\{ {\cal O}_1 \, {}_{\lambda_1} \, \dots \, {\cal O}_{n-1} \, {}_{\lambda_{n-1}} \,\partial \mathcal{O}_n\}
=
(\partial+\lambda_1+\dots+\lambda_{n-1}) \{ {\cal O}_1 \, {}_{\lambda_1} \, \dots \, {\cal O}_{n-1} \, {}_{\lambda_{n-1}} \, {\cal O}_{n} \}\,.
\end{equation}
Together the translation invariance and shift-property imply the sesquilinearity of our $\lambda$-bracket.

The multi-brackets are graded-symmetric in all the inputs, with a grading shifted by $H+T$, under permutations of the pairs $({\cal O}_i, \lambda_i)$ for $i<n$. The symmetry can be extended to the last argument by formally defining
\begin{equation}
    \lambda_n = - \partial - \sum_{i<n} \lambda_i\,,
\end{equation}
where $\partial$ acts on the whole bracket, as seen above.

Most importantly, the quadratic identities for the $I$ imply that the brackets satisfy associativity relations of the form 
\begin{equation}\label{eq:quadraticIdentity}
    \sum_{k} (-1)^{\dots}
    \{
    \calO_1 \,{}_{\lambda_1}\, \dots \, 
    \calO_{n-k} \, {}_{\lambda_{n-k}} \{     
        \calO_{n-k+1} \,{}_{\lambda_{n-1}}\, \dots \,{}_{\lambda_{n}}\, \calO_{n+1}
    \}\} = 0
\end{equation}
acting on $n+1$ arguments and depending on $n$ parameters $\lambda_i$. The sum runs over all the ways to split the $n+1$ arguments in two groups and the parameters in the brackets are selected so that an insertion of $(\partial + \lambda_i) {\cal O}_i$ as the $i$-th argument gives $0$. This is perfectly analogous to the way $\lambda$ is split into $\lambda[S]$ and $\lambda(S)$ in the quadratic identity for $I$. We illustrate these quadratic identities in some examples in Appendix \ref{app:Associativity}.

These are the expected axioms for the $\lambda$-brackets of a holomorphic-topological field theory. We propose that these are indeed the $\lambda$-brackets for a theory of free semi-chiral fields valued in $V$ with kinetic term $\eta \, \dd$.

\subsection{Composite BRST Anomalies and Regularized Products}\label{sec:AuxBracket}
To probe a larger collection of possible interactions, we can add additional fields to our original theory i.e. couple it to an auxiliary theory, as described briefly in Section \ref{sec:furtherStructures}. 

We can break down the brackets in the combined theory into a sum of terms, characterized by the pattern of Wick contractions between fields in the auxiliary theory. Denote the pattern of contractions as a subgraph $\gamma \subset \Gamma$. From the point of view of the original theory, we are putting regularized propagators along edges of $\gamma$ by hand. This leads to the definition of a modified bracket  
\begin{equation}
    \{ {\cal O}_1 \, {}_{\lambda_1} \, \dots \, {\cal O}_{n-1} \, {}_{\lambda_{n-1}} \, {\cal O}_{n} \}^{\gamma,z}\,,
\end{equation}
where $\gamma$ is a graph with vertices which are entries of the bracket. We use the same $I_\Gamma$'s as before, but do not do Wick contractions along $\gamma \subset \Gamma$, and treat the shifts $z_e$ for $e \in \gamma$ as parameters of the bracket. Of course, are interested in a bracket with auxiliary Wick contraction pattern $\gamma$, we only sum over $\Gamma$ containing $\gamma$ to compute the corresponding brackets.

\begin{figure}[t]
\centering
\begin{minipage}{0.31\textwidth}
\centering
\begin{tikzpicture}
	[
	baseline={(current bounding box.center)},
	line join=round
	]
        \def\gS{0.9};
	\coordinate (pd1) at (1.*\gS,0.*\gS);
	\coordinate (pd2) at (-1.*\gS,0.*\gS);
        \coordinate (label) at (0*\gS,2*\gS);
        \draw (label) node {$\{
            \calO_1 \,{}_{\lambda_1}\,  
            \calO_2
            \}^{{\pick{1.5ex}{segment}},{z_{12}}}$};
	\draw[GraphEdge, red] (pd1) -- (pd2) node[midway, above] {$z_{12}$};
	\draw (pd1) node[GraphNode] {} node[right] {$\lambda_{1}$};
	\draw (pd2) node[GraphNode] {} node[left] {$\lambda_{2}$};
\end{tikzpicture}
\end{minipage}
\hspace{0.25cm}
\begin{minipage}{0.31\textwidth}
\centering
 \begin{tikzpicture}
    [
	baseline={(current bounding box.center)},
	line join=round
	]
    \def\gS{1.5};
	\coordinate (pd1) at (-0.866*\gS,-0.5*\gS);
	\coordinate (pd2) at (0.*\gS,1.*\gS);
	\coordinate (pd3) at (0.866*\gS,-0.5*\gS);
        \coordinate (label) at (0*\gS,2*\gS);
        \draw (label) node {$\{
            \calO_1 \,{}_{\lambda_1}\,  
            \calO_2 \,{}_{\lambda_2}\, \calO_3
            \}^{{\pick{1.5ex}{segment}},z_{13}}$};
	\draw[GraphEdge] (pd1) -- (pd2) node[midway, left] {$z_{12}$};
	\draw[GraphEdge,red] (pd1) -- (pd3) node[midway, above] {$z_{13}$};
	\draw[GraphEdge] (pd2) -- (pd3) node[midway, right] {$z_{23}$};
	\draw (pd1) node[GraphNode] {} node[left] {$\lambda_{1}$};
	\draw (pd2) node[GraphNode] {} node[above] {$\lambda_{2}$};
	\draw (pd3) node[GraphNode] {} node[right] {$\lambda_{3}$};
   \end{tikzpicture}
\end{minipage}
\hspace{0.25cm}
\begin{minipage}{0.31\textwidth}
\centering
 \begin{tikzpicture}
    [
	baseline={(current bounding box.center)},
	line join=round
	]
    \def\gS{1.5};
	\coordinate (pd1) at (-0.866*\gS,-0.5*\gS);
	\coordinate (pd2) at (0.*\gS,1.*\gS);
	\coordinate (pd3) at (0.866*\gS,-0.5*\gS);
        \coordinate (label) at (0*\gS,2*\gS);
        \draw (label) node {$\{
            \calO_1 \,{}_{\lambda_1}\,  
            \calO_2 \,{}_{\lambda_2}\, \calO_3
            \}^{{\pick{1.5ex}{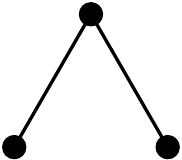}},{z_{12},z_{23}}}$};
	\draw[GraphEdge,red] (pd1) -- (pd2) node[midway, left] {$z_{12}$};
	\draw[GraphEdge] (pd1) -- (pd3) node[midway, above] {$z_{13}$};
	\draw[GraphEdge,red] (pd2) -- (pd3) node[midway, right] {$z_{23}$};
	\draw (pd1) node[GraphNode] {} node[left] {$\lambda_{1}$};
	\draw (pd2) node[GraphNode] {} node[above] {$\lambda_{2}$};
	\draw (pd3) node[GraphNode] {} node[right] {$\lambda_{3}$};
   \end{tikzpicture}
\end{minipage}
\caption{Above we mark the propagators of the original theory in black and the auxiliary propagators in red, above each diagram is the bracket that they contribute to. Left, $\gamma = \Gamma$ is a propagator, which just leads to the regularized product of the input vertices. When $H+T=1$, there are many graphs with just two vertices and the regularized product also has higher loop corrections. Middle, one edge of the 2-Laman loop diagram is auxiliary. The graph corresponds to the leading corrections to the regularized product. Right, a triangle with two coloured edges is the associator for the regularized product at free and leading order.}
\label{fig:auxPatterns}
\end{figure}
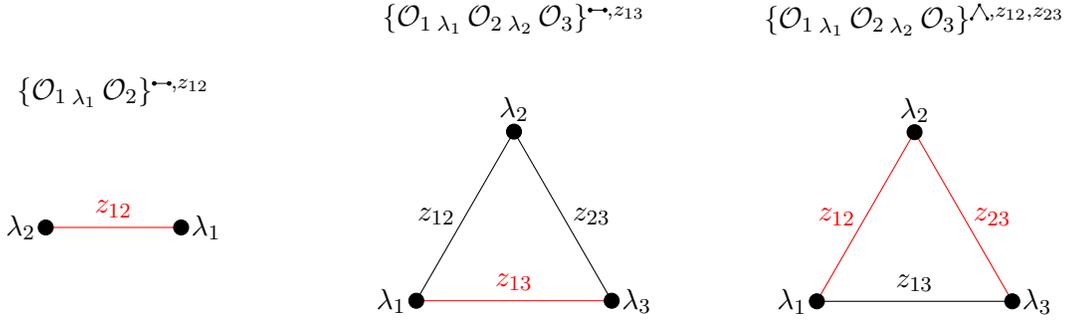

The corresponding quadratic axioms for these modified brackets can be written with a bit of patience. Each axiom is labelled by some pattern $\gamma$, and we sum over all possible ways to distribute the edges of $\gamma$ between the inner and outer brackets. Schematically, these are sums of the form 
\begin{equation}
    \sum_{\gamma_{\mathrm{in}}}\sum_{k} (-1)^{\dots}
    \{
    \calO_1 \,{}_{\lambda_1}\, \dots \, 
    \calO_{n-k} \, {}_{\lambda_{n-k}} \{     
        \calO_{n-k+1} \,{}_{\lambda_{n-1}}\, \dots \,{}_{\lambda_{n}}\, \calO_{n+1}
    \}^{\gamma_{\mathrm{in}},z_{\mathrm{in}}}\}^{\gamma_{\mathrm{out}},z_{\mathrm{out}}} = 0\,.
\end{equation}
We have decomposed the edges of $\gamma$ into an inner and outer set $\gamma_1 = \gamma_{\mathrm{in},1} \cup \gamma_{\mathrm{out},1}$ above, and chosen the last entry of the outer bracket to correspond to the ``new vertex'' obtained by shrinking $\gamma_{\mathrm{in}}$ (following our usual shrinking subgraph procedure).


The simplest example of such a modified bracket would be a regularized product, with $\gamma = \Gamma$ being the segment with two vertices: this returns $e^{\lambda z} O_1(z) O_2(0)$. If $H+T=1$, there are extra higher loop contributions to the regularized product of two operators. See Figure \ref{fig:auxPatterns} for some additional examples.

\subsection{Brackets as BRST Anomalies}
We now return to our examples to stress the large amount of information in an HT theory captured by BRST anomaly brackets, possibly with the help of extra auxiliary spectator fields. In simple examples, the information includes essentially all the HT factorization algebra information, i.e. all the `generalized OPE'' information for the theories. It would be very interesting to know how much information is captured by BRST anomaly brackets in general $H$ and $T$, and how general the choice of ``spectator'' theories need to be.

\subsubsection{Example: \texorpdfstring{$H=0$}{H=0}, \texorpdfstring{$T=1$}{T=1} Brackets in Topological Quantum Mechanics}\label{sec:H0T1Example}
As anticipated in Section \ref{sec:exampleH0T1}, a free topological quantum mechanics has action
\begin{equation}
    S =\int_{\mathbb{R}} \left[p \,\dd q+ \psi \,\dd\psi \right] = \int_{\mathbb{R}} \left[p^{(0)} \dot q^{(0)}+ \psi^{(0)} \dot \psi^{(0)} \right]dx\, .
\end{equation}
An interaction $\cI$ introduces an extra BRST differential acting on local operators as an (anti)commutator with $\cI$:
\begin{equation}
    Q_0\Phi = [\cI, \Phi] \,.
\end{equation}
This extra differential is well-defined if $\cI^2=0$, so the BRST anomaly is controlled by the anti-commutator $[\cI_i, \cI_j]$ of the interaction operators. 

If we combine two theories $\mathcal{T}_1$ and $\mathcal{T}_2$, the anti-commutators contain enough information to reconstruct the full operator algebra of the topological QM $\mathcal{T}_1 \times \mathcal{T}_2$, i.e. the Moyal star product for $\calT_1 \times \calT_2$. For example, if we add an auxiliary fermion $\psi$ and look at interactions of the form $\psi \cI_1+\psi \cI_2$, where $\mathcal{I}_i$ is an operator of $\mathcal{T}_i$, the BRST anomaly with the auxiliary system becomes proportional to a graded {\it anti}-commutator
\begin{equation}
    [\psi \cI_i, \psi \cI_j] = (-1)^{|\cI_i|}(\cI_i \cI_j + (-1)^{|\cI_i||\cI_j|} \cI_j \cI_i)\,.
\end{equation}
We can combine the BRST anomaly for the system coupled to the auxiliary fermion $[\psi \cI_i, \psi \cI_j]$ with the original BRST anomaly $[\cI_i, \cI_j]$ to recover the product $\cI_i \cI_j$.

We will verify that this is indeed what our regularized Feynman diagrams compute. First of all, recall that 1-Laman graphs only have two vertices, so indeed we only obtain BRST brackets with two entries. We already computed the integrals with no auxiliary propagators, to get some numbers $I_N$, which match the coefficients of the Moyal commutator: 
\begin{equation}
\begin{aligned}
    \{f,g\}_{\mathrm{ours}}=[f,g]_{\star} 
        &\define \frac{1}{\hbar}(f\star g-g\star f)\\
        &= \eta^{i j}\left(\partial_i f\right)\left(\partial_j g\right)+ \frac{\hbar^2}{24}  \eta^{i j} \eta^{k m} \eta^{nl}\left(\partial_i \partial_k \partial_n f\right)\left(\partial_j \partial_m \partial_l g\right) + \dots
\end{aligned}\label{eq:moyalCom}
\end{equation}
where the numerical coefficients are $I_N/N!$ and the factor of $\hbar^{N-1}$ counts the number $N-1$ of loops in the diagram. The vanishing of odd-loop terms in the Moyal commutator is consistent with our explicit calculations in Section \ref{sec:exampleH0T1} and result in Section \ref{sec:projective}.

If we insert an auxiliary propagator between the two vertices in order to compute an ``auxiliary bracket'' as in Section \ref{sec:AuxBracket}, we are precisely 
reproducing the $[\psi \cI_i, \psi \cI_j]$ calculation above: the extra propagator comes from the auxiliary $\psi$ and we obtain the graded anti-commutator from our bracket. 

A more elegant procedure, which generalizes to any 1d topological (sub)system, is to combine the standard and auxiliary brackets with arbitrary numbers of 1d propagators and auxiliary propagators to get integrals where the insertions on the 1d line are {\it ordered} along the line. The combined brackets and auxiliary brackets, with ordered insertions, define a binary operation with coefficients
\begin{equation}
    I^\star_{N} =  2^{-N}\,.
\end{equation}
This coincides with Moyal's star product for the Poisson bracket using $\eta^{-1}$. Indeed, Moyal's star product using $\eta$ is precisely given by
\begin{equation}
    f \star g = m \circ e^{\frac{\eta}{2}} (f \otimes g)\,,
\end{equation}
i.e. the $N$'th term is computed by contracting $f \otimes g$ with $N$ copies of $\eta$ in all possible ways, multiplying by $I^\star_{N}$, and then multiplying together the surviving powers of $\phi_a$. In other words, computing BRST brackets and BRST brackets with the auxiliary fermion allows us to recover the full Moyal star product in addition to the Moyal commutator.

Up to this point, we were studying purely 1d topological systems, so the local operators were described by a dga (or dgla) given by the Moyal star product (or Moyal commutator). In a general 1d topological subsystem of a higher-dimensional system, the 1d subsystem is equipped with a collection of $L_\infty$-brackets (generalizing quantum mechanical commutators) controlling the deformations of the subsystem. This perspective with the auxiliary degrees of freedom similarly enhances the $L_\infty$-brackets to full $A_\infty$-operations (generalizing quantum mechanical products) which control the coupling of the 1d subsystem to any auxiliary 1d quantum-mechanical system. We can understand this enhancement by promoting interactions $\cI$ to a matrix acting on the auxiliary Hilbert space, and write a non-commutative MC equation valued in such matrices:
\begin{equation}\label{eq:NCMCE}
    \{\cI, \cI\}_{A_\infty} + \{\cI, \cI, \cI\}_{A_\infty} + \cdots = 0\,.
\end{equation}
We will return to such examples in Section \ref{sec:LineDefects}.

It is amusing to observe that the $A_\infty$-brackets computed from Feynman diagrams of the full system should naively be hard to compute, due to the ordering of positions ruining the Gaussian nature of the position integrals. Once we write the $A_\infty$-operations as linear combinations of the $\{\cI, \dots\}^\gamma$ brackets, though, the Gaussian nature of individual terms is restored as the 1d propagators (step functions) are recast as integrals over auxiliary Schwinger times. 

\subsubsection{Example: \texorpdfstring{$H=1$}{H=1}, \texorpdfstring{$T=0$}{T=0} Brackets in 2d Chiral Algebras}
We can also revisit the case of 2d chiral algebras 
\begin{equation}
    S = \int_{\mathbb{C}} \left[ \psi\bar{\partial}\psi + \beta \bar{\partial} \gamma \right] \dd z\,.
\end{equation}
As before, ``interactions'' do not actually change the action of physical fields (i.e. zero form components), but rather add an extra BRST differential $Q_0$ with BRST current $\cI(\beta, \gamma, \psi)$:
\begin{equation}
    [Q_0,\Phi(z)] = \{\cI, \Phi\}(z) = \oint_{|w-z|=\epsilon} \frac{dw}{2 \pi i}\,\cI(w)\, \Phi(z)\,.
\end{equation}
The differential may fail to be nilpotent, leading to a quadratic BRST anomaly controlled by the simple pole in the $\cI$-$\cI$ OPE. More generally, position-dependent deformations give access to the whole ``$\lambda$-bracket''
\begin{equation}
    \{\cI {}_\lambda \Phi\}(z) = \oint_{|w-z|=\epsilon} \frac{dw}{2 \pi i}e^{\lambda (w-z)} \,\cI(w)\, \Phi(z)\,.
\end{equation}
The associated BRST anomalies thus ``catch'' the entire singular part of the OPE of two interactions.

In order to capture the whole chiral algebra structure, the singular part of the OPE is not quite enough. For example, some naive composite operators may actually vanish. As with the previous example, combining two systems gives information about the full OPE. For example, if we add a spectator fermion $\psi$ and look at interactions of the form $\psi \cI$, the BRST anomaly with the auxiliary system is controlled by 
\begin{align}
    \oint_{|w-z|=\epsilon} \frac{dw}{2 \pi i}\,\psi(w) \cI_i(w)\, \psi(z) \cI_j(z) 
    &= \oint_{|w-z|=\epsilon} \frac{dw}{2 \pi i}\,\frac{1}{z-w} \cI_i(w)\, \cI_j(z) \cr 
    &+\oint_{|w-z|=\epsilon} \frac{dw}{2 \pi i}\, \cI_i(w) \NORD{\psi(w)\psi(z)} \cI_j(z) \,,
\end{align}
which contains the regular part of the $\cI_i(w) \cI_j(z)$ OPE as well. A chiral algebra can be reconstructed fully from the $\lambda$-bracket and the regular term in the OPE:
\begin{equation}
    ( f, g ) \define \oint \frac{dw}{2 \pi i} \frac{1}{w} f(w) \cdot g(0)\,.
\end{equation}

We will verify that our regularized Feynman diagrams indeed compute all of this data. In this example we can easily identify the standard brackets. First, promote $\eta$ to a pairing
\begin{equation}
    \eta': \quad \phi_a(w_a)\otimes \phi_b(w_b) \mapsto \frac{\eta_{ab}}{w_a - w_b}\,,
\end{equation}
then Wick contractions 
\begin{equation}
     m \circ e^{\eta'} ( f(w) \otimes g(0) )
\end{equation}
are just a standard normal-ordering for the OPE following 
\begin{equation}
    \phi_a(w) \cdot \phi_b(0)\sim \frac{\eta_{ab}}{w}\,.
\end{equation}
More generally, a term with $N$ Wick contractions will involve a product of the form 
\begin{equation}
    \prod_{e=1}^N \frac{1}{w+z_e} \,,
\end{equation}
for $z_e = u_a - u_b$ parameters associated to the contraction of $\phi_a(w+u_a)\otimes \phi_b(u_b)$. Using the definition of $I_N$ from Section \ref{sec:H1T0ExampleDiagrams}, we can compute the $\lambda$-brackets in the $\beta \gamma$ system, which collect the singular parts of the OPE:
\begin{equation}
    \{ f \, {}_\lambda \, g \} \define \oint \frac{dw}{2 \pi i} e^{\lambda w} e^{\eta} \left(f(w) \otimes g(0)\right)\,.
\end{equation}
This, of course, is compatible with this theory being a free $\beta \gamma$ system. 

We can also add $n$ auxiliary fermion propagators to the calculation, this simply adds a negative power $w^{-n}$ to the integral:
\begin{equation}
    \{ f \, {}_\lambda \, g \}_{n} \define \oint \frac{dw}{2 \pi i} \frac{1}{ w^n} e^{\lambda w} e^{\eta} \left(f(w) \otimes g(0)\right)\,.
\end{equation}
This clearly contains the regular part of the OPE, albeit in a somewhat redundant form. All together, our BRST brackets, with and without the auxiliary fermions, compute the data of the chiral algebra.

With a bit of patience, one can match the quadratic axioms satisfied formally by our general brackets with the known axioms of for $\lambda$-brackets and regularized products in a chiral algebra. Specifically, the known axioms for $\lambda$-brackets for $H=1$ and $T=0$ form what is called a (graded) {\it Lie conformal superalgebra}. A Lie conformal algebra is a vector space $\cV$ equipped with an endomorphism $\del \colon \cV \to \cV$ and a $\lambda$-bracket $\{A \, {}_\lambda \, B\}$ satisfying \cite{Bakalov_2003, secretKac} (see also \cite{wang2022n} and \cite{Oh:2019mcg, Garner:2022its, Budzik:2022mpd, Budzik:2023xbr} for physics discussions):
\begin{enumerate}
    \itemsep 0em
	\item \textbf{Sesquilinearity}. $\{\partial A\,{}_\lambda\,B\} = - \lambda \{A\,{}_\lambda\,B\}$ and $\{ A\,{}_\lambda\,\partial B\} = (\partial + \lambda) \{A\,{}_\lambda\,B\}$.
	\item \textbf{(Graded) Skew-Symmetry}. $\{A\,{}_\lambda\,B\} = - (-1)^{|A||B|} \{B\,{}_{-\lambda- \partial }\,A\}$.
	\item \textbf{$\lambda$-Jacobi Identity}. $\{A\,{}_\lambda\, \{B\,{}_{\lambda'}\,C\}\} - (-1)^{|A||B|}\{B\,{}_{\lambda'} \, \{A\,{}_\lambda\,C\}\} = \{\{A\,{}_\lambda\, B\} \,{}_{\lambda+\lambda'} \,C\}$.
\end{enumerate}
Here we denote the fermion parity of operators by $\abs{\,\cdot\,}$ and accounted for the potential Grassmann nature of the operators. 

The manipulations used to prove these properties of $\lambda$-brackets are precisely the sort of integration contour reorganizations we employed to derive the identities satisfied by $I_N$. Conversely, the identities satisfied by $I_N$ guarantee that the bracket associated to them is a $\lambda$-bracket in the usual sense (see e.g the identity in \eqref{eq:IdentityH1T0}).

The axioms satisfied by the regularized product and the $\lambda$-bracket are somewhat lengthy. The combination we encountered above as the auxiliary bracket with one extra propagator:
\begin{equation}
    ( f \, {}_\lambda \, g ) \define \{ f \, {}_\lambda \, g \}_{1} =  \oint \frac{dw}{2 \pi i} \frac{1}{w} e^{\lambda w} f(w) \cdot g(0)\,,
\end{equation}
is a repackaging of the regularized product
\begin{equation}
    ( f \, {}_\lambda \, g )=(f,g) + \int_0^\lambda d \lambda' \{ f {}_{\lambda'} g\}\,,
\end{equation}
and is constructed in the same manner as the $\lambda$-bracket, except one would replace $I_N(\lambda;z_e)$ with $I_{N+1}(\lambda;z_e)$, i.e. inserting an extra auxiliary propagator between operator insertion points by hand. Its quadratic axioms imply the properties for the regularized product.

\subsubsection{Example: \texorpdfstring{$H=2$}{H=2}, \texorpdfstring{$T=0$}{T=0} Brackets in 4d Holomorphic Twists}
We include here some explicit axioms for the case of holomorphic theories in two complex dimensions (see also \cite{Budzik:2022mpd, Budzik:2023xbr}).

The free limit of the holomorphic twist of a four-dimensional $\cN=1$ theory has full space of BV fields given by $\Omega^{0,*}(\mathbb{C}^2)\otimes W$, where $W$ is a graded vector space equipped with a skew-symmetric pairing $\omega$ of cohomological degree $+1$. In this example, $V = W^*$ which is equipped with the degree $-1$ pairing $\eta \define \omega^{-1}$. The $\lambda$-bracket will carry cohomological degree $-1$.

The situation is formally similar to the usual $\beta\gamma$ system for $T=0,H=1$. We can collect the singular part of the OPE via the following $\lambda$ bracket
\begin{equation}
    \{f \,_\lambda \, g\} \define \oint_{S^3} \omega_{\text{BM}}(w) e^{\lambda \cdot w} e^{\eta} \left(f(w) \otimes g(0) \right)
\end{equation}
where $w_1$ and $w_2$ are coordinates of $\mathbb{C}^2$, $\lambda \cdot w \define \lambda_1 w_1 + \lambda_2 w_2$, and
\begin{equation}
    \omega_{\text{BM}}(w) = \frac{d^2 w}{(2\pi i)^2} \frac{\Bar{w}_1 \d \Bar{w}_2 - \Bar{w}_2 \d \Bar{w}_1}{|w|^4} 
\end{equation}
is the Bochner--Martinelli integral kernel.

As explained in Section \ref{sec:PropertiesofLambdaBracket}, quadratic identities of the ${I}_{\Gamma}$ integrals leads to associativity relations satisfied by the brackets. We now list the first few examples of the associativity relations whose detailed derivations can be found in Appendix \ref{app:Associativity}. The first one is simply the generalization of Jacobi identity:
\begin{equation}
    \{\mathcal{O}_1  \, {}_{\lambda_1} \{ \mathcal{O}_2\, {}_{\lambda_2} \,\mathcal{O}_3 \}\} - (-1)^{(|\mathcal{O}_1|+1)(|\mathcal{O}_2|+1)}\{\mathcal{O}_2  \, {}_{\lambda_2} \{ \mathcal{O}_1\, {}_{\lambda_1} \,\mathcal{O}_3 \}\} +(-1)^{|\mathcal{O}_1|} \{\{ \mathcal{O}_1\, {}_{\lambda_1} \,\mathcal{O}_2 \} \, {}_{\lambda_1+\lambda_2} \mathcal{O}_3   \}=0\,.
\end{equation}
The second one involves a bracket of $2$ arguments and a bracket of $3$ arguments
\begin{align}
     0 = &\{\{ \mathcal{O}_1\, {}_{\lambda_1} \,\mathcal{O}_2 \} \, {}_{\lambda_1+\lambda_2} \,\mathcal{O}_3  \, {}_{\lambda_3}\, \mathcal{O}_4
     \} 
        + (-1)^{|\cO_2||\cO_3|}\{\{ \mathcal{O}_1\, {}_{\lambda_1} \,\mathcal{O}_3 \} \, {}_{\lambda_1+\lambda_3} \,\mathcal{O}_2  \, {}_{\lambda_2}\, \mathcal{O}_4
     \} \nonumber
     \\
     + &(-1)^{|\cO_1|+|\cO_2|}
     \{ \mathcal{O}_1\, {}_{\lambda_1} \,\mathcal{O}_2  \, {}_{\lambda_2} \,\{ \mathcal{O}_3  \, {}_{\lambda_3}\, \mathcal{O}_4\}
     \} 
     + (-1)^{|\cO_1|+(|\cO_2|+1)|\cO_3|}
     \{ \mathcal{O}_1\, {}_{\lambda_1} \,\mathcal{O}_3  \, {}_{\lambda_3} \,\{ \mathcal{O}_2  \, {}_{\lambda_2}\, \mathcal{O}_4\}
     \} \nonumber
     \\
     + &(-1)^{|\cO_1|}
     \{ \mathcal{O}_1\, {}_{\lambda_1} \, \{\mathcal{O}_2  \, {}_{\lambda_2} \, \mathcal{O}_3 \} \, {}_{\lambda_2+\lambda_3}\, \mathcal{O}_4
     \} \nonumber
     \\
     + &\left(\frac{1+(-1)^{|\cO_1|}}{2}(-1)^{|\cO_2|+|\cO_3|}+\frac{1-(-1)^{|\cO_1|}}{2}\right)
     \{ \mathcal{O}_2\, {}_{\lambda_2} \,\mathcal{O}_3  \, {}_{\lambda_3} \,\{ \mathcal{O}_1  \, {}_{\lambda_1}\, \mathcal{O}_4\}
     \}
     \\ 
     + &\{\{ \mathcal{O}_1\, {}_{\lambda_1} \,\mathcal{O}_2  \, {}_{\lambda_2} \, \mathcal{O}_3 \} \, {}_{\lambda_1 + \lambda_2+\lambda_3}\, \mathcal{O}_4
     \}
     +(-1)^{|\cO_1|}\{ \mathcal{O}_1\, {}_{\lambda_1}\,\{ \mathcal{O}_2  \, {}_{\lambda_2} \, \mathcal{O}_3  \, {}_{\lambda_3}\, \mathcal{O}_4\}
     \} \nonumber
     \\
     +&(-1)^{(|\cO_1|+1)|\cO_2|} \{ \mathcal{O}_2\, {}_{\lambda_2}\,\{ \mathcal{O}_1  \, {}_{\lambda_1} \, \mathcal{O}_3  \, {}_{\lambda_3}\, \mathcal{O}_4\}
     \}
     +(-1)^{(|\cO_1|+|\cO_2|+1)|\cO_3|} \{ \mathcal{O}_3\, {}_{\lambda_3}\,\{ \mathcal{O}_1  \, {}_{\lambda_1} \, \mathcal{O}_2 \, {}_{\lambda_2}\, \mathcal{O}_4\}
     \} \nonumber \,.
\end{align}
and so on.

\subsection{Interacting Brackets}
In our formalism, an interacting theory described perturbatively by an interaction $\cI$ can be deformed further by looking at theory with perturbative interaction $\cI + \cI'$. As a result, the brackets which control the extra BRST anomalies 
from introducing $\cI'$ can be written as 
\begin{equation}
    \{\cO_1, \dots, \cO_n\}_\cI \define \{\cO_1, \dots, \cO_n\} + \{\cI, \cO_1, \dots, \cO_n\}+ \frac12 \{\cI, \cI, \cO_1, \dots, \cO_n\}+ \dots\,.
\end{equation}
More generally, any set of brackets is defined in a similar manner, e.g. the brackets for auxiliary fields become
\begin{equation}
    \{\cO_1, \dots, \cO_n\}^\gamma_\cI \define \{\cO_1, \dots, \cO_n\}^\gamma + \{\cI, \cO_1, \dots, \cO_n\}^\gamma+ \frac12 \{\cI, \cI, \cO_1, \dots, \cO_n\}^\gamma+ \dots\,,
\end{equation}
and so on. Thus the perturbative factorization algebra brackets on the interacting theory can be recovered systematically from those of the free theory. 

\section{Defect Integrals}\label{sec:DefectIntegrals}
We can extend our analysis to a variety of extended defects by repeating analogous procedures to all of our preceding bulk analyses. The simplest scenario is that we select some subsets of directions in the bulk, $H^\partial$ and $T^\partial$ say, and add some standard collection of free superfields $\phi^\partial$ which are holomorphic in the $H^\partial$ directions and topological in the $T^\partial$ directions. We can then couple the bulk and defect fields by interactions $\cI^\partial$ integrated along the defect directions only. By locality, this construction can only add new BRST anomalies at the defect.

Geometrically, we can think about the space of deformations of the ``bulk $+$ defect'' system as being fibered over the space of deformations of the bulk system. Correspondingly, locality means the nilpotent odd vector field $\bm{\eta}$ which describes the action of the BRST transformations on the space of deformations preserves the fibration: it is the sum of the standard bulk vector field and a vector field $\bm{\eta}^\partial$ which only changes the defect couplings in a manner dependent on \textit{both} sets of couplings. Dually, the coefficients of $\bm{\eta}^\partial$ define brackets mapping a collection of defect and bulk interactions to defect interactions. We can add holomorphic momenta $\lambda^\partial$ in the $H^\partial$ directions and also define generalized operations where some bulk or defect propagators are inserted by hand. 

The corresponding Feynman diagrams will now have both bulk vertices and defect vertices. They will also have edges associated to bulk fields and defect fields. Using (the reduced) translation symmetry, we can fix the position of one interaction vertex in the direction parallel to the defect only. Likewise, the defect vertices will carry holomorphic momenta along the defect only (see e.g. Figure \ref{fig:bulk-defectT2}).

The counting of form degrees now depend on both the exact bulk $n=H+T$ and defect $n^\partial = H^\partial + T^\partial$ dimensionalities. In particular, the integrated bulk and defect vertices will have degrees $n$ and $n^\partial$ respectively, and the propagators will subtract $n-1$ and $n^\partial-1$ respectively. The $n$-Laman conditions are adjusted accordingly by the same logic as in Section \ref{sec:GraphCombinatorics}.

The change of variables to $s_e$ or $y_e$ also proceeds in the same manner and one can define $\Delta_\Gamma$ regions and derive quadratic relations in the usual way. When $\Gamma$ has both bulk and defect vertices, the axioms may include sums of compositions of two defect operations or one defect and one bulk operation, depending on $S$ having defect vertices or not. As in the bulk case, once we know that the answers are finite, and satisfy the correct axioms, we can re-state them in geometric terms or as Schwinger time integrals. 

Geometrically, we can identify a representative for the quotient of the configuration space by a uniform twisted scaling of bulk and defect coordinates, as in Section \ref{sec:twistedScaling}. In terms of Schwinger times, the integral over positions can be performed independently in the various directions, resulting in a product of forms, just as with the bulk case in Section \ref{sec:projective}. The integrals in directions \textit{parallel} to the defect are formally identical to those for bulk defects only and give the usual $\alpha_\Gamma$ or $\rho_\Gamma[\lambda;z]$ forms. However, the integrals in directions \textit{perpendicular} to the defect will be different, as the positions of some vertices are fixed. This will result in new ``perpendicular forms''  $\alpha^\partial_\Gamma$ or $\rho^\partial_\Gamma[\lambda;z]$. 

By the preceding factorization arguments for forms, and our claim in Section \ref{sec:projective} that $\alpha_\Gamma^2=0$, we see that defects with at least two topological directions $T^\partial \geq 2$ will not have loop contributions to the BRST brackets. In the following sections we will study some examples that do have loop contributions to the BRST brackets and we will end with a proof of our non-renormalization theorem in Section \ref{sec:MismatchingTwists}.

\subsection{Line Defects \texorpdfstring{$T^\partial = 1$}{T**delta = 1} and \texorpdfstring{$A_\infty$}{A-infty}}\label{sec:LineDefects}
The case of topological line defects, i.e. $T^{\partial}=1$, is particularly simple in our setup because the defect edge variables $s_e^\partial$ are only constrained to have the same sign as the differences in $x^{\mathbb{R}}_\partial$ coordinates of the defect vertices. The $s_e^\partial$ Gaussian integral can thus be performed immediately, and affects the integrals over bulk edge variables by inserting signs depending on the relative position of the defect vertices.

Another important simplification in $T=1$ is that 1d TFTs are really simple: they are a choice of a space of states ${\cal H}$ and a dg algebra of observables $A$. Accordingly, one can readily discuss topological defects defined by coupling a higher-dimensional system to a 1d TFT with a finite-dimensional space of states in a manner akin to a Wilson line: pick a matrix-valued interaction $\cI$ and path-order exponentiate it along the line. 

If we imagine computing BRST anomalies, say for a free bulk theory,  the brackets are now associated with integrals over regions constrained by the condition that the defect vertices have a specific order along the line, i.e. we must path-order the non-commuting variables. These brackets endow the algebra of defect local operators restricted to the line with the structure of an $A_\infty$-algebra (or $E_1$-algebra, see Appendix \ref{sec:operadsAndQFT}), in the same way described in Section \ref{sec:H0T1Example}, and the non-commutative Maurer Cartan equaton
\begin{equation}
    \{\cI, \cI\}_{A_\infty} + \{\cI, \cI, \cI\}_{A_\infty} + \cdots = 0\,,
\end{equation}
should be thought of as a ``path-ordered computation.'' As before, if the ordering of defect vertices is not important as in e.g. computing commutators, we will restrict our attention to just an $L_\infty$ algebra inside of the $A_\infty$ algebra.

There is a helpful geometric/dual way to think of this enhancement from $L_\infty$ to $A_\infty$-algebras. Recall from Section \ref{sec:QFT} that the $L_\infty$ structure coefficients (and therefore brackets) arose as the coefficients of an odd nilpotent vector field $\bm{\eta}$, associated with BRST symmetry, on the formal pointed dg-supermanifold of deformed theories. In the case that our couplings are matrices, we are instead studying a \textit{non-commutative} pointed dg supermanifold. Essentially by definition, the coefficients of an odd nilpotent vector field $\bm{\eta}$ on such a manifold are the structure constants of an $A_\infty$-algebra.

Additionally, we have operations involving ordered vertices along the line and any number of bulk vertices. Among other things, these will contribute to the $A_\infty$-algebra controlling line defects in the (perturbatively) interacting theory. From this perspective, the brackets are a map from the bulk algebra of operators to the Hochschild cohomology of the defect algebra, i.e. cochains in the Hochschild cohomology of the defect algebra are collections of brackets partially-filled by bulk local operators.

These $A_\infty$-like operations can be employed to define the coupling of the bulk to an abstract quantum-mechanical system, by replacing the Moyal product with another abstract algebra, denoted as $A_{\text{QM}}$.

\subsubsection{Example: \texorpdfstring{$H=0$}{H=0} \texorpdfstring{$T=2$}{T=2} Topological Line Defect in 2d TQFT}
Consider the case with $T=2$, $H=0$  in the bulk, and $T^\partial =1$ for the defect. Denote the coordinates on the spacetime $\mathbb{R}^2$ by $x_{\perp}$ and $x_{\parallel}$, with the topological line defect positioned at $x_\perp = 0$.

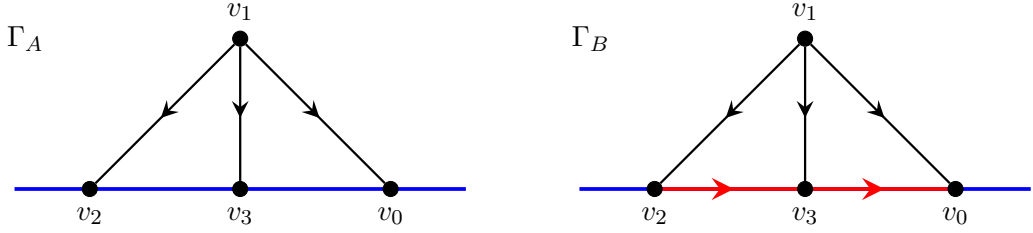
\begin{figure}
    \centering
\begin{tikzpicture}[scale=1, auto,swap,
baseline=(current bounding box.center)
]
    \node[vertex, fill = black, label={above:$v_1$}] (1) at (0,2) {};
    \node[vertex, fill = black,label=below:$v_2$] (2) at (-2,0) {};
    \node[vertex, fill = black,label=below:$v_3$] (3) at (0,0) {};
    \node[vertex, fill = black,label=below:$v_0$] (4) at (2,0) {};
    
    \draw[lineShimmy] (1) -- (2);
    \draw[lineShimmy] (1) -- (3);
    \draw[lineShimmy] (1) -- (4);

    \node[text width=2cm] at (-2.1,2) {$\Gamma_A$};

    \begin{scope}[on background layer]
    \draw[blue, line width = 0.5mm] ([xshift=-1cm]2.center) -- ([xshift=1cm]4.center);
    \end{scope}
\end{tikzpicture}
\hspace{1cm} 
\begin{tikzpicture}[scale=1, auto,swap,
baseline=(current bounding box.center)
]

    \node[vertex,fill = black, label={above:$v_1$}] (1) at (0,2) {};
    \node[vertex,fill = black,label=below:$v_2$] (2) at (-2,0) {};
    \node[vertex,fill = black,label=below:$v_3$] (3) at (0,0) {};
    \node[vertex,fill = black,label=below:$v_0$] (4) at (2,0) {};

    \draw[lineShimmy] (1) -- (2);
    \draw[lineShimmy] (1) -- (3);
    \draw[lineShimmy] (1) -- (4);
    \draw[lineShimmy, red, line width=0.5mm] (2) -- (3);
    \draw[lineShimmy, red, line width=0.5mm] (3) -- (4);

    \node[text width=2cm] at (-2.1,2) {$\Gamma_B$};
    
    \begin{scope}[on background layer]

    \draw[blue, line width = 0.5mm] ([xshift=-1cm]2.center) -- ([xshift=1cm]4.center);
    \end{scope}
\end{tikzpicture}
\caption{Graphs in $\mathbb{R}^2$ where we fix the position of the line defect (blue horizontal line) to be $x_{\perp} = 0$. So coordinates of the defect insertions are $(x^{\partial}_2,0)$, $(x^{\partial}_3,0)$ and $(0,0)$ and the bulk insertion is at $(x_{1,\parallel}, x_{1,\perp})$. Left, $\Gamma_A$: three bulk propagators (black arrows). Right, $\Gamma_B$ three bulk propagators with two additional defect propagators (red arrows).
}
\label{fig:bulk-defectT2}
\end{figure}

A nontrivial graph occurs with three defect insertions ($v_2$, $v_3$ and $v_0$), and one bulk insertion ($v_1$) connecting to the three defect insertions. This is shown in $\Gamma_A$ of Figure \ref{fig:bulk-defectT2}, where we fix the defect insertion $v_0$ to be at the origin. It is helpful to perform a simple degree counting for the integral \eqref{eq:FeynIntegralP}:
\begin{itemize}
    \itemsep 0em
    \item A bulk propagator is a 1-form. So the product of three propagators with the differential outside is a 4-form.
    \item Each bulk insertion introduces an integral over $\bbR^2$, lowering the degree by 2.
    \item Each boundary defect introduces an integral over $\bbR$, lowering the degree by 1, except the vertex $v_0$ which is fixed by translation symmetry.
\end{itemize}
Thus we have 4 integrals and a 4-form to integrate.

As before, we associate a Schwinger time $t_e$ and $x_{e} := x_{e(0)}-x_{e(1)}$ to each edge $e \in \Gamma_1$. The $s$ coordinates are:
\begin{equation}
    s_{a,e}= \frac{x_{a,e}}{t^{1/2}_e}, \quad a = \parallel, \perp,\quad  e\in \Gamma_1\,,
\end{equation}
and the bulk propagator is given by
\begin{equation}
    \mathcal{P}_e = \pi^{-2/2} e^{-(s^2_{\parallel,e} + s^2_{\perp,e})} \, ds_{\parallel,e} ds_{\perp,e}\,.
\end{equation}

As in \eqref{eq:FeynIntegralcalP}, the Feynman integral is given by the product of propagators
\begin{equation}
   I_{\Gamma_A} = \int_{\mathbb{RP}^2_{> }} \int_{\mathbb{R}^2\times \mathbb{R}\times \mathbb{R}}  \mathcal{P}_{12}\mathcal{P}_{13}\mathcal{P}_{10} =: \int_{\mathbb{RP}^2_{\geq }} \omega_{\Gamma_A}\, ,
   \label{eq:T2H0Td1IGamma}
\end{equation}
where $\mathcal{P}_{ij}$ is the propagator associated to the $ij$ edge. The Schwinger integrand $\omega_{\Gamma_A}$ is obtained after we perform the Gaussian integral over the spacetime. A straightforward calculation yields
\begin{equation}
    \omega_{\Gamma_A} = -\frac{1}{8\pi}\frac{t_{12}t_{13}t_{10}}{(t_{12}t_{13}+ t_{12}t_{10} + t_{13}t_{10} )^{3/2}} \left(- \frac{dt_{12}}{t_{12}} \frac{dt_{13}}{t_{13}} +\frac{dt_{12}}{t_{12}} \frac{dt_{10}}{t_{10}} -\frac{dt_{13}}{t_{13}} \frac{dt_{10}}{t_{10}}\right)\,. \label{eq:T2H0T1omega}
\end{equation}
The further integration over the Schwinger time can be readily done by using the identity
\begin{equation}
    \int_{x\geq 0}\int_{y\geq 0} \left(x+y+xy
    \right)^{-3/2} dxdy = 2\pi\,.
\end{equation}
The Feynman integral in \eqref{eq:T2H0Td1IGamma} appears in the ternary bracket of the defect $L_{\infty}$-algebra. Note that it is an $L_\infty$-algebra because we are integrating over the entire worldline of the defect without any ordering of the defect vertices.

To calculate the $A_{\infty}$-bracket on the defect, we can enforce an ordering of the defect vertices by introducing additional defect propagators:
\begin{equation}
    \mathcal{P}_{\partial,e} = \pi^{-1/2} e^{-s_{\partial}^2} ds_{\partial}\,,
\end{equation}
where we analogously define $s_e^{\partial} = {x_e^{\partial}}/{t^{1/2}_e}$. A single defect propagator between $v_2$ and $v_3$ or between $v_3$ and $v_0$ will make the integral vanish due to the reflection symmetry. If we introduce both, as shown in the graph $\Gamma_B$ on the right of Figure \ref{fig:bulk-defectT2}, the integral reads
\begin{equation}
   I_{\Gamma_B} = \int_{\mathbb{RP}^4_{> }} \int_{\mathbb{R}^2\times \mathbb{R}\times \mathbb{R}}   \mathcal{P}_{12}\mathcal{P}_{13}\mathcal{P}_{10} \mathcal{P}_{\partial,23} \mathcal{P}_{\partial, 30} =: \int_{\mathbb{RP}^4_{\geq }} \omega_{\Gamma_B}\,.
\end{equation}
A direct computation yields
\begin{equation}
\omega_{\Gamma_B} = \frac{t_{12} t_{13}^2 t_{10} t_{23} t_{30}  d\mathrm{Vol}_t}{64 \pi^2\left(\left(t_{13} t_{10}+t_{12}\left(t_{13}+t_{10}\right)\right)\left(t_{23}\left(t_{10}+t_{30}\right)+t_{12}\left(t_{13}+t_{10}+t_{30}\right)+t_{13}\left(t_{10}+t_{23}+t_{30}\right)\right)\right)^{3 / 2}}
\end{equation}
where $d\mathrm{Vol}_t$ is the volume form on $\mathbb{RP}^4_{>}$
\begin{equation}
\begin{aligned}
d\mathrm{Vol}_t = &  -\frac{dt_{12}dt_{13}dt_{10}d t_{23}}{t_{12} t_{13} t_{10} t_{23}}+\frac{dt_{12}dt_{13}d t_{10}dt_{30}}{t_{12} t_{13} t_{10} t_{30}} \\
& -\frac{dt_{12}dt_{13}dt_{23}dt_{30}}{t_{12} t_{13} t_{23} t_{30}}+\frac{dt_{12}dt_{10}dt_{23}dt_{30}}{t_{12} t_{10} t_{23} t_{30}}-\frac{dt_{13}dt_{10}dt_{23}dt_{30}}{t_{13} t_{10} t_{23} t_{30}}\,.
\end{aligned}
\label{eq:dVolRP4}
\end{equation}

We observe that $\omega_{\Gamma_B}$ neatly factorizes into
\begin{equation}
    \omega_{\Gamma_B}= \omega_{\Gamma_B,\parallel} \wedge \omega_{\Gamma_B, \perp}\,,
\end{equation}
where
\begin{align}
\omega_{\Gamma_B, \parallel} 
    =& \frac{1}{8\pi}\Big[ \left(t_{10}+t_{30}\right)\left(dt_{12}  dt_{13}-dt_{13}  dt_{23}\right)+\left(t_{12}+t_{23}\right)\left(dt_{13}  dt_{10}+dt_{13}  dt_{30}\right)\nonumber\\
    &-t_{13}\left(dt_{12}  dt_{10}+dt_{12}  dt_{30}-dt_{10}  dt_{23}+dt_{23}  dt_{30}\right)\Big] \\
    &\Big(t_{23}\left(t_{10}+t_{30}\right) +t_{12}\left(t_{13}+t_{10}+t_{30}\right)+t_{13}\left(t_{10}+t_{23}+t_{30}\right)\Big)^{-3 / 2}\nonumber\,,\\
\omega_{\Gamma_B, \perp}  =& -\frac{1}{8\pi}\frac{t_{12}t_{13}t_{10}}{(t_{12}t_{13}+ t_{12}t_{10} + t_{13}t_{10} )^{3/2}} \left(- \frac{dt_{12}}{t_{12}} \frac{dt_{13}}{t_{13}} +\frac{dt_{12}}{t_{12}} \frac{dt_{10}}{t_{10}} -\frac{dt_{13}}{t_{13}} \frac{dt_{10}}{t_{10}}\right)\,.
\end{align}
Notice that $\omega_{\Gamma_B, \perp}$ coincides with $\omega_{\Gamma_A}$ in \eqref{eq:T2H0T1omega}. This is a straightforward generalization of the observation in \eqref{eq:omegafactorize}: both bulk propagators and defect propagators are factorized over individual spacetime dimensions. We will observe such factorization repeatedly in the next few examples.

\subsubsection{Example: \texorpdfstring{$H=1$}{H=1} \texorpdfstring{$T=1$}{T=1} Topological Line Defect in 3d HT Theory}
As another example, suppose the bulk is $\mathbb{R}\times \mathbb{C}$ with $T=1$, $H=1$ and coordinates $(x^{\mathbb{R}}, x^{\mathbb{C}}, \bar{x}^{\mathbb{C}})$. Consider a topological line defect $T^\partial =1$ along the topological direction and positioned at the origin of the holomorphic plane.

\begin{figure}[t]
    \centering
\begin{tikzpicture}[scale=1, auto,swap,
baseline=(current bounding box.center)
]
    \node[vertex, fill = black, label={above:$v_1$}] (1) at (0,2) {};
    \node[vertex, fill = black,label=below:$v_2$] (2) at (-2,0) {};
    \node[vertex, fill = black,label=below:$v_3$] (3) at (0,0) {};
    \node[vertex, fill = black,label=below:$v_0$] (4) at (2,0) {};
    
    \draw[lineShimmy] (1) -- (2);
    \draw[lineShimmy] (1) -- (3);
    \draw[lineShimmy] (1) -- (4);

    \node[text width=2cm] at (-2.1,2) {$\Gamma_C$};

    \begin{scope}[on background layer]
    \draw[blue, line width = 0.5mm] ([xshift=-1cm]2.center) -- ([xshift=1cm]4.center);
    \end{scope}
\end{tikzpicture}
\hspace{1cm} 
\begin{tikzpicture}[scale=1, auto,swap,baseline=(current bounding box.center)]

    \node[vertex,fill = black, label={above:$v_1$}] (1) at (0,2) {};
    \node[vertex,fill = black,label=below:$v_2$] (2) at (-2,0) {};
    \node[vertex,fill = black,label=below:$v_3$] (3) at (0,0) {};
    \node[vertex,fill = black,label=below:$v_0$] (4) at (2,0) {};

    \draw[lineShimmy] (1) -- (2);
    \draw[lineShimmy] (1) -- (3);
    \draw[lineShimmy] (1) -- (4);
    \draw[lineShimmy, red, line width=0.5mm] (2) -- (3);
    \draw[lineShimmy, red, line width=0.5mm] (3) -- (4);

    \node[text width=2cm] at (-2.1,2) {$\Gamma_D$};
    
    \begin{scope}[on background layer]

    \draw[blue, line width = 0.5mm] ([xshift=-1cm]2.center) -- ([xshift=1cm]4.center);
    \end{scope}
\end{tikzpicture}
\caption{Graphs in $\mathbb{R}\times \mathbb{C}$ where we fix the position of the line defect (blue horizontal line) to be $(x^{\mathbb{C}}, \bar{x}^{\mathbb{C}}) = (0,0)$. The coordinates of the defect insertions are $(x_{\partial,2}^{\mathbb{R}},0,0)$, $v_3 = (x_{\partial,3}^{\mathbb{R}},0,0)$, $v_0 = (0,0,0)$ and the bulk insertion is at $(x^{\mathbb{R}}_1, x^{\mathbb{C}}_1, \bar{x}^{\mathbb{C}}_1)$. Left, $\Gamma_C$.  Right, $\Gamma_D$.
}
\label{fig:bulk-defectT1H1}
\end{figure}
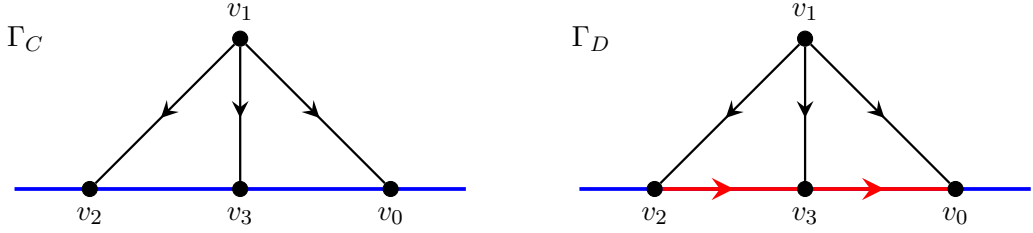

Nontrivial graphs take the same form as before, see Figure \ref{fig:bulk-defectT1H1}. The first has one bulk vertex at $v_1 = (x^{\mathbb{R}}_1, x^{\mathbb{C}}_1, \bar{x}^{\mathbb{C}}_1)$ and three defect vertices at
\begin{equation}
    v_2 =  (x_{\partial,2}^{\mathbb{R}},0,0), \quad v_3 = (x_{\partial,3}^{\mathbb{R}},0,0), \quad v_0 = (0,0,0)\,.
\end{equation}
Passing to the $s$ and $y$ coordinates
\begin{equation}
    s_e = \frac{x_e^{\mathbb{R}}}{t^{1/2}_e},\quad s_{\partial,e} = \frac{x_{\partial,e}^{\mathbb{R}}}{t^{1/2}_e}, \quad y = \frac{\bar{x}_1^{\mathbb{C}}}{t}\,,
\end{equation}
we have the bulk propagator
\begin{equation}
    \mathcal{P}_e = \pi^{-1/2} e^{-(s^2_e + x^{\mathbb{C}}_ey_e)} dy_e ds_e\,.
\end{equation}
The Feynman integral for $\Gamma_C$ reads
\begin{equation}
    I_{\Gamma_C} = \int_{\mathbb{RP}_{> }^2}\int_{\mathbb{R} \times \mathbb{C}\times \mathbb{R}\times \mathbb{R}} \frac{dx^{\mathbb{C}}}{2\pi i} e^{\lambda x^{\mathbb{C}}} \mathcal{P}_{12} \mathcal{P}_{13} \mathcal{P}_{10} =: \int_{\mathbb{RP}_{> }^2}\omega_{\Gamma_C}\,.
\end{equation}
Upon integrating the Gaussian integral over the spacetime $\mathbb{C}\times \mathbb{R} \times \mathbb{R} \times \mathbb{R}$, we find the Schwinger integrand
\begin{equation}
    \omega_{\Gamma_C} = \lambda^2\frac{-\left(t_{12}t_{13}t_{10}\right)^2}{\left(t_{12}t_{13}+ t_{12}t_{10} + t_{13}t_{10} \right)^3} \left( -\frac{dt_{12}}{t_{12}} \frac{dt_{13}}{t_{13}} +\frac{dt_{12}}{t_{12}} \frac{dt_{10}}{t_{10}} -\frac{dt_{13}}{t_{13}} \frac{dt_{10}}{t_{10}}\right)
\end{equation}
which is again a convergent integral using
\begin{equation}
    \int_{x\geq 0}\int_{y\geq 0} \frac{xy}{\left(x+y+xy
    \right)^{3}} dxdy= \frac12\,.
\end{equation}

We can include additional defect propagators to compute Feynman integrals relevant to the $A_\infty$-structure on the defect. The defect propagator is
\begin{equation}
    \mathcal{P}_{\partial,e} = \pi^{-1/2} e^{-s_{\partial,e}^2} ds_{\partial,e}\,.
\end{equation}
A single defect propagator between $v_2$ and $v_3$ or between $v_3$ and $v_0$ will make the integral vanish due to the reflection symmetry again. If we introduce both, as shown in the graph $\Gamma_D$ on the right of Figure \ref{fig:bulk-defectT1H1}, the integral reads
\begin{equation}
   I_{\Gamma_D} = \int_{\mathbb{RP}^4_{> }} \int_{\mathbb{R} \times \mathbb{C} \times \mathbb{R}\times \mathbb{R}}   \mathcal{P}_{12}\mathcal{P}_{13}\mathcal{P}_{10} \mathcal{P}_{\partial,23} \mathcal{P}_{\partial, 30} =: \int_{\mathbb{RP}^4_{\geq }} \omega_{\Gamma_D}\,.
\end{equation}
By direct evaluation, we find:
\begin{equation*}
\omega_{\Gamma_D} = \frac{  t_{12}^2 t_{13}^3 t_{10}^2 t_{23} t_{30}\ \lambda^2 d\mathrm{Vol}_t }{8\pi\left(t_{13} t_{10}+t_{12}\left(t_{13}+t_{10}\right)\right)^3\left[t_{23}\left(t_{10}+t_{30}\right)+t_{12}\left(t_{13}+t_{10}+t_{30}\right)+t_{13}\left(t_{10}+t_{23}+t_{30}\right)\right]^{3 / 2}}
\end{equation*}
where $d\mathrm{Vol}_t$ is the volume form \eqref{eq:dVolRP4} on $\mathbb{RP}^4_{\geq }$. 

It again factorizes as expected
\begin{equation}
    \omega_{\Gamma_D}= \omega_{\Gamma_D,\parallel} \wedge \omega_{\Gamma_D, \perp}\,, 
\end{equation}
where $\omega_{\Gamma_D,\parallel} = \omega_{\Gamma_B, \perp}$ and $\omega_{\Gamma_D,\perp} = \omega_{\Gamma_C}$.

\subsection{Holomorphic Surface Defects \texorpdfstring{$H^\partial=1$}{H**partial=1} and Chiral \texorpdfstring{$A_\infty$}{A-infty}}
Similar considerations apply for holomorphic surface defects, with $T^\partial=0$ and $H^\partial=1$. The $y_{\partial,e}$ integrals can be performed first, effectively producing standard free field chiral algebra contractions into the integral for the bulk degrees of freedom. 

One is thus led to consider integrals for the bulk degrees of freedom enriched by inverse powers of the $x_i^{\mathbb{C}} -x_j^{\mathbb{C}}$ holomorphic coordinate along the defect. These ``higher vertex algebra operations'' can be combined with any 2d chiral algebra, even interacting chiral algebras, to produce a holomorphic surface defect.

\subsubsection{Example: \texorpdfstring{$H=1$}{H=1} \texorpdfstring{$T=1$}{T=1} Holomorphic Surface Defect in 3d HT Theory} 
Consider a holomorphic surface defect in a holomorphic topological bulk $\mathbb{R}\times \mathbb{C}$, i.e. $T=1$, $H=1$ and $H^\partial =1$. The surface defect is positioned at $x^{\mathbb{R}} = 0$. 

\begin{figure}
    \centering
    \begin{tikzpicture}[scale=0.8, auto,swap,baseline=(current bounding box.center)]

    \draw (0,0) -- (0,4) -- (2,6) -- (2,2)-- cycle;
    
    \node[vertex, fill = black, label={above:$v_1$}] (1) at (5, 4) {};
    \node[vertex, fill = black,label=below:$v_2$] (2) at (1.7, 5) {};
    \node[vertex, fill = black,label=below:$v_3$] (3) at (0.8,3.6) {};
    \node[vertex, fill = black,label=below:$v_0$] (4) at (1,2) {};
    
    \node[text width=2cm] at (0.3,5) {$\Gamma_E$};
    
    \draw[lineShimmy] (1) -- (2);
    \draw[lineShimmy] (1) -- (3);
    \draw[lineShimmy] (1) -- (4);
\end{tikzpicture}
\hspace{1cm}
\begin{tikzpicture}[scale=0.8, auto,swap,baseline=(current bounding box.center)]

    \draw (0,0) -- (0,4) -- (2,6) -- (2,2)-- cycle;
    
    \node[vertex, fill = black, label={above:$v_1$}] (1) at (5, 4) {};
    \node[vertex, fill = black,label=below:$v_2$] (2) at (1.7, 5) {};
    \node[vertex, fill = black,label=west:$v_3$] (3) at (0.8,3.6) {};
    \node[vertex, fill = black,label=below:$v_0$] (4) at (1,2) {};
    
    \node[text width=2cm] at (0.3,5) {$\Gamma_F$};
    
    \draw[lineShimmy] (1) -- (2);
    \draw[lineShimmy] (1) -- (3);
    \draw[lineShimmy] (1) -- (4);
    
    \draw[lineShimmy,red] (2) -- (3);
    \draw[lineShimmy,red] (3) -- (4);
\end{tikzpicture}
    \caption{Graphs in $\mathbb{R}\times \mathbb{C}$ where we fix the position of the holomorphic surface defect to be $x^\mathbb{R} = 0$. The coordinates of four vertices are $v_1 = (x^{\mathbb{R}}_1, x^{\mathbb{C}}_1, \bar{x}^{\mathbb{C}}_1)$,  $v_2 = (0,x^{\mathbb{C}}_{\partial, 2} ,\bar{x}^{\mathbb{C}}_{\partial, 2} )$,  $v_3 = (0,x^{\mathbb{C}}_{\partial, 3} ,\bar{x}^{\mathbb{C}}_{\partial, 3} )$,  $v_0 = (0,0,0)$.
    }
    \label{fig:bulk-defectT1H1H1}
\end{figure}
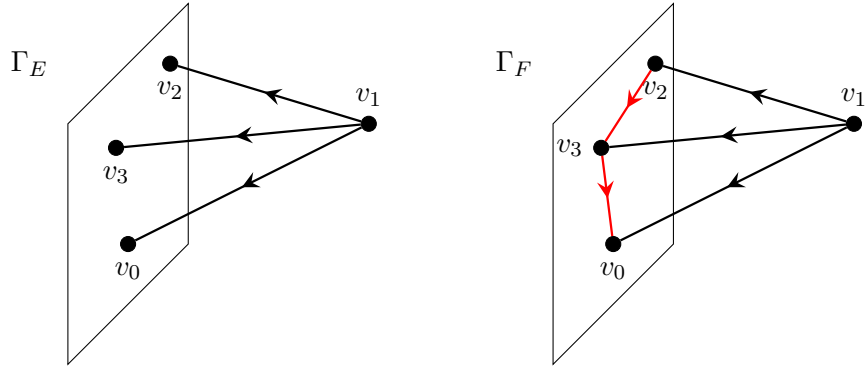

We can consider similar graphs to the previous examples, depicted in Figure \ref{fig:bulk-defectT1H1H1}. Consider the graph $\Gamma_E$ with bulk vertex at $v_1= (x^{\mathbb{R}}_1, x^{\mathbb{C}}_1, \bar{x}^{\mathbb{C}}_1)$ and three defect vertices at
\begin{equation}
   v_2 = (0,x^{\mathbb{C}}_{\partial, 2} ,\bar{x}^{\mathbb{C}}_{\partial, 2} ), \quad v_3 = (0,x^{\mathbb{C}}_{\partial, 3} ,\bar{x}^{\mathbb{C}}_{\partial, 3} ), \quad v_0 = (0,0,0)\,.
\end{equation}
Passing to the $s$ and $y$ coordinates
\begin{equation}
    s_e = \frac{x_e^{\mathbb{R}}}{t^{1/2}},\quad y_{\partial,e} = \frac{\bar{x}_{\partial,e}^{\mathbb{C}}}{t^{1/2}_e}, \quad y = \frac{\bar{x}_{e}^{\mathbb{C}}}{t_e}\,,
\end{equation}
we have the bulk propagator
\begin{equation}
    \mathcal{P}_e = \pi^{-1/2} e^{-(s_e^2 -x_e^{\mathbb{C}}y_e)} dy_e ds_e\,.
\end{equation}
The Feynman integral for the graph $\Gamma_E$ is 
\begin{equation}
   I_{\Gamma_E}= \int_{\mathbb{RP}_{>}^2}\int_{\mathbb{R}}\int_{\mathbb{C}} \frac{dx^{\mathbb{C}}_1}{2\pi i} e^{\lambda_1 x_1^{\mathbb{C}}} \int_{\mathbb{C}} \frac{dx^{\mathbb{C}}_{\partial, 2}}{2\pi i} e^{\lambda_2 x_{\partial, 2}^{\mathbb{C}}}
    \int_{\mathbb{C}} \frac{dx^{\mathbb{C}}_{\partial, 3}}{2\pi i} e^{\lambda_3 x_{\partial, 3}^{\mathbb{C}}}
    \ \mathcal{P}_{12} \mathcal{P}_{13} \mathcal{P}_{10} =: \int_{\mathbb{RP}_{>}^2} \omega_{\Gamma_E}\,,
\end{equation}
where direct calculation gives
\begin{equation}
    \omega_{\Gamma_E} = -\frac{1}{8\pi}\frac{t_{12}t_{13}t_{10}}{(t_{12}t_{13}+ t_{12}t_{10} + t_{13}t_{10} )^{3/2}} \left(- \frac{dt_{12}}{t_{12}} \frac{dt_{13}}{t_{13}} +\frac{dt_{12}}{t_{12}} \frac{dt_{10}}{t_{10}} -\frac{dt_{13}}{t_{13}} \frac{dt_{10}}{t_{10}}\right)\,.
\end{equation}
This coincides with $\omega_{\Gamma_A}$ in \eqref{eq:T2H0T1omega}, as expected. 

We can include additional defect propagators
\begin{equation}
    \mathcal{P}_{\partial, e} = e^{-x^{\mathbb{C}}_ey_e} dy_e \,,
\end{equation}
as in graph $\Gamma_F$ in Figure \ref{fig:bulk-defectT1H1H1}. The Feynman integral now becomes
\begin{equation*}
   I_{\Gamma_F}= \int_{\mathbb{RP}_{>}^4}\int_{\mathbb{R}}\int_{\mathbb{C}} \frac{dx^{\mathbb{C}}_1}{2\pi i} e^{\lambda_1 x_1^{\mathbb{C}}} \int_{\mathbb{C}} \frac{dx^{\mathbb{C}}_{\partial, 2} }{2\pi i}e^{\lambda_2 x_{\partial, 2}^{\mathbb{C}}}
    \int_{\mathbb{C}} \frac{dx^{\mathbb{C}}_{\partial, 3}}{2\pi i} e^{\lambda_3 x_{\partial, 3}^{\mathbb{C}}}
    \ \mathcal{P}_{12} \mathcal{P}_{13} \mathcal{P}_{10} \mathcal{P}_{\partial, 23} \mathcal{P}_{\partial,30}
    =: \int_{\mathbb{RP}_{>}^4} \omega_{\Gamma_F}\,,
\end{equation*}
and the Schwinger integrand is found to be
\begin{equation}
\begin{aligned}
& \omega_{\Gamma_F} =-\frac{1}{8\pi} \sqrt{t_{12} t_{13} t_{10} t_{23} t_{30}}
\Big[2 t_{12}\left(t_{10}+t_{30}\right) \lambda_2+t_{13}\left(2 t_{12} \lambda_2+t_{10}\left(\lambda_1+2 \lambda_2\right)-2 t_{30} \lambda_3\right)\Big] \\
& \Big[t_{10}\left(t_{13}+t_{23}\right)\left(\lambda_1+2 \lambda_2\right)+2\left(t_{10} t_{23}+t_{13}\left(t_{10}+t_{23}\right)\right) \lambda_3+t_{12}\left(2 t_{13}\left(\lambda_2+\lambda_3\right)+t_{10}\left(\lambda_1+2\left(\lambda_2+\lambda_3\right)\right)\right)\Big] \\
& \Big(t_{13} t_{10}+t_{12}\left(t_{13}+t_{10}\right)\Big)^{-3 / 2}
\Big[t_{23}\left(t_{10}+t_{30}\right)+t_{12}\left(t_{13}+t_{10}+t_{30}\right)+t_{13}\left(t_{10}+t_{23}+t_{30}\right)\Big]^{-5 / 2}
d\mathrm{Vol}_t
\end{aligned}
\end{equation}
where $d\mathrm{Vol}_t$ is the volume form \eqref{eq:dVolRP4} on $\mathbb{RP}^4_{\geq }$. It factorizes into the form
\begin{equation}
    \omega_F = \omega_{F,\parallel}\wedge \omega_{A}
\end{equation}
The full form of $\omega_{F,\parallel}$, along with the details of the calculations in this section, can be found at \cite{omega_integrand}.

\subsection{Generalizations to Boundaries and Beyond}
There are larger collections of defects which can be treated perturbatively as deformations of defects in a free theory. Essentially, we get one such collection for each reference ``free'' defect in a free theory. 

As an example, consider a boundary condition replacing a topological direction with $\bR^+$. A bulk superfield brought to the boundary can be decomposed into $0$ and $1$ form components 
in the direction perpendicular to the boundary $x_{\perp}$:
\begin{equation}
    \Phi = \Phi_{\parallel} + \Phi_{\perp} dx_{\perp}\,.
\end{equation}
Likewise, $\dd$ decomposes as
\begin{equation}
    \dd = dx_{\perp} \frac{\partial}{\partial x_\perp} + \dd_{\parallel}\,,
\end{equation}
so that the normal component of the kinetic term only involves $0$-form components $\Phi_{\parallel}$, i.e.
\begin{equation}
    (\Phi, dx_{\perp} \partial_\perp \Phi)
        = (\Phi_{\parallel}, dx_{\perp} \partial_\perp \Phi_{\parallel})\,.
\end{equation}
As the bulk free action is first order in normal derivatives, we can pick a Lagrangian splitting of the bulk superfields and set the corresponding half of the $0$-form components to $0$ at the boundary. The $0$-form component of the other half will survive as a boundary superfield which can participate in boundary interactions.

Even in the absence of extra boundary degrees of freedom or couplings, a bulk interaction can induce a boundary anomaly for these free boundary conditions. Classically, the $\dd \cI$ anomalous contribution, which integrates by parts to zero in the absence of a boundary, will give the integral of $\cI|_\partial$ at the boundary. Unless the boundary conditions set $\cI$ to zero at the boundary, we get a BRST anomaly linear in the bulk interaction.\footnote{This is the origin of the matrix factorization construction of boundary conditions in twisted theories with $T=2$, $H=0$ \cite{Kapustin:2002bi} or $T=1$, $H=1$ \cite{Costello:2020ndc} in the presence of a bulk super-potential $W$: boundary interactions are selected so that the quadratic part of the MC equation cancels $W_\partial$.}

At the level of Feynman diagrams, the bulk-bulk propagators will be modified by reflected images of one of the bulk points,
\begin{equation}
    \widetilde{P}(x_{\parallel}, x_{\perp}; x_{\parallel}', x_{\perp}') = P(x_{\parallel}, x_{\perp}; x_{\parallel}', x_{\perp}') \pm  P(x_{\parallel}, x_{\perp}; x_{\parallel}', -x_{\perp}')
\end{equation}
with a negative sign for the fields which are set to zero at the boundary and a positive sign for the ones which are not.\footnote{This is a convenient gauge choice, ensuring e.g. that a trivial interface can be treated as a diagonal boundary condition.} 

Notice that the bulk normal ordering used to define interactions only subtracts the unmodified bulk propagators. We will thus have to worry about contractions \textit{within a single bulk operator}. As an extra regulation step, we can turn off the interaction in a layer of width $\epsilon_\parallel$ from the boundary. As we compute perturbative corrections to the BRST anomaly, mimicking the calculations in Section \ref{sec:BRSTAnomaly}, the step where we integrate by parts to move the differential $\dd$ from fields to propagators produces an extra boundary term. We thus get two types of contributions to the anomaly: either one of the interactions is placed at distance $\epsilon_\parallel$ or one of the Schwinger times is set to $\epsilon$. As the propagators only depend on the projective combinations $y$, $s$, which now include differences of positions and their reflected images, is it easy to see that all the contributions combine into a single integral over a region $\Delta_\Gamma^\partial$ in the space of $y$ and $s$ coordinates and \textit{both $\epsilon$ and $\epsilon_\partial$ drop out of the final expression.}

For example, if we go back to the case with no Wick contractions, we have a single vertex and no propagators (i.e. no $s$ or $y$ variables). Thus, the region is necessarily a point and we get the anomaly $\cI|_\partial$. At the next order, we can Wick contract two fields in a single bulk $\cI$ interaction, employing the reflected propagator for the self-contraction. There is a single $s$ variable now, which is positive. Thus $\Delta_\Gamma^\partial = \bR^+$, leading to a single Gaussian integral. The overall sign depends on the sign of the reflected propagator. 

We leave a detailed analysis of boundaries and other free disorder defects to future work. 

\subsection{Mismatching Twists and a Non-Renormalization Theorem}\label{sec:MismatchingTwists}
There are many situations where some translations which are exact in the bulk may not be exact at the defect. For example, one may consider a holomorphic boundary condition for a 3d topological field theory. 

One potential approach to this problem is to ``forget'' some of the structure of the bulk theory. For example, two topological directions could be traded for a holomorphic direction. In this case, the topological superfield $\Phi$ will split into two fields in the holomorphic theory, a $(0,*)$-form and a $(1,*)$-part:
\begin{equation}
    \Phi = \Phi^{(0)} + \Phi^{(1)} dz\,.
\end{equation}
Likewise, the topological superfield condition splits in terms of the new (less topological) $\dd$:
\begin{equation}
    (Q+\dd_{\mathrm{old}}) \Phi = (Q + \dd)\Phi + \partial \Phi = 0\,.
\end{equation}
The $\partial \Phi$ term is now interpreted as a BRST anomaly due to the holomorphic part of the kinetic term. i.e. the term
\begin{equation}
    (\Phi, \partial \Phi)\,,
\end{equation}
is an interaction in the holomorphic theory. In other words, we can view the topological theory as a holomorphic theory with an odd symmetry ``$\partial$''  (the homotopy for an actual holomorphic translation), deformed by the current for that symmetry. We can then proceed with calculations in the topological theory as if the system is holomorphic.

The strategy of converting two topological directions into a holomorphic direction also gives important insight on the 
$\alpha_\Gamma^2=0$ identity from Section \ref{sec:projective}. For example, any calculation involving the diagram $\Gamma$ in an $(H,T\geq2)$-theory, can be converted into an equivalent calculation, involving a sequence of diagrams built from $\Gamma$, in an $(H+1,T-2)$-theory. If we can find a mismatch between the relevant Laman conditions for the $(H,T)$ and $(H+1,T-2)$ theories, that would imply a vanishing contribution from Feynman graph of the topology $\Gamma$.\footnote{This argument was suggested to us by K. Costello, inspired by the structure of Kontsevich original proof.}

To this end, consider a theory with $(H,T\geq 2)$ and let $n=H+T$. Any calculation in the $(H,T)$-theory involving the $n$-Laman graph $\Gamma$ can be replaced by a collection of graphs in the alternative presentation as an $(H+1,T-2)$ theory. The graphs for the same calculation in the $(H+1,T-2)$ theory will be constructed by substituting some edges $e_i\in\Gamma_1$ with chains of edges $\{f_{i,1}, \dots, f_{i,m_i}\}$ with $(\Phi,\partial\Phi)$ insertions at all of the new vertices. There can also be ``dangling'' chains of edges glued onto various vertices. In other words, the graphs of the $(H+1,T-2)$-theory built from $\Gamma$ involve breaking up edges of $\Gamma$ by inserting the ``two-point interaction'' $(\Phi,\partial\Phi)$ everywhere. Each of the new $(H+1,T-2)$ graphs will therefore have $\abs{\Gamma_1}+k$ edges, and $\abs{\Gamma_0}+k$ vertices for some $k\geq 0$.

If the new graphs replacing $\Gamma$ are going to be non-vanishing, they must be $(n-1)$-Laman graphs. Thus we have two constraints from the global Laman conditions:
\begin{align}
    n \abs{\Gamma_0} 
        &= (n-1) \abs{\Gamma_1} + n + 1\\
    (n-1)(\abs{\Gamma_0}+k) 
        &= (n-2) (\abs{\Gamma_1}+k) + (n-1) + 1\,,
\end{align}
for some $k\geq0$. The solutions are
\begin{equation}
    \abs{\Gamma_1} = 1 - kn\,, \quad
    \abs{\Gamma_0} = 1+k+ \abs{\Gamma_1}\,.
\end{equation}
Such conditions imply that $\Gamma$ must have been a tree. i.e. if the equivalent computations in the $(H+1,T-2)$-theory are going to be non-vanishing, they must be replacing a tree computation in the original theory. So all loop graphs in $(H,T\geq2)$-theories must have vanishing contributions. 

Another way to understand this inequality is to note that each term in the original Lagrangian contains exactly one $(1,*)$ superfield. The holomorphic kinetic term $(\Phi, \partial \Phi)$, viewed as an interaction, only contains $(0,*)$ superfields. Thus each edge of a graph $\Gamma$ uses up a $(1,*)$ superfield, and the final answer must contain at least a $(1,*)$ superfield.

It is interesting to note that this proof used only facts about the integrand/Feynman graph combinatorics, not the Feynman integrals. In principle, we could have imagined such a result was true, but with careful cancellations after integrations, possibly between many collections of graphs at the same loop order. But no such delicate integral cancellations were required.

We leave the interesting problem concerning combinatorial identities of $(\alpha_{\Gamma}^{\partial})^2$ to future work.

\section{Point-Splitting Regularization and BRST Anomalies}\label{sec:PointSplittingRevisted}
In this section we will illustrate how BRST anomalies and Maurer-Cartan equations appear in a point-splitting regularization of perturbation theory for a generic topological(ly twisted) field theory. The contents of this section are largely independent of the rest of the paper. They serve as an illustration of how our physical perspective on BRST anomalies and associated operations relates to other notions of higher operations in TFTs, and mathematical notions in the theory of (topological) factorization algebras.

We will set up the problem by describing a very general space of possible point-splitting regularization schemes in QFTs. The language of disk operads, reviewed in Appendix \ref{sec:operadsAndQFT}, will emerge naturally in the process. Specializing to the case of TFTs, the BRST anomalies will be expressed geometrically as (scheme-dependent) integrals which generalize the point-splitting expression in Section \ref{sec:PointSplitting} for $\{\cI, \cI\}$ and match the conventional mathematical presentation of higher operations in cohomological TFTs. 

\subsection{Point-Splitting Schemes for Composite Operators}
The notion of local operator is central to our understanding of QFTs. Local operators are an idealization, representing observations/operations supported on an infinitesimal neighbourhood of a point in spacetime. The notion of local operator is necessarily relaxed in any specific renormalization scheme. For example, introducing a UV cutoff $\Lambda$ makes all of spacetime slightly non-local and ``blows up'' composite operators or point interactions to sizes $\sim\!\!\Lambda^{-1}$. A point-splitting regularization of a local operator manifestly breaks down a composite operator into a collection of elementary fields at separate points.\footnote{In a gauge theory, one may want to include Wilson lines for a gauge-invariant point-splitting regularization. In perturbation theory, these are just complicated linear combinations of collections of elementary fields at separate locations.} The relative locations of the elementary fields are part of the definition of a specific regularization scheme. 

The notion of ``local observable'' is also very useful in a QFT. A local observable represents an observation/operation supported on some open set $U$ in spacetime. We will denote as $\mathrm{Obs}(U)$ local observables supported on $U$. In most regularization schemes, local operators are effectively replaced by a \textit{family} of local observables with increasingly small support.\footnote{The concept of a local observable is also better defined in more scenarios. For example, we shouldn't consider too seriously point observables in an effective field theory, or really any local observable supported on a region smaller than the EFT cutoff scale.} Accordingly, we will explicitly replace local operators $\calO(x)$ with families $\calO(x,\mu) \in \mathrm{Obs}(B_\mu(x))$ of local observables supported on balls of radius $\mu$, defined explicitly for any $(x,\mu) \in \bbR^d \times \bbR^+$. In a point-splitting regularization scheme, we will make sure that the supports of different observables never overlap. 

Given a collection of regularized local operators $\calO_1, \dots, \calO_n$, we often wish to construct a new composite local operator $[\calO_1,\dots,\calO_n]$. Keeping up with our conventions, we need to give a specific point-splitting definition of $[\calO_1,\dots,\calO_n](x,\mu)$ for any $x$ and $\mu$. Concretely, that means selecting locations and sizes for the individual constituents $(x_i, \mu_i)$, in such a manner that the $B_{\mu_i}(x_i)$ supports do not overlap and lie within the desired support $B_\mu(x)$.\footnote{Here we implicitly use the axiom that a collection of observables with disjoint support can be composed. i.e. there exists a map 
\begin{equation}
    m^{\{\mu_i; x_i\}}_{\mu,x}: \mathrm{Obs}(B_{\mu_1}(x_1)) \otimes \cdots \otimes \mathrm{Obs}(B_{\mu_n}(x_n)) \to \mathrm{Obs}(B_\mu(x))\,,
\end{equation} 
Mathematically, composition of disjoint observables is called the ``factorization product'' of the underlying factorization algebra.}  For simplicity, we define a scheme for $(x=0,\mu=1)$ and then apply a translation and rescaling to the configuration of disks
\begin{equation}
    (x_i, \mu_i) \to (\mu x_i + x, \mu \mu_i)\,
\end{equation}
to define a composite operator for general $(x,\mu)$.\footnote{Of course, one may prefer a regularization scheme where the definition of composite operators is given independently (but continuously) for different $\mu$'s. In this case, one works with a ``coloured disk operad'' structure. It is easy to adjust the definitions below accordingly.}

For any $n \in \mathbb{N}^0$ introduce the topological space $D_d(n)$ consisting of all configurations of $n$ \textit{non-overlapping} disks inside a $d$-dimensional disk of radius $1$. Then each of the $n$-ary point-splitting prescriptions above corresponds to a point $p\in D_d(n)$. Thus there are a huge number of schemes, labelled by $p$, to produce a fattened composite local operator from the $\calO_i$, each corresponding to a particular arrangement of little disks inside a large disk. We denote them as $[\calO_1,\dots,\calO_n]_p$ where the point
\begin{equation}
    p=\{(x_1,\mu_1),\dots,(x_n,\mu_n)\} \qquad \text{is pictorially} \qquad  
    \begin{tikzpicture}[thick, baseline={(current bounding box.center)}]
    \def\gS{2};
    \coordinate (C) at (0,0);
    \coordinate (c1) at (-0.5*\gS,-0.0*\gS);
    \coordinate (c2) at (-0.1*\gS,0.7*\gS);
    \coordinate (c3) at (0.5*\gS,-0.3*\gS);
    \coordinate (dots) at (0.3*\gS, 0.3*\gS);
    
    \draw (C) circle (1*\gS);
    \draw (c1) circle (0.4*\gS);
    \draw (c2) circle (0.25*\gS);
    \draw (c3) circle (0.3*\gS);

    \draw (C) node[above] {$0$};
    \draw (c1) node[below] {$x_1$};
    \draw (c2) node[below] {$x_2$};
    \draw (c3) node[below] {$x_n$};
    \draw (dots) node {$\dotsb$};

    \draw (C) -- node[midway, left] {$1$} (-90:\gS);
    \draw (c1) -- ({-0.5*\gS+0.4*\gS*cos(45)},{-0.0*\gS+0.4*\gS*sin(45)}) node[midway, above left = -0.15cm] {$\mu_1$};
    \draw (c2) -- ({-0.1*\gS+0.25*\gS*cos(45)},{0.7*\gS+0.25*\gS*sin(45)}) node[midway, above left = -0.15cm] {$\mu_2$};
    \draw (c3) -- ({0.5*\gS+0.3*\gS*cos(45)},{-0.3*\gS+0.3*\gS*sin(45)}) node[midway,above left = -0.15cm] {$\mu_n$};
\end{tikzpicture}
\,\,\,.
\end{equation}
We thus get a family of ``operations'' $[\calO_1,\dots,\calO_n]_p$ parameterized by $D_d(n)$. These operations map local operators $\calO_i$ to local operators
\begin{equation}\label{eq:compositeFamily}
    [\calO_1,\dots,\calO_n]_{p}(x,\mu) \define \prod_i \calO_i\left(\mu x_i + x,\mu \mu_i\right)\,, 
\end{equation}\
described explicitly as families of observables supported on $B_\mu(x)$.

These operations can be safely composed. A composition gives an operation of the same type: 
\begin{equation}
    [\calO_1,\dots,[\calO_{k+1},\dots,\calO_{k+m}]_{p'},\dots,\calO_{n+m-1}]_p = [\calO_{1},\dots,\calO_{n+m-1}]_{p\circ_k p'}
\end{equation}
where $p\circ_k p'$ is a configuration of disks obtained by shrinking the unit disk for $p'$ to size $\mu_k$ and inserting it into the $k$-th disk in $p$. The axiomatics of general compositions can be expressed as $[\,\cdot\,,\dots,\,\cdot\,]_p$ defining an ``algebra'' over the ``$E_d$-operad'' of little disks, see Appendix \ref{sec:operadsAndQFT}. We can also view the composite operator $[\calO_1,\dots,\calO_n]$ as a function of $p \in D_d(n)$. 

In a topological(ly twisted) theory, both translations and scale transformations should be BRST exact, i.e. homotopically trivial. So we can pick a (scheme dependent) identification between operators defined at different points and scales, allowing us to speak about placing an operator $\calO$ at position $x$ and scale $\mu$ up to BRST-exact terms. In other words, $\calO(x,\mu)$ represents a family of observables which limits to the point operator $\calO$, where changes of $(x,\mu)$ are $Q$-exact. 

In order to keep track of BRST exact quantities, we once again employ extended superfields $\cO$ which are forms on $\bR^d \times \bR^+$. They satisfy an enhanced descent relation: 
\begin{equation}\label{eq:topDescent}
    (Q + \dd) \cO(x,\mu)=0\,,
\end{equation}
where $\dd$ is the de Rham differential in $x$ and $\mu$, aka in the space of disks. The composite operator 
\begin{equation}
    [\calO_1,\dots,\calO_n]_p(x,\mu)
\end{equation}
becomes a form on $D_d(n)$ rather than just a function of $p$. Furthermore, it is annihilated by $Q+\tilde{\dd}$ where $\tilde{\dd}$ \textit{acts on all of the variables}: the point $p \in D_d(n)$ as well as $x$ and $\mu$.

We can see the descent relation satisfied by the composite local operator explicitly. Introduce the notation:
\begin{align}
    \dd 
        &= dx \frac{\partial}{\partial x} + d\mu \frac{\partial}{\partial \mu}\,,\\
    \dd^{(i)} 
        &= dx_i \frac{\partial}{\partial x_i} + d\mu_i \frac{\partial}{\partial \mu_i}\,,
\end{align}
so that
\begin{equation}
    \tilde{\dd} = \dd + \sum_i \dd^{(i)}\,,
\end{equation}
and use the shorthand
\begin{equation}
    y_i := \mu x_i + x\,, \quad \nu_i := \mu \mu_i\,.
\end{equation}
The descent relation in \eqref{eq:topDescent} applied to a constituent local operator $\calO_i(x_i,\mu_i)$ (i.e. family of local observables) is written explicitly as:
\begin{equation}
    Q \mathcal{O}_i(x_i,\mu_i) + dx_i \mathcal{O}_i^{(1,0)}(x_i,\mu_i) + d\mu_i \mathcal{O}_i^{(0,1)}(x_i,\mu_i) = 0\,.
\end{equation}
Then we see that the shifted and scaled family of local observables $\calO_i(y_i,\nu_i)$ are annihilated by $(Q+\tilde{\dd})$:
\begin{align}
    (Q+\tilde{\dd})\calO_i(y_i,\nu_i)
        &=(Q + \dd + \dd^{(i)})\calO_i(y_i,\nu_i)\\
        &={Q\calO_i(y_i,\nu_i)}
            +(d\mu x_i + \mu dx_i + dx) \calO^{(1,0)}(y_i,\nu_i)\nonumber\\
        &\hphantom{ = Q\calO_i(y_i,\nu_i)}
            \,\,+(d\mu \mu_i + \mu d\mu_i) \calO^{(0,1)}(y_i,\nu_i)\\
        &=Q\calO_i(y_i,\nu_i) + dy_i \calO_i^{(1,0)}(y_i,\nu_i) + d\nu_i \calO_i^{(0,1)}(y_i,\nu_i)\\
    (Q+\tilde{\dd})\calO_i(y_i,\nu_i) &=0\,.    
\end{align}
Then the composite local observable in \eqref{eq:compositeFamily} is annihilated by $(Q+\tilde{d})$ using the Leibniz rule:
\begin{equation}
    (Q+\tilde{\dd})[\calO_1,\dots,\calO_n]_p(x,\mu) 
        = (Q+\tilde{\dd}) \prod_{i=1}^n \calO_i(y_i, \nu_i) = 0\,.
\end{equation}

At this point, it becomes natural to also consider integrals of composite operators over contours (better, chains) $\gamma \subset D_d(n)$:
\begin{equation}
    [\calO_1,\dots,\calO_n]_\gamma(x,\mu) \define \int_\gamma [\calO_1,\dots,\calO_n]_p(x,\mu)\,.
\end{equation}
Note that this integral is only over the internal $(x_i, \mu_i)$, the output is still a local operator at $x$ parametrized by a $\mu$-dependent family of observables. Then an integration by parts gives 
\begin{align}
    (Q + \dd)[\calO_1,\dots,\calO_n]_\gamma(x,\mu)
        &= \int_\gamma (Q + \tilde{\dd} - \sum_i \dd^{(i)})[\calO_1,\dots,\calO_n]_p(x,\mu)\\
        &= -\int_{\partial\gamma} [\calO_1,\dots,\calO_n]_p(x,\mu)\\
        &= -[\calO_1,\dots,\calO_n]_{\partial \gamma}(x,\mu)\,.
\end{align}
If the inputs do not satisfy the descent relations, the right hand side will have extra terms involving $(Q + \tilde{\dd}) \cO_i(y_i,\nu_i)$. Note that a priori, $[\calO_1,\dots,\calO_n]_\gamma(x,\mu)$ does not satisfy the descent relation \eqref{eq:topDescent} for a $Q$-closed operator at $x$, even when it is composed from $Q$-closed operators. It's failure to do so is controlled by the boundary of $\gamma$. 

In the same way that we ``compose'' points in $D_d$, we can compose paths $\gamma_n \circ_k \gamma_m$ by shrinking a path and inserting it into another. The corresponding composed brackets are just an integral over the composed paths. As we will see, the final answer for the BRST anomaly brackets, in point-splitting regularization, will take precisely this form for a judiciously chosen collection of chains $\gamma_n$ of dimension $d(n-1)-1$ (codimension 1) in $D_d(n)$.

In the literature, higher brackets are often described in terms of compact integration contours in the configuration space of distinct points in $\bbR^d$:
\begin{equation}
    \mathrm{Conf}_n(\bbR^d) = \{(x_1,\dots,x_n)\in (\bbR^d)^n \,\vert \text{ such that  $x_i\neq x_j$ if $i\neq j$}\}\,.
\end{equation}
As the sizes of disks in $D_d(n)$ are contractible, $\mathrm{Conf}_n$ captures enough information about the topology of $D_d(n)$ to label integration contours up to homotopy. However, an actual choice of contour $\gamma_n \subset D_d(n)$ is needed to prescribe any specific renormalization scheme. 

With this in mind, if we pass to $Q$-cohomology, we see that there is a distinct operation for each $\gamma \in H_*(\mathrm{Conf}_n)$. The homology of this space is well understood \cite{sinha2006homology}, and it turns out that all $n$-ary operations coming from closed chains in $H_*(\mathrm{Conf}_n)$ are (homologous to) nested $2$-ary operations \cite{Beem:2018fng}. This fact, albeit important, is not directly relevant for our discussion, since we are looking for the coefficients of Maurer-Cartan equations which control deformations \textit{beyond the linearized order}.

\subsection{Computing the BRST Anomaly Brackets}
Correlation functions of a theory $\calT$ deformed by $\calI$ are (in perturbation theory) correlation functions of $\calT$ with additional insertions of the naive form:
\begin{equation}
    O(g^n) \sim \int \calI(x_1) \cdots \calI(x_n)\,.
\end{equation}
To avoid UV divergences from colliding interaction vertices, one must regulate the above interactions. In Section \ref{sec:QFT} we discussed a point-splitting regularization scheme at the second order in perturbation theory, by excising small balls around interactions. In Section \ref{sec:PointSplitting}, we saw how interaction terms which are classically BRST-invariant could still give ``quantum corrections'' to the BRST anomaly, captured by the bracket $\{\calI,\calI\}$ of the undeformed theory $\calT$. 

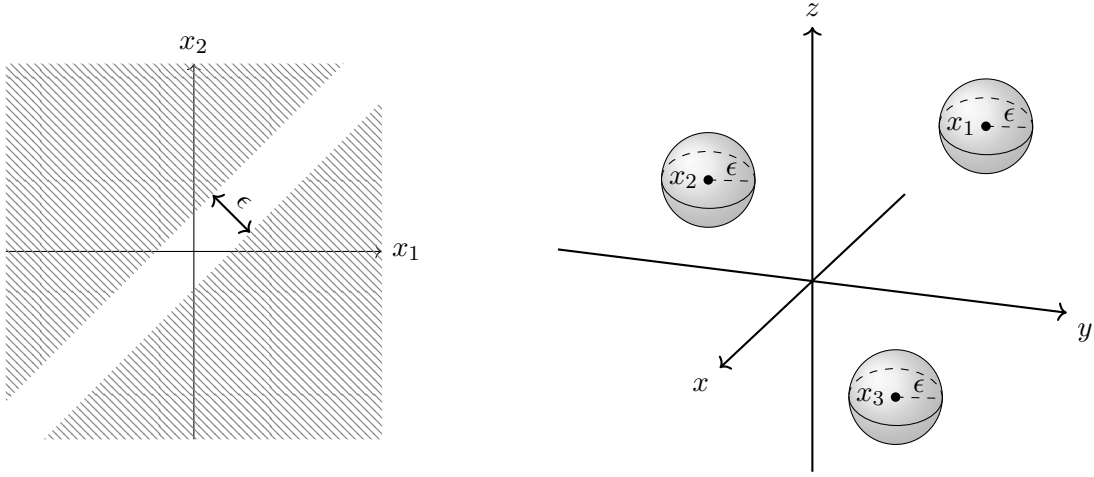
\begin{figure}[t]
\centering
\begin{minipage}{0.45\textwidth}
\centering
\begin{tikzpicture}[thick, baseline={(current bounding box.center)}]
  \def\planeSize{5}
  \def\stripWidth{4.5}
  \def\sepSlide{0.5}

  \draw[thin, ->] (0, -\planeSize/2) -- (0, \planeSize/2) node[above] {$x_2$};
  \draw[thin, ->] (-\planeSize/2, 0) -- (\planeSize/2, 0) node[right] {$x_1$};

  \fill[pattern=north west lines, pattern color=gray] (-\planeSize/2, \planeSize/2) -- (-\planeSize/2 + \stripWidth, \planeSize/2) -- (-\planeSize/2, \planeSize/2 - \stripWidth) -- cycle;

  \fill[pattern=north west lines, pattern color=gray] (\planeSize/2, -\planeSize/2) -- (\planeSize/2 - \stripWidth, -\planeSize/2) -- (\planeSize/2, -\planeSize/2 + \stripWidth) -- cycle;

  \draw[<->, thick] (-\planeSize/2 + \stripWidth/2 + \sepSlide, \planeSize/2 - \stripWidth/2 + \sepSlide) -- (\planeSize/2 - \stripWidth/2 + \sepSlide, -\planeSize/2 + \stripWidth/2 + \sepSlide) node[midway, above, sloped] {$\epsilon$};
\end{tikzpicture}
\end{minipage}
\hspace{0.75cm}
\begin{minipage}{0.45\textwidth}
\centering
\tdplotsetmaincoords{70}{110}
\begin{tikzpicture}[tdplot_main_coords, scale=1.8]

    \def\eps{0.35}

  \draw[thick,->] (-2,0,0) -- (2,0,0) node[anchor=north east]{$x$};
  \draw[thick,->] (0,-2,0) -- (0,2,0) node[anchor=north west]{$y$};
  \draw[thick,->] (0,0,-1.5) -- (0,0,2) node[anchor=south]{$z$};

  \coordinate (x1) at (-1,1,1);
  \coordinate (x2) at (-0.5,-1,0.5);
  \coordinate (x3) at (1.5,1.2,-0.25);

  \tdplotsetrotatedcoordsorigin{(x1)}
  \tdplotsetrotatedcoords{30}{70}{85}
  \begin{scope}[tdplot_rotated_coords]
    \shade[ball color = gray!40, opacity = 0.4] (0,0,0) circle (\eps);
      \draw (0,0,0) circle (\eps);
      \draw (-\eps,0,0) arc (180:360:{\eps} and 0.6*\eps);
      \draw[dashed] (\eps,0,0) arc (0:180:{\eps} and 0.6*\eps);
      \draw[dashed] (0,0,0) -- node[above = -0.06cm]{$\epsilon$} (\eps,0,0);
  \end{scope}

  \tdplotsetrotatedcoordsorigin{(x2)}
  \begin{scope}[tdplot_rotated_coords]
    \shade[ball color = gray!40, opacity = 0.4] (0,0,0) circle (\eps);
      \draw (0,0,0) circle (\eps);
      \draw (-\eps,0,0) arc (180:360:{\eps} and 0.6*\eps);
      \draw[dashed] (\eps,0,0) arc (0:180:{\eps} and 0.6*\eps);
      \draw[dashed] (0,0,0) -- node[above = -0.06cm]{$\epsilon$} (\eps,0,0);
  \end{scope}

  \tdplotsetrotatedcoordsorigin{(x3)}
  \begin{scope}[tdplot_rotated_coords]
    \shade[ball color = gray!40, opacity = 0.4] (0,0,0) circle (\eps);
      \draw (0,0,0) circle (\eps);
      \draw (-\eps,0,0) arc (180:360:{\eps} and 0.6*\eps);
      \draw[dashed] (\eps,0,0) arc (0:180:{\eps} and 0.6*\eps);
      \draw[dashed] (0,0,0) -- node[above = -0.06cm]{$\epsilon$} (\eps,0,0);
  \end{scope}

    \fill (x1) circle (1pt) node[anchor=east] {$x_1$};
    \fill (x2) circle (1pt) node[anchor=east] {$x_2$};
    \fill (x3) circle (1pt) node[anchor=east] {$x_3$};

\end{tikzpicture}
\end{minipage}
\caption{Left, in gray, the full configuration space $\mathrm{Conf}_2^{\epsilon}(\bbR)$, where two points in $\bbR$ are kept apart by a distance at least $\epsilon$. The boundary can be factored into a compact sphere and $\bR^d$. Right, a \textit{point} in the configuration space $\mathrm{Conf}^{2\epsilon}_{3}(\bbR^3)$ of points which keep a distance of at least $2 \epsilon$ from each other. This condition is equivalent to the requirement that balls of radius $\epsilon$ around each point do not overlap. The boundary of such a regularized integration region is more complicated. Our objective in the main text is to design an improved integration region with a well-controlled boundary.}
\label{fig:Conf2Disks}
\end{figure}

In a sharp cutoff regularization scheme, the introduction of the regulator 
\begin{equation}
    \int \calI(x_1) \calI(x_2) \mapsto \int f^{(2)}_{\epsilon}(x_1,x_2)\calI(x_1) \calI(x_2)
\end{equation}
(discussed in Section \ref{sec:PointSplitting}) is equivalent to changing the region of integration from $(\bbR^d)^2$ to the configuration space
\begin{equation}
    \mathrm{Conf}_2^{\epsilon}(\bbR^d) = \{(x_1,x_2)\in (\bbR^d)^2 \,\vert \text{ such that $\abs{x_1-x_2}\geq \epsilon$}\}\,,
\end{equation}
see Figure \ref{fig:Conf2Disks}. We can also think of this space as the configuration space of two non-overlapping disks of size $\epsilon/2$ in $\bbR^d$. The resulting BRST anomaly for the regularized integral is localized on the boundary of this region, which is essentially $\bbR^d \times S^{d-1}$,
and is rewritten as the integral on $\bbR^d$ of $\{\calI,\calI\}$.

As we point-split regulate the interactions themselves to families $\cI(x_i,\mu_i)$, it is natural to postulate a potential regularization of the interaction terms as an $nd$-dimensional contour $\gamma^\infty_n$ in an enlarged configuration space $D^\infty_d(n)$ of non-overlapping disks in $\bR^d$. Given such a contour, we can integrate the composition of $n$ interactions $\cI(x_i,\mu_i)$ over it to obtain a regularized interaction 
\begin{equation}
    (\cI, \dots, \cI)_{\gamma^\infty_n} \define \int_{\gamma_n^\infty} \prod_i \cI(x_i,\mu_i)\,.
\end{equation}
Our goal in the remainder of this section will be to argue for the existence of this contour $\gamma_n^\infty$.

The space $D^\infty_d(n)$ and the associated brackets $(\,\cdot\,, \dots, \,\cdot\,)_{\gamma_n^\infty}$ bear a close relationship to $D_d(n)$ and  $[\,\cdot\,, \dots, \,\cdot\,]_{\gamma_n}$: we cannot compose two points in $D^\infty_d(n)$, but we can insert a configuration (point) in $D_d(m)$ into any of the disks in $D^\infty_d(n)$. In an operadic language, $D^\infty_d$ is a module for the $E_d$-operad. This geometric construction describes the insertion of a composite interaction into the bracket: 
\begin{equation}
    (\calO_1,\dots,[\calO_{k+1},\dots,\calO_{k+m}]_{\gamma_m},\dots,\calO_{n+m-1})_{\gamma_n^\infty} = (\calO_{1},\dots,\calO_{n+m-1})_{\gamma_n^\infty\circ_k \gamma_m}\,,
\end{equation}
and allows us to neatly express the BRST anomaly of the regularized interaction:
\begin{equation}
   Q (\cI, \dots, \cI)_{\gamma^\infty_n} = -(\cI, \dots, \cI)_{\partial \gamma^\infty_n}\,.
\end{equation}

In order to formulate a BRST anomaly cancellation condition, we want to find a collection of contours $\gamma^\infty_n$ in $D_d^\infty(n)$ with a ``nice'' boundary, where nice means satisfying:
\begin{equation}
    \partial\gamma^\infty_{n}+ \sum_m \sum_k (-1)^{\cdots} \gamma^\infty_{n-m+1} \circ_k \gamma_m =0\,,
\end{equation}
because on brackets this would imply  
\begin{equation}
       Q (\cI, \dots, \cI)_{\gamma^\infty_n} + \sum_m \sum_k (-1)^{\dots} (\cI, \dots, [\cI, \dots, \cI]_{\gamma_m},\dots, \cI)_{\gamma^\infty_{n-m+1}} =0\,.
\end{equation}
Summing over $n$, we find that the BRST variation of the regularized exponentiated interaction is equivalent to a shift of the interaction by a Maurer-Cartan element (recall equation \eqref{eq:MCE}), corresponding to the sum of the BRST anomaly brackets,
\begin{equation}
    \sum_m [\cI, \dots, \cI]_{\gamma_m} = 0\,,
\end{equation}
as desired.

Can we build such a collection of cycles? The condition $\partial^2=0$ applied to $\gamma^\infty_n$ effectively imposes a consistency condition 
\begin{equation}
    \partial\gamma_{n}+ \sum_m \sum_k (-1)^{\cdots} \gamma_{n-m+1} \circ_k \gamma_m = 0\,,
\end{equation}
which is equivalent to the expected quadratic relations for the BRST brackets if\footnote{Note, BRST anomalies are local and correspond to a local operator, so the BRST bracket $\{\,\cdot\,,\dots,\,\cdot\,\}$ correctly corresponds to an operator bracket $[\,\cdot\,,\dots,\,\cdot\,]$. Recall that interactions are local operators modulo $(Q_{\mathrm{BRST}}+d)$, so in this topologically twisted scenario interactions and local operators are identifiable anyway.}
\begin{equation}
    \{\cI_1, \dots, \cI_n\} = [\cI_1, \dots, \cI_n]_{\gamma_n}\,.
\end{equation}
We should thus first build contours (chains) $\gamma_n$ and then $\gamma_n^\infty$, with the boundary condition that $\gamma^\infty_n$ should approach the unregulated contour on $\bR^d$ with $\mu_i=\epsilon$ when disks are well-separated. 

An explicit construction goes beyond the scope of this text. A recursive strategy seems appropriate. Schematically, the right hand sides of the relations we need to solve are $\partial$ closed because of the previous steps in the recursion. If one can show that there is no non-trivial homology of the appropriate dimension (and asymptotic boundary conditions for $\gamma_n^\infty$) in $D_d$ and  $D^\infty_d$, one can pick any solution and proceed to the next step (see e.g. \cite{sinha2006homology} and references within).

\acknowledgments
We would like to thank Kevin Costello, Diego Delmastro, Leon Liu, and Surya Raghavendran for useful conversations and feedback on the draft. We would like to thank Kasia Budzik, Brian Williams, and Matthew Yu for contributions at early stages of the project. This research was supported in part by a grant from the Krembil Foundation. DG is supported by the NSERC Discovery Grant program and by the Perimeter Institute for Theoretical Physics. JK acknowledges the support of the NSERC CGS-D and PDF programs. The work of JW is supported by the UKRI Frontier Research Grant, underwriting the ERC Advanced Grant titled “Generalized Symmetries in Quantum Field Theory and Quantum Gravity” and by the Simons Foundation Collaboration on “Special Holonomy in Geometry, Analysis, and Physics”, Award ID: 724073, Schafer-Nameki. Research at Perimeter Institute is supported in part by the Government of Canada through the Department of Innovation, Science and Economic Development Canada and by the Province of Ontario through the Ministry of Colleges and Universities.

\appendix
\section{Vector Field Calculations} \label{app:VFCalcs}
All calculations use Einstein summation convention unless otherwise stated.

\subsection{Odd Nilpotent Vector Fields \texorpdfstring{$\bm{\eta}^2 = 0$}{eta**2=0}}\label{app:eta2}
Consider the formal vector field $\bm{\eta}$ associated with any odd nilpotent symmetry, as introduced in Section \ref{sec:QFT}. $\bm{\eta}$ takes the generic form given in \eqref{eq:Eta}; in condensed notation
\begin{equation}
    \bm{\eta} = \eta^i(g) \partial_i \,.
\end{equation}
For consistency, the vector field must satisfy $\bm{\eta}^2 = 0$. When $\eta$ is induced by BRST transformations, this is a form of Wess-Zumino consistency condition. 

We will confirm that the term quadratic in $\partial_i$'s disappears in $\bm{\eta}^2$, producing \eqref{eq:WZConsistency}. Expanding in component form
\begin{align}
    0 = \bm{\eta}^2 
        &= \eta^i\partial_i \eta^j\partial_j + (-1)^{p(i)(p(j)+1)}\eta^i\eta^j\partial_i\partial_j\\
        &= \eta^i\partial_i\eta^j\partial_j
\end{align}
Here $p(i)$ denotes the Grassmann parity of $\eta^i$, which is opposite of $\partial_i$. The second term in the expression can be seen to vanish as follows: When $p(i)=p(j)$ then the $\eta$-coefficients are symmetric/anti-symmetric and the derivatives are anti-symmetric/symmetric respectively, and so contract to 0. Conversely, when $p(i) \neq p(j)$ then $\eta^i\eta^j\partial_i\partial_j$ is invariant under $i\leftrightarrow j$, but then $(-1)^{p(i)(p(j)+1)} = -(-1)^{p(j)(p(i)+1)}$, so the cross-terms will cancel in the summation. Thus we are always left with just the single-derivative term, as we would expect for a vector field generating isometries on a (super)manifold.

\subsection{Position Dependent Couplings}\label{app:posDep}
Consider formal deformations to the space of position dependent interactions of the form $x^{\mu_1} \cdots x^{\mu_k} {\cal I}_i$ modulo total derivatives as in Section \ref{sec:pBracket}. Our odd nilpotent vector field is of the form \eqref{eq:bigEta}:
\begin{equation}
    \bm{\eta} = \sum_{k} \eta^i_{\mu_1\cdots\mu_k}(g) \partial_i^{\mu_1 \cdots \mu_k}\,.
\end{equation}
Likewise, the even vector field corresponding to translations takes the form:
\begin{equation}
    \bm{P}_\mu = \sum_k (P_\mu)^i_{\mu_1 \cdots \mu_k} \partial_i^{\mu_1 \cdots \mu_k}\,.
\end{equation}

Assuming $[\bm{P}_\mu,\bm{\eta}] = 0$ (or similarly for any vector field) gives strong constraints intertwining the coefficients:
\begin{align}
0 = [\bm{P}_\mu, \bm{\eta}] f(g)
    &= \sum_{\ell, k}\left((P_\mu)^j_{\nu_1\cdots\nu_{\ell}} \partial_j^{\nu_1\cdots\nu_\ell}(\eta^i_{\mu_1\cdots\mu_k}\partial_i^{\mu_1\cdots \mu_k}f(g))\right.\nonumber\\
    &\hphantom{= \sum_{\ell, k}}-\left.\eta^i_{\mu_1\cdots\mu_{k}} \partial_i^{\mu_1\cdots\mu_k}((P_\mu)^j_{\nu_1\cdots\nu_{\ell}} \partial_j^{\nu_1\cdots\nu_\ell}f(g))\right)\\
    &= \sum_{\ell, k}\left((P_\mu)^j_{\nu_1\cdots\nu_{\ell}} \partial_j^{\nu_1\cdots\nu_\ell}\eta^i_{\mu_1\cdots\mu_k}\partial_i^{\mu_1\cdots \mu_k}f(g)\right.\nonumber\\
    &\hphantom{= \sum_{\ell, k}}-\left.\eta^i_{\mu_1\cdots\mu_{k}} \partial_i^{\mu_1\cdots\mu_k}(P_\mu)^j_{\nu_1\cdots\nu_{\ell}} \partial_j^{\nu_1\cdots\nu_\ell}f(g)\right)\,.
\end{align}
Since it must be true for any $f(g)$, we can set $f(g) = g^m_{\rho_1\cdots\rho_n}$ arbitrary. Then using $\partial_i^{\mu_1\cdots \mu_k} g^m_{\rho_1\cdots \rho_n} = \delta_i^m \delta^{\mu_1\cdots\mu_k}_{\rho_1\cdots\rho_n}$, we find:
\begin{equation}
    \sum_{\ell} (P_\mu)^i_{\nu_1 \cdots \nu_\ell} \partial^{\nu_1\cdots\nu_\ell}_i \eta^m_{\rho_1 \cdots \rho_n}
    =
    \sum_{\ell} \eta^i_{\nu_1 \cdots \nu_\ell} \partial^{\nu_1\cdots\nu_\ell}_i (P_\mu)^m_{\rho_1 \cdots \rho_n}\,.
\end{equation}
The further assumption that $\bm{P}_\mu$ receives no corrections collapses this expression to \eqref{eq:PfromEta}.

\section{Schwinger Parametrization}\label{app:SchwingerParametrization}
A convenient example of UV regulator is the heat kernel regularization, 
which replaces the source with
\begin{equation}
    \dd P_\epsilon(x) = K_\epsilon(x) \define  \frac{1}{(\pi  \epsilon)^{\frac{d}{2}}}  e^{- \epsilon^{-1}\left((x^{\bR})^2 + x^{\bC} \bar{x}^\bC\right) } \,d^H\!\bar{x}^\bC 
         \,d^R\!x^\bR\,. 
\end{equation}
We introduce the Laplacian
\begin{equation}
    {\nabla}^2 = \frac{1}{4}\partial^2_{x^{\bbR}} +  \partial_{x^{\bbC}}\partial_{\bar{x}^{\bbC}}\,,
\end{equation}
and the codifferential $\dd^*$, defined so that
\begin{equation}
    \dd \dd^* = \nabla^2\,.
\end{equation}
Then we see that
\begin{equation}
    \nabla^2 K_\epsilon(x) = \partial_t K_t(x)\,,
\end{equation}
so that the truncated Schwinger time presentation of the regularized propagator is:
\begin{equation}
    P_\epsilon(x) = \int_{t=\epsilon}^\infty \dd^* K_t(x) \,dt\,.
\end{equation}

As seen in Section \ref{sec:FeynmanIntegrals}, a general Feynman integral is a (sum over) products of propagators $P(x)$ with one cut propagator $K(x)$. Hence we are faced with collections of integrands of the form
\begin{equation}
    \sum_{e \in \Gamma_1} (-1)^{\sigma(\Gamma)} K_{\epsilon}(x_e) \dvol_{x_e} \prod_{e' \in \Gamma_1}^{e' \neq e}  \int_{\epsilon}^\infty dt_{e'} \dd^* K_{t_e'}(x_{e'}) \,,
\end{equation}
possibly IR regulated to
\begin{equation}
\sum_{e \in \Gamma_1} (-1)^{\sigma(\Gamma)} K_{\epsilon}(x_e) \dvol_{x_e} \prod_{e' \in \Gamma_1}^{e' \neq e} \int_{\epsilon}^L dt_{e'} \dd^* K_{t_e'}(x_{e'}) \,. 
\end{equation}
We have introduced the symbol $\sigma(\Gamma)$ for the many signs based on our choice of $H$, $T$, and ordering conventions on the graph.

With the IR regulator on, we can safely interchange the integral over positions and Schwinger times. The position integrals are Gaussian and can be performed explicitly, leaving an integral of the form 
\begin{equation}
\sum_{e \in \Gamma_1} (-1)^{\sigma(\Gamma)} \left[\prod_{e' \in \Gamma_1}^{e' \neq e} \int_{\epsilon}^L dt_{e'}\right] \omega_e\,,
\end{equation}
where $\omega_e$ is a rational function of the $t$'s, with $t_e = \epsilon$. At this stage, we can safely send $L \to \infty$. We can envision this as the integral of some rational form $\omega$ defined on $\mathbb{R}^{\abs{\Gamma_1}}$ where $\abs{\Gamma_1}$ is the number of propagators. This form is integrated along the sum of regions $\epsilon < t_{e'}<\infty$ for all $e' \neq e$ at $t_e=\epsilon$. 

In concrete calculations we find a surprising fact: $\omega$ is the pull-back of a form defined on the projective space parameterized by the $t$'s modulo scale transformations. We will sketch a general proof of this observation. 

Consider the positive part $\mathbb{RP}_>$ of real projective space, parameterized by real positive $t$'s modulo rescaling, and divide it into cells where one of the $t_e$ is smaller than all other $t_{e'}$. We can gauge-fix the $t$'s to $\epsilon< t_{e'}<\infty$ for all $e' \neq e$ at $t_e=\epsilon$. If $\omega$ is actually the pull-back of a form on $\mathbb{RP}_>$, then the integral over the above regions can be rewritten as an integral of $\omega$ over the whole $\mathbb{RP}_>$. In particular, the $\epsilon$ dependence drops out.\footnote{There is a neat variant of this where 
we introduce independent UV cutoffs $\epsilon_e$ for each edge. The resulting regions still combine into $\mathbb{RP}_>$ as regions where $t_e/\epsilon_e$ is smaller than all other $t_{e'}/\epsilon_{e'}$. The $\epsilon_e$ dependence then drops out. Averaging this statement over choices of $\epsilon_a$ allows one to make contact with other UV regularizations of the propagators.}

In order to argue that $\omega$ comes from a form on $\mathbb{RP}_>$, we can work with forms defined on the combination of spacetime and the space of Schwinger parameters. For example, we could define something like 
\begin{equation}
    {\cal P}(x;t) = dt \, \dd^* K_t(x)  + K_t(x)\, \,d^H\!\bar{x}^\bC  \,d^T\!x^\bR\,,
\end{equation}
so that the full answer is an integral over product of ${\cal P}(x;t)$'s. As 
discussed at length in \cite{Budzik:2022mpd}, the propagator greatly simplifies in terms of auxiliary variables: 
\begin{equation}
    {\cal P}(x;t) 
        =  \pi^{-\frac{T}{2}} e^{- s^2 - x^\bC \cdot y} \prod_{a=1}^H d y^a \prod_{b=1}^T d s^b \,,
    \qquad 
    y 
        = \frac{\bar x^\bC}{t} \,, 
    \qquad
    s 
        = \frac{x^\bR}{t^{\frac12}} \,.
\end{equation}
After a mild analytic continuation, making $\bar x_\bC$ real and $x_\bC$ imaginary, this change of variables maps the original integral to a manifestly finite integral over a region $\Delta_\Gamma$ in the space of $y_e$ and $s_e$
as well as the $x_\bC$ integral. The $\Delta_\Gamma$ regions satisfy quadratic identities which we imply associativity, as described in Sections \ref{sec:fey} and \ref{sec:brackets}.

\section{Associativity Relations from Quadratic Identities}\label{app:Associativity}
Recall from {Section \ref{sec:BRSTAnomaly}} (see also \cite{Budzik:2022mpd}) that the $\lambda$-bracket is computed via Wick contraction of the free fields. For example, in order to compute $\{\{ \mathcal{O}_1 \, {}_{\lambda_1} \mathcal{O}_2 \}\, {}_{\lambda_1+\lambda_2}, \mathcal{O}_3 \}$, we start with the inner bracket and Wick contract once between $\mathcal{O}_1$ and $\mathcal{O}_2$, yielding the integral $\mathcal{I}_{\pick{1.5ex}{segment}}[\lambda_1,z_{12}]$ multiplying the product of the remaining fields in $\mathcal{O}_1$ and $\mathcal{O}_2$. Then, we perform a second Wick contraction with $\mathcal{O}_3$, which can contract either with remaining fields in $\mathcal{O}_1$ or in $\mathcal{O}_2$. 

Graphically, we can denote this pattern of contractions by the following sum of two graphs
\begin{equation}
    \{\{ \mathcal{O}_1 \, {}_{\lambda_1} \mathcal{O}_2 \}\, {}_{\lambda_1+\lambda_2}, \mathcal{O}_3 \} =  \begin{tikzpicture}
	\begin{pgfonlayer}{nodelayer}
		\node [style=red dot] (5) at (-3, 0) {};
		\node [style=red dot] (6) at (-5, 0) {};
		\node [style=black dot] (8) at (-1, 0) {};
		\node [style=none] (15) at (-5, 0.5) {$1$};
		\node [style=none] (17) at (-3, 0.5) {$2$};
		\node [style=none] (19) at (-1, 0.5) {$3$};
	\end{pgfonlayer}
	\begin{pgfonlayer}{edgelayer}
		\draw [style=red edge] (6) to (5);
		\draw [style=black edge] (5) to (8);
	\end{pgfonlayer}
\end{tikzpicture}
+
\begin{tikzpicture}
	\begin{pgfonlayer}{nodelayer}
		\node [style=red dot] (5) at (-3, 0) {};
		\node [style=red dot] (6) at (-5, 0) {};
		\node [style=black dot] (8) at (-1, 0) {};
		\node [style=none] (15) at (-5, 0.5) {$2$};
		\node [style=none] (17) at (-3, 0.5) {$1$};
		\node [style=none] (19) at (-1, 0.5) {$3$};
	\end{pgfonlayer}
	\begin{pgfonlayer}{edgelayer}
		\draw [style=red edge] (6) to (5);
		\draw [style=black edge] (5) to (8);
	\end{pgfonlayer}
\end{tikzpicture}
\label{eq:JacobiTerm1}\,.
\end{equation}
{In the picture, the operator $\calO_i$ is associated with the $i$'th node and Wick contractions are denoted by edges connecting the nodes. We color the first contraction (associated to the inner bracket) in red. The inner bracket corresponds to a Laman subgraph $\Gamma[S]$, as expected, and evaluates to $\mathcal{I}_{\Gamma[S]}(\lambda_1,z_{12})$.} Likewise, the outer bracket is given by the graph $\Gamma(S)$, obtained by shrinking the red subgraph $\Gamma[S]$ into a single vertex. {Altogether, in the graphical notation introduced in \cite{Budzik:2022mpd}, the first term in \eqref{eq:JacobiTerm1} is the product of:
\begin{itemize}
    \item The remaining fields that survive one Wick contraction between $\mathcal{O}_1$ and $\mathcal{O}_2$, and a second Wick contraction between fields in $\mathcal{O}_2$ and $\mathcal{O}_3$
    \item And, the Feynman integral $\mathcal{I}_{\Gamma[S]}$ acting on $\mathcal{I}_{\Gamma(S)}$:
    \begin{equation}
        \begin{tikzpicture}
		[line join=round,baseline={([yshift=-.5ex] current bounding box.center)},vertex/.style={anchor=base,circle,fill=black!25,minimum size=18pt,inner sep=2pt}]
		\coordinate (pd1) at (1.*\gS,0.*\gS);
		\coordinate (pd2) at (-1.*\gS,0.*\gS);
		\draw (pd1) node[GraphNode,red] {} node[above] {$2$} node[below] {};
		\draw (pd2) node[GraphNode,red] {} node[above] {$1$} node[below] {$\lambda_1$};
		\draw[GraphEdge,red] (pd1) -- (pd2); 
	\end{tikzpicture}
 \begin{tikzpicture}
		[line join=round,baseline={([yshift=-.5ex]current bounding box.center)},vertex/.style={anchor=base,
    circle,fill=black!25,minimum size=18pt,inner sep=2pt}]
		\coordinate (pd1) at (1.*\gS,0.*\gS);
		\coordinate (pd2) at (-1.*\gS,0.*\gS);
            \draw[GraphEdge,thick] (pd1) -- (pd2) ;
		\draw (pd1) node[GraphNode] {} node[above] {$3$};
		\draw (pd2) node[GraphNode,color = red] {} node[below] {$\lambda_1+\lambda_2$};
\end{tikzpicture} 
    \end{equation}\,.
\end{itemize}
}

Similarly, we have
\begin{align}
    \{\mathcal{O}_2  \, {}_{\lambda_2} \{ \mathcal{O}_1\, {}_{\lambda_1} \,\mathcal{O}_3 \}\} = 
\begin{tikzpicture}
	\begin{pgfonlayer}{nodelayer}
		\node [style=red dot] (5) at (-3, 0) {};
		\node [style=red dot] (6) at (-5, 0) {};
		\node [style=black dot] (8) at (-1, 0) {};
		\node [style=none] (15) at (-5, 0.5) {$1$};
		\node [style=none] (17) at (-3, 0.5) {$3$};
		\node [style=none] (19) at (-1, 0.5) {$2$};
	\end{pgfonlayer}
	\begin{pgfonlayer}{edgelayer}
		\draw [style=red edge] (6) to (5);
		\draw [style=black edge] (5) to (8);
	\end{pgfonlayer}
\end{tikzpicture}
+
\begin{tikzpicture}
	\begin{pgfonlayer}{nodelayer}
		\node [style=red dot] (5) at (-3, 0) {};
		\node [style=red dot] (6) at (-5, 0) {};
		\node [style=black dot] (8) at (-1, 0) {};
		\node [style=none] (15) at (-5, 0.5) {$3$};
		\node [style=none] (17) at (-3, 0.5) {$1$};
		\node [style=none] (19) at (-1, 0.5) {$2$};
	\end{pgfonlayer}
	\begin{pgfonlayer}{edgelayer}
		\draw [style=red edge] (6) to (5);
		\draw [style=black edge] (5) to (8);
	\end{pgfonlayer}
\end{tikzpicture}\label{eq:JacobiTerm2}
\\
 \{\mathcal{O}_1  \, {}_{\lambda_1} \{ \mathcal{O}_2\, {}_{\lambda_2} \,\mathcal{O}_3 \}\}  = \begin{tikzpicture}
	\begin{pgfonlayer}{nodelayer}
		\node [style=red dot] (5) at (-3, 0) {};
		\node [style=red dot] (6) at (-5, 0) {};
		\node [style=black dot] (8) at (-1, 0) {};
		\node [style=none] (15) at (-5, 0.5) {$2$};
		\node [style=none] (17) at (-3, 0.5) {$3$};
		\node [style=none] (19) at (-1, 0.5) {$1$};
	\end{pgfonlayer}
	\begin{pgfonlayer}{edgelayer}
		\draw [style=red edge] (6) to (5);
		\draw [style=black edge] (5) to (8);
	\end{pgfonlayer}
\end{tikzpicture}
+
\begin{tikzpicture}
	\begin{pgfonlayer}{nodelayer}
		\node [style=red dot] (5) at (-3, 0) {};
		\node [style=red dot] (6) at (-5, 0) {};
		\node [style=black dot] (8) at (-1, 0) {};
		\node [style=none] (15) at (-5, 0.5) {$3$};
		\node [style=none] (17) at (-3, 0.5) {$2$};
		\node [style=none] (19) at (-1, 0.5) {$1$};
	\end{pgfonlayer}
	\begin{pgfonlayer}{edgelayer}
		\draw [style=red edge] (6) to (5);
		\draw [style=black edge] (5) to (8);
	\end{pgfonlayer}
\end{tikzpicture}
\label{eq:JacobiTerm3}
\end{align}
\begin{equation}
    \{\mathcal{O}_1  \, {}_{\lambda_1} \{ \mathcal{O}_2\, {}_{\lambda_2} \,\mathcal{O}_3 \}\} - (-1)^{(|\mathcal{O}_1|+1)(|\mathcal{O}_2|+1)}\{\mathcal{O}_2  \, {}_{\lambda_2} \{ \mathcal{O}_1\, {}_{\lambda_1} \,\mathcal{O}_3 \}\} +(-1)^{|\mathcal{O}_1|} \{\{ \mathcal{O}_1\, {}_{\lambda_1} \,\mathcal{O}_2 \} \, {}_{\lambda_1+\lambda_2} \mathcal{O}_3   \}=0\,.
\end{equation}
{This is the $\lambda$-Jacobi identity for the associated 2-ary $\lambda$-bracket to an $H=2$ theory.} To see this identity, notice that the sum of the first term in \eqref{eq:JacobiTerm1} and the second term in \eqref{eq:JacobiTerm3}, with appropriate signs, is zero since they precisely make up a ``bi-segment diagram.'' {As explained in Section 4.2 of \cite{Budzik:2022mpd}, the bi-segment integral vanishes identically; it is the only degree-1 almost 2-Laman graph.} Likewise, the second term of \eqref{eq:JacobiTerm1} cancels precisely with the second term of \eqref{eq:JacobiTerm2}, and the first term of \eqref{eq:JacobiTerm2} cancels precisely with the first term of \eqref{eq:JacobiTerm3}. {So we see very explicitly how the diagrammatic identities correspond to associativity relations for the brackets.}

The next associativity relation involve four arguments of the form
\begin{equation}
    \sum_{6\ \mathrm{terms}} \{\{-,- \},-,-\} + \sum_{4\ \mathrm{terms}} \{\{-,-,-\},- \} = 0\,. \label{eq:associativity4form}
\end{equation}
For example, one of the terms in the first sum looks like
\begin{equation}
    \{\{ \mathcal{O}_1\, {}_{\lambda_1} \,\mathcal{O}_2 \} \, {}_{\lambda_1+\lambda_2} \,\mathcal{O}_3  \, {}_{\lambda_3}\, \mathcal{O}_4
     \}  =  \begin{tikzpicture} [
	baseline={(current bounding box.center)},
	line join=round]
	\begin{pgfonlayer}{nodelayer}
		\node [style=red dot] (0) at (-1, 0) {};
		\node [style=red dot] (1) at (0, 0) {};
		\node [style=black dot] (2) at (1, 0.5) {};
		\node [style=black dot] (3) at (1, -0.5) {};
		\node [style=none] (4) at (-1, 0.5) {$1$};
		\node [style=none] (5) at (0, 0.5) {$2$};
		\node [style=none] (6) at (1, 1) {$3$};
		\node [style=none] (7) at (1, -1) {$4$};
	\end{pgfonlayer}
	\begin{pgfonlayer}{edgelayer}
		\draw [style=red edge] (0) to (1);
		\draw [style=black edge] (1) to (2);
		\draw [style=black edge] (2) to (3);
		\draw [style=black edge] (1) to (3);
	\end{pgfonlayer}
\end{tikzpicture}
+
\begin{tikzpicture} [
	baseline={(current bounding box.center)},
	line join=round]
	\begin{pgfonlayer}{nodelayer}
		\node [style=red dot] (0) at (-1, 0) {};
		\node [style=red dot] (1) at (0, 0) {};
		\node [style=black dot] (2) at (1, 0.5) {};
		\node [style=black dot] (3) at (1, -0.5) {};
		\node [style=none] (4) at (-1, 0.5) {$2$};
		\node [style=none] (5) at (0, 0.5) {$1$};
		\node [style=none] (6) at (1, 1) {$3$};
		\node [style=none] (7) at (1, -1) {$4$};
	\end{pgfonlayer}
	\begin{pgfonlayer}{edgelayer}
		\draw [style=red edge] (0) to (1);
		\draw [style=black edge] (1) to (2);
		\draw [style=black edge] (2) to (3);
		\draw [style=black edge] (1) to (3);
	\end{pgfonlayer}
\end{tikzpicture}
+\begin{tikzpicture} [
	baseline={(current bounding box.center)},
	line join=round]
	\begin{pgfonlayer}{nodelayer}
		\node [style=red dot] (0) at (0, 1) {};
		\node [style=red dot] (1) at (0, 0) {};
		\node [style=black dot] (2) at (1, 1) {};
		\node [style=black dot] (3) at (1, 0) {};
		\node [style=none] (4) at (0, 1.5) {$1$};
		\node [style=none] (5) at (0, -0.5) {$2$};
		\node [style=none] (6) at (1, 1.5) {$3$};
		\node [style=none] (7) at (1, -0.5) {$4$};
	\end{pgfonlayer}
	\begin{pgfonlayer}{edgelayer}
		\draw [style=red edge] (0) to (1);
		\draw [style=black edge] (1) to (3);
		\draw [style=black edge] (3) to (2);
		\draw [style=black edge] (2) to (0);
	\end{pgfonlayer}
\end{tikzpicture}+\begin{tikzpicture} [
	baseline={(current bounding box.center)},
	line join=round]
	\begin{pgfonlayer}{nodelayer}
		\node [style=red dot] (0) at (0, 1) {};
		\node [style=red dot] (1) at (0, 0) {};
		\node [style=black dot] (2) at (1, 1) {};
		\node [style=black dot] (3) at (1, 0) {};
		\node [style=none] (4) at (0, 1.5) {$2$};
		\node [style=none] (5) at (0, -0.5) {$1$};
		\node [style=none] (6) at (1, 1.5) {$3$};
		\node [style=none] (7) at (1, -0.5) {$4$};
	\end{pgfonlayer}
	\begin{pgfonlayer}{edgelayer}
		\draw [style=red edge] (0) to (1);
		\draw [style=black edge] (1) to (3);
		\draw [style=black edge] (3) to (2);
		\draw [style=black edge] (2) to (0);
	\end{pgfonlayer}
\end{tikzpicture}\,,
\end{equation}
and one of the terms in the second sum takes the form
\begin{align}
      \{\{ \mathcal{O}_1\, {}_{\lambda_1} \,\mathcal{O}_2  \, {}_{\lambda_2} \, \mathcal{O}_3 \} \, {}_{\lambda_1 + \lambda_2+\lambda_3}\, \mathcal{O}_4\}
     \} =& \begin{tikzpicture}[
	baseline={(current bounding box.center)},
	line join=round]
	\begin{pgfonlayer}{nodelayer}
		\node [style=red dot] (0) at (-1, 0.5) {};
		\node [style=red dot] (1) at (-1, -0.5) {};
		\node [style=red dot] (2) at (0, 0) {};
		\node [style=black dot] (3) at (1, 0) {};
		\node [style=none] (4) at (-1, 1) {$1$};
		\node [style=none] (5) at (-1, -1) {$2$};
		\node [style=none] (6) at (0, 0.5) {$3$};
		\node [style=none] (7) at (1, 0.5) {$4$};
	\end{pgfonlayer}
	\begin{pgfonlayer}{edgelayer}
		\draw [style=red edge] (0) to (2);
		\draw [style=red edge] (1) to (2);
		\draw [style=red edge] (0) to (1);
		\draw [style=black edge] (2) to (3);
	\end{pgfonlayer}
\end{tikzpicture}+\begin{tikzpicture}[
	baseline={(current bounding box.center)},
	line join=round]
	\begin{pgfonlayer}{nodelayer}
		\node [style=red dot] (0) at (-1, 0.5) {};
		\node [style=red dot] (1) at (-1, -0.5) {};
		\node [style=red dot] (2) at (0, 0) {};
		\node [style=black dot] (3) at (1, 0) {};
		\node [style=none] (4) at (-1, 1) {$3$};
		\node [style=none] (5) at (-1, -1) {$1$};
		\node [style=none] (6) at (0, 0.5) {$2$};
		\node [style=none] (7) at (1, 0.5) {$4$};
	\end{pgfonlayer}
	\begin{pgfonlayer}{edgelayer}
		\draw [style=red edge] (0) to (2);
		\draw [style=red edge] (1) to (2);
		\draw [style=red edge] (0) to (1);
		\draw [style=black edge] (2) to (3);
	\end{pgfonlayer}
\end{tikzpicture}
+\begin{tikzpicture}[
	baseline={(current bounding box.center)},
	line join=round]
	\begin{pgfonlayer}{nodelayer}
		\node [style=red dot] (0) at (-1, 0.5) {};
		\node [style=red dot] (1) at (-1, -0.5) {};
		\node [style=red dot] (2) at (0, 0) {};
		\node [style=black dot] (3) at (1, 0) {};
		\node [style=none] (4) at (-1, 1) {$3$};
		\node [style=none] (5) at (-1, -1) {$2$};
		\node [style=none] (6) at (0, 0.5) {$1$};
		\node [style=none] (7) at (1, 0.5) {$4$};
	\end{pgfonlayer}
	\begin{pgfonlayer}{edgelayer}
		\draw [style=red edge] (0) to (2);
		\draw [style=red edge] (1) to (2);
		\draw [style=red edge] (0) to (1);
		\draw [style=black edge] (2) to (3);
	\end{pgfonlayer}
\end{tikzpicture}\,.
\end{align}
Therefore \eqref{eq:associativity4form} involves a sum of $12$ graphs of the form \begin{tikzpicture} [
	line join=round,scale=0.3, baseline=-3pt]
	\begin{pgfonlayer}{nodelayer}
		\node [style=red dot] (0) at (-1, 0) {};
		\node [style=red dot] (1) at (0, 0) {};
		\node [style=black dot] (2) at (1, 0.5) {};
		\node [style=black dot] (3) at (1, -0.5) {};
	\end{pgfonlayer}
	\begin{pgfonlayer}{edgelayer}
		\draw [style=red edge] (0) to (1);
		\draw [style=black edge] (1) to (2);
		\draw [style=black edge] (2) to (3);
		\draw [style=black edge] (1) to (3);
	\end{pgfonlayer}
\end{tikzpicture} and $12$ graphs of the form \begin{tikzpicture}[
	baseline=-3pt,
	line join=round,scale = 0.3]
	\begin{pgfonlayer}{nodelayer}
		\node [style=red dot] (0) at (-1, 0.5) {};
		\node [style=red dot] (1) at (-1, -0.5) {};
		\node [style=red dot] (2) at (0, 0) {};
		\node [style=black dot] (3) at (1, 0) {};
	\end{pgfonlayer}
	\begin{pgfonlayer}{edgelayer}
		\draw [style=red edge] (0) to (2);
		\draw [style=red edge] (1) to (2);
		\draw [style=red edge] (0) to (1);
		\draw [style=black edge] (2) to (3);
	\end{pgfonlayer}
\end{tikzpicture}
which precisely cancels pairwise due to the quadratic identity of the form:
\begin{equation}
0\,=\,\,
    \begin{tikzpicture}[
	baseline={(current bounding box.center)},
	line join=round]
	\begin{pgfonlayer}{nodelayer}
		\node [style=red dot] (0) at (0, 0) {};
		\node [style=red dot] (1) at (-1, 0.5) {};
		\node [style=red dot] (2) at (-1, -0.5) {};
		\node [style=black dot] (4) at (2, 0) {};
		\node [style=red dot] (5) at (1, 0) {};
	\end{pgfonlayer}
	\begin{pgfonlayer}{edgelayer}
		\draw [style=red edge] (1) to (0);
		\draw [style=red edge] (0) to (2);
		\draw [style=red edge] (2) to (1);
		\draw [style=black edge] (5) to (4);
	\end{pgfonlayer}
\end{tikzpicture}
\quad-\quad
\begin{tikzpicture}[
	baseline={(current bounding box.center)},
	line join=round]
	\begin{pgfonlayer}{nodelayer}
		\node [style=red dot] (0) at (-1, 0) {};
		\node [style=red dot] (1) at (0, 0) {};
		\node [style=red dot] (4) at (1, 0) {};
		\node [style=black dot] (5) at (2, 0.5) {};
		\node [style=black dot] (6) at (2, -0.5) {};
	\end{pgfonlayer}
	\begin{pgfonlayer}{edgelayer}
		\draw [style=red edge] (0) to (1);
		\draw [style=black edge] (4) to (5);
		\draw [style=black edge] (5) to (6);
		\draw [style=black edge] (6) to (4);
	\end{pgfonlayer}
\end{tikzpicture}\,,
\end{equation}
obtained from the sliding graph:
\begin{tikzpicture}
[baseline=+2pt,
    	line join=round,scale = 0.3
    	]
        \def\gS{1.35};
    	\coordinate (pd1) at (1.1141*\gS,0.4234*\gS);
    	\coordinate (pd2) at (2.0313*\gS,0.*\gS);
    	\coordinate (pd3) at (2.0313*\gS,0.8471*\gS);
    	\coordinate (pd4) at (0.*\gS,0.4235*\gS);
    
    	\draw (pd1) node[GraphNode] {};
    	\draw (pd2) node[GraphNode] {} ;
    	\draw (pd3) node[GraphNode] {} ;
    	\draw (pd4) node[GraphNode] {} ;
    
    	\draw[GraphEdge] (pd1) -- (pd2) ;
    	\draw[GraphEdge] (pd1) -- (pd3);
    	\draw[GraphEdge] (pd1) -- (pd4) ;
    	\draw[GraphEdge] (pd2) -- (pd3) ;
    \end{tikzpicture}. 
The rest of $12$ graphs of the form \begin{tikzpicture} [
	baseline=+2pt,
	line join=round,scale =0.3]
	\begin{pgfonlayer}{nodelayer}
		\node [style=red dot] (0) at (0, 1) {};
		\node [style=red dot] (1) at (0, 0) {};
		\node [style=black dot] (2) at (1, 1) {};
		\node [style=black dot] (3) at (1, 0) {};
	\end{pgfonlayer}
	\begin{pgfonlayer}{edgelayer}
		\draw [style=red edge] (0) to (1);
		\draw [style=black edge] (1) to (3);
		\draw [style=black edge] (3) to (2);
		\draw [style=black edge] (2) to (0);
	\end{pgfonlayer}
\end{tikzpicture}
cancel as well, by using the quadratic identity of the form (see Section 4.4 of \cite{Budzik:2022mpd}):
\begin{equation}
    0 \,\,=\,\, \sum_{4\ \mathrm{terms}}\,\,
\begin{tikzpicture}
    [
	baseline={(current bounding box.center)},
	line join=round
	]
    \def\gS{0.9};
	\coordinate (pd1) at (1.*\gS,0.*\gS);
	\coordinate (pd2) at (-1.*\gS,0.*\gS);
	\draw (pd1) node[GraphNode,red] {} node[above] {};
	\draw (pd2) node[GraphNode,red] {} node[above] {};
	\draw[GraphEdge,red] (pd1) -- (pd2) node[midway, above] {};
\end{tikzpicture}
\quad
\begin{tikzpicture}
    [   
	baseline={(current bounding box.center)},
	line join=round
	]
	\def\gS{1.1};
	\coordinate (pd1) at (-0.866*\gS,-0.5*\gS);
	\coordinate (pd2) at (0.*\gS,1.*\gS);
	\coordinate (pd3) at (0.866*\gS,-0.5*\gS);
	\draw[GraphEdge] (pd1) -- (pd2) node[midway, left] {};
	\draw[GraphEdge] (pd1) -- (pd3) node[midway, above] {};
	\draw[GraphEdge] (pd2) -- (pd3) node[midway, right] {};
	\draw (pd1) node[GraphNode,red] {} node[left] {};
	\draw (pd2) node[GraphNode] {} node[left,above] {};
	\draw (pd3) node[GraphNode] {} node[right] {};
\end{tikzpicture}\label{eq:quadraticIdentitysquare}
\end{equation}
obtained from the degree-1 almost 2-Laman graph: 
\begin{tikzpicture}
        [
    	baseline=-3pt,
    	line join=round,scale =0.3
    	]
        \def\gS{0.9};
    	\coordinate (pd1) at (-1.*\gS,0.*\gS);
    	\coordinate (pd2) at (0.*\gS,1.*\gS);
    	\coordinate (pd3) at (0.*\gS,-1.*\gS);
    	\coordinate (pd4) at (1.*\gS,0.*\gS);
    
     	\draw (pd1) node[GraphNode] {};
    	\draw (pd2) node[GraphNode] {} ;
    	\draw (pd3) node[GraphNode] {} ;
    	\draw (pd4) node[GraphNode] {};
    
    	\draw[GraphEdge] (pd1) -- (pd2) ;
    	\draw[GraphEdge] (pd1) -- (pd3) ;
    	\draw[GraphEdge] (pd2) -- (pd4);
    	\draw[GraphEdge] (pd3) -- (pd4) ;
    \end{tikzpicture}.
Taking the $\lambda_i$ parameters and the fermion parity into account, we arrive at the following {higher-associativity relation} for \eqref{eq:associativity4form}:
\begin{align}
     0 = &\{\{ \mathcal{O}_1\, {}_{\lambda_1} \,\mathcal{O}_2 \} \, {}_{\lambda_1+\lambda_2} \,\mathcal{O}_3  \, {}_{\lambda_3}\, \mathcal{O}_4
     \} 
        + (-1)^{|\cO_2||\cO_3|}\{\{ \mathcal{O}_1\, {}_{\lambda_1} \,\mathcal{O}_3 \} \, {}_{\lambda_1+\lambda_3} \,\mathcal{O}_2  \, {}_{\lambda_2}\, \mathcal{O}_4
     \} \nonumber
     \\
     + &(-1)^{|\cO_1|+|\cO_2|}
     \{ \mathcal{O}_1\, {}_{\lambda_1} \,\mathcal{O}_2  \, {}_{\lambda_2} \,\{ \mathcal{O}_3  \, {}_{\lambda_3}\, \mathcal{O}_4\}
     \} 
     + (-1)^{|\cO_1|+(|\cO_2|+1)|\cO_3|}
     \{ \mathcal{O}_1\, {}_{\lambda_1} \,\mathcal{O}_3  \, {}_{\lambda_3} \,\{ \mathcal{O}_2  \, {}_{\lambda_2}\, \mathcal{O}_4\}
     \} \nonumber
     \\
     + &(-1)^{|\cO_1|}
     \{ \mathcal{O}_1\, {}_{\lambda_1} \, \{\mathcal{O}_2  \, {}_{\lambda_2} \, \mathcal{O}_3 \} \, {}_{\lambda_2+\lambda_3}\, \mathcal{O}_4
     \} \nonumber
     \\
     + &\left(\frac{1+(-1)^{|\cO_1|}}{2}(-1)^{|\cO_2|+|\cO_3|}+\frac{1-(-1)^{|\cO_1|}}{2}\right)
     \{ \mathcal{O}_2\, {}_{\lambda_2} \,\mathcal{O}_3  \, {}_{\lambda_3} \,\{ \mathcal{O}_1  \, {}_{\lambda_1}\, \mathcal{O}_4\}
     \}
     \\ 
     + &\{\{ \mathcal{O}_1\, {}_{\lambda_1} \,\mathcal{O}_2  \, {}_{\lambda_2} \, \mathcal{O}_3 \} \, {}_{\lambda_1 + \lambda_2+\lambda_3}\, \mathcal{O}_4
     \}
     +(-1)^{|\cO_1|}\{ \mathcal{O}_1\, {}_{\lambda_1}\,\{ \mathcal{O}_2  \, {}_{\lambda_2} \, \mathcal{O}_3  \, {}_{\lambda_3}\, \mathcal{O}_4\}
     \} \nonumber
     \\
     +&(-1)^{(|\cO_1|+1)|\cO_2|} \{ \mathcal{O}_2\, {}_{\lambda_2}\,\{ \mathcal{O}_1  \, {}_{\lambda_1} \, \mathcal{O}_3  \, {}_{\lambda_3}\, \mathcal{O}_4\}
     \}
     +(-1)^{(|\cO_1|+|\cO_2|+1)|\cO_3|} \{ \mathcal{O}_3\, {}_{\lambda_3}\,\{ \mathcal{O}_1  \, {}_{\lambda_1} \, \mathcal{O}_2 \, {}_{\lambda_2}\, \mathcal{O}_4\}
     \} \nonumber \,.
\end{align}

More generally for the associativity relation of $n$ arguments, each summand is a bracket of $k$ arguments, followed by a bracket of $n+1-k$ arguments with $k = 2, \dots, n-1$ (see \eqref{eq:quadraticIdentity}). The evaluation of the inner bracket results in an integral $\mathcal{I}_{\Gamma[S]}$  for a Laman graph $\Gamma[S]$ whose nodes are given by the set $S$ of $k$ arguments and edges are given by Wick contractions. The evaluation of the outer bracket results in an integral $\mathcal{I}_{\Gamma(S)}$ for another Laman graph $\Gamma(S)$, which has $n+1-k$ nodes including the remaining $n-k$ arguments and the output of the inner bracket. The edges are again given by Wick contractions. The key observation is that $\Gamma[S]$ and $\Gamma(S)$ can be thought of as the two induced graphs by shrinking a Laman subgraph $\Gamma[S]$ of a sliding graph of $n$ nodes. Different ways of combining $\Gamma[S]$ and $\Gamma(S)$ lead to different summands in the associativity relation, which follow from the quadratic identities derived from all possible sliding graphs with $n$ nodes. 

\section{Operads and QFT}\label{sec:operadsAndQFT}
In the following we will give some semi-rigorous and fairly general mathematical definitions on the theory of ``operads.'' While our definitions are moderately abstract, treatments of the subject are simultaneously too broad and too deep for us to recap entirely. Instead, we will briefly define the relevant algebraic gadgets in a way accessible to high energy physicists, outline the intuitions that the abstract definitions capture, and then describe how they naturally appear in quantum field theory. Our formal definitions for operads follow \cite{vallette2012algebra+, claudiaNotes}, while our formal definitions for factorization algebras follow \cite{CG1, CG2, Costello:2023knl} (see also \cite{Amabel:2023hzs}). This technology in terms of operads is not needed to understand the bulk of the paper, but underlies the discussion as we will see.

\subsection{Basic Definitions}

We start with the formal definition of a (planar) operad in a monoidal category $\calC$. It is common for authors to just pick a particular monoidal category $\calC$ and define operads for that specific category. Common choices for the category $\calC$ include: the category of sets $(\mathrm{Set}, \times)$, vector spaces $(\mathrm{Vec},\otimes)$, or topological spaces $(\mathrm{Top}, \times)$.

\begin{definition}[Operad]
    Fix a monoidal category $\calC$.\footnote{Note: we see the necessity for a monoidal category $\calC$ so that we can make sense of tensor products of objects in $\calC$ and $\circ$ as a morphism in $\calC$ for the composition.} An \textit{operad $\calO$ in $\calC$} is a collection of objects $\{\calO(n)\}_{n\in \mathbb{N}^0}$ in $\calC$. An element $m \in \calO(n)$ is called an \textit{$n$-ary operation}.\footnote{Note: $\calO(n)$ may not have any ``elements'' at all, since $\calO(n)$ is just an object in $\calC$. However, objects in $\calC = \mathrm{Vec}, \mathrm{Set}, \mathrm{Top}$, etc. all have ``elements inside of them'' and are more concrete, so we will present the axioms with such cases in mind. For general $\calC$ we can proceed with all the operad axioms using abstract $\mathrm{Hom}$s; the unit axiom becomes the existence of a Hom from $\mathbf{1}_\calC \to \calO(1)$, etc, but we will refrain from such abstraction.\label{footnote:OperadWarning}} The $n$-ary operations are subject to the following conditions:
\begin{itemize}
    \itemsep0em 
    \item \textbf{Composition}. Given elements $m \in \calO(n)$, $m_1 \in \calO(k_1)$, $\dots$, $m_n \in \calO(k_n)$, we have a composition operation:
    \begin{equation}
        \circ: \calO(n) \otimes \calO(k_1) \otimes \cdots \otimes \calO(k_n) \to \calO(k_1 + \dots + k_n)\,,
    \end{equation}
    i.e. we can form the element
    \begin{equation}
        m \circ (m_1, \dots, m_n) \in \calO(k_1 + \dots + k_n)\,.
    \end{equation}
    \item \textbf{Unit}. There is a unary operation $I \in \calO(1)$, called the unit, such that for all $m$:
    \begin{equation}
        m \circ (I, \dots, I) = m = I \circ m\,.
    \end{equation}
    \item \textbf{Associativity}. An associativity condition on the composition function:\footnote{We note that the associativity condition \textit{does not} mean that the $n$-ary operations are associative, but only that the composition function is.}
    \begin{align}
        m \circ [m_1 \circ (&m_{11}, \dots, m_{1k_1}), \dots, m_n \circ (m_{n1}, \dots, m_{nk_n})] \nonumber\\
            &= [m\circ (m_1,\dots, m_n)] \circ(m_{11},\dots, m_{1k_1},\dots, m_{n1},\dots, m_{nk_n})\,.
    \end{align}
\end{itemize}
\end{definition} 

The idea of an operad is to capture properties of abstract families of composable functions and coherence relations between them. Such families are not exotic. For example, the multiplication law of any algebra $A$ is a $2$-ary operation $m: A\times A \to A$ by $m(a,b) = ab$. But the existence of a single $2$-ary operation implies the existence of arbitrarily many higher operations, e.g. the $3$-ary operations:
\begin{align}
    [m \circ (m,I)](a,b,c) 
        &= (ab)c\quad\text{and}\quad
    [m \circ (I,m)](a,b,c) 
        = a(bc)\,.
\end{align}
So the algebra $A$ comes equipped with many $n$-ary operations generated from the $2$-ary operation $m$. Note again that $\circ$ is associative, but at no point do we assume that the $2$-ary operation $m$ in the algebra is associative. Associativity of the algebra would be encoded in the $3$-ary operations by enforcing:
\begin{equation}
    m \circ (m,I) \stackrel{!}{=} m \circ (I,m)\,, \quad \text{for an associative algebra $A$.}
\end{equation}

The utility of an operad is that rhyming properties of distinct algebraic gadgets may be described by one operad. We can make this example and intuition more precise by concretizing operads through algebras over operads. To do so, we must first introduce the concept of operad morphisms.

\begin{definition}[Operad Morphism]
    An \textit{operad morphism} $f: \calO \to \mathcal{P}$ in a category $\calC$ is a collection of maps $\{f_n: \calO(n) \to \mathcal{P}(n)\}_{n\in\mathbb{N}^0}$ between two operads in $\calC$ that are compatible with the composition and unit axioms above. Explicitly, we have $f_1(I_{\calO}) = I_{\mathcal{P}}$ and 
    \begin{equation}
        f_{k_1+\dots+k_n}(m \circ (m_1, \dots, m_n)) = f_n(m) \circ (f_{k_1}(m_1),\dots, f_{k_n}(m_n))\,.
    \end{equation}
\end{definition}

Let $\Gamma$ be an object in a monoidal category $\calD$ with unit $\mathbf{1}_{\calD} \in \calD$. We can build the ``endomorphism operad'' $\mathrm{End}_\Gamma$ by taking the $n$-ary operations to be $\mathrm{Hom}_{\calD}(\Gamma^{\otimes n}, \Gamma)$, and the operad identity to be $\mathrm{id}_{\Gamma} \in \mathrm{Hom}_{\calD}(\Gamma, \Gamma)$. In the category of Sets, this is essentially an $n$-ary operation for each map $\Gamma^n \to \Gamma$, and in the category $\mathrm{Vec}$ this is an $n$-ary operation for each linear map between the vector spaces $\Gamma^{\otimes n} \to \Gamma$. The composition operation is just given by the obvious block-decomposition of the endomorphisms.

\begin{definition}[Algebra over an Operad]
    Fix an operad $\calO$ in $\calC$ and an object $\Gamma$ in a monoidal category $\calD$. An \textit{algebra over $\calO$} or \textit{$\calO$-algebra on a $\calD$-object $\Gamma$} is an operad morphism $\calO \to \mathrm{End}_\Gamma$. The category of all $\calO$-algebras is $\mathrm{Alg}_{\calO}(\calD)$.
\end{definition}

The same way that we can concretize the abstract presentation of an algebra by encoding it in a collection of matrices, an $\calO$-algebra concretizes the operad $\calO$ by encoding each abstract $n$-ary operation $m \in \calO(n)$ into a morphism in $\mathrm{Hom}_{\calD}(\Gamma^{\otimes n}, \Gamma)$. In the case that the monoidal category $\calD = \mathrm{Vec}$, it is quite literally encoding the $n$-ary operations into matrices. We can also think of this roughly as giving $\mathrm{End}_{\Gamma}$ a left-$\calO$ module structure (although this is not strictly correct)
\begin{equation}
    \calO(n) \otimes \Gamma^n \to \Gamma\,.
\end{equation}

Note that $\calD$ has to be a $\calC$-module category for the notion of $\calC$ acting on $\calD$ to make sense. Moreover, $\calD$ must be enriched in $\calC$. This is because a priori $\mathrm{Hom}(\Gamma^{\otimes n},\Gamma)$ is just a set, not an object in $\calC$. Saying $\calD$ is enriched in $\calC$ means that for any $d_1,d_2 \in \calD$ there exists a $\underline{\mathrm{Hom}}(d_1,\d_2)\in\calC$ so that ``$\mathrm{Hom}$'s in $\calD$ look like objects in $\calC$'' in a suitable way. An example is $\calC = (\mathrm{Vec},\otimes)$ and $\calD = (\mathrm{Vec},\otimes)$. In this case, for any $V_1,V_2\in \calD$, $\mathrm{Hom}(V_1,V_2)$ is canonically a vector space, not just a set.

As a concrete example to complete the previous intuition, we can define the \textit{associative operad} $\mathrm{As}$ as the operad in $(\mathrm{Vec}, \otimes)$ with $\mathrm{As}(n) = \bbC$ for all $n \in \bbN^0$ and composition given by multiplication of scalars $\bbC \otimes \cdots \otimes \bbC \to \bbC$. An $\mathrm{As}$-algebra on a vector space $V$ equips $V$ with the structure of a unital associative algebra. In other words, $\mathrm{As}$-algebras are associative algebras.

It is helpful to encode the $n$-ary operations of any operad in tree diagrams. For example, if we have a $2$-ary operation $m_2 \in \calO(2)$ and $3$-ary operation $m_3 \in \calO(3)$, we can denote:
\begin{equation}
    m_2 \circ (1, m_3) = 
    \begin{tikzcd}
	  {} & {} & {} & {} \\
	1 && {m_3} \\
	& {m_2} \\
	& {}
	\arrow[no head, from=2-3, to=1-2]
	\arrow[no head, from=2-3, to=1-4]
	\arrow[no head, from=2-3, to=1-3]
	\arrow[no head, from=2-1, to=1-1]
	\arrow[no head, from=2-1, to=3-2]
	\arrow[no head, from=2-3, to=3-2]
	\arrow[no head, from=3-2, to=4-2]
    \end{tikzcd} \in \calO(4)\,.
\end{equation}
The associativity axiom for the operad essentially declares that we do not need to insert brackets into the tree diagrams that capture function composition. i.e. we do not have to be concerned with the order of building tall trees out of subtrees.

The operads defined above are called ``planar,'' and encode just associativity. However, a more common species of operad is a ``symmetric operad,'' which naturally has a commutative composition rule and is valued in a \textit{symmetric} monoidal category. We will mainly be focused on these symmetric operads from here out.
\begin{definition}[Symmetric Operad]
    We call an operad in a symmetric monoidal category \textit{symmetric} if there is a compatible right $S_n$ action on the objects $\calO(n)$. Moreover, the compositions must be compatible with the symmetric action.
\end{definition}

The data of a operad can be presented by just giving rules for \textit{partial compositions}:
\begin{equation}
    \circ_i: \calO(m) \otimes \calO(n) \to \calO(m+n-1)\,,
\end{equation}
for all $1\leq i \leq m$, subject to associativity and unit coherence relations (see Section 2.4 of \cite{vallette2012algebra+}). Practically, $\circ_i$ should be thought of as inserting an $n$-ary operation into the $i$'th argument of the $m$-ary operation. In the language of trees, this defines the operad composition by its rule for attaching a sub-tree to a terminal leg $i$ (or $i$'th ``leaf'') of a tree.

We saw the definition for an algebra over an operad above, in particular we saw that it looked like a (special kind of) left-module for the operad. We can generalize this idea \cite{Markl:1994by}:
\begin{definition}[Module over an Operad]
    Fix a (symmetric) operad $\calO$ in a (symmetric) monoidal category $\calC$ and a collection of objects $M := \{M(n)\}_{n\in\mathbb{N}^0}$ in $\calD$. Then $M$ is a right (resp. left) \textit{module} over $\calO$ if $\calD$ is a right (left) $\calC$ module, and there are partial composition maps:
    \begin{equation}
        \circ_i^R: M(m) \otimes \calO(n) \to M(m+n-1)\,,
    \end{equation}
    coherent with the desired operad axioms (and similarly for left-modules with $\circ_i^L$). We should demand a compatible $S_n$ action if we are studying a symmetric operad.
\end{definition}

Finally, we would like to enrich the definition of an operad to allow for a larger collection of objects (indexed by a ``colour set'' $\mathrm{Col}(\calO)$) to correspond to $n$-ary operations.
\begin{definition}[Coloured Operad]
    Fix a symmetric monoidal category $\calC$ and a set $\mathrm{Col}(\calO)$ of ``colours.'' A \textit{coloured operad in $\calC$} a collection of objects $\{\calO(c_1,\dots, c_n;c)\}_{n\in \mathbb{N}^0}$ in $\calC$, where each $c, c_i \in \mathrm{Col}(\calO)$. An element of $\calO(c_1,\dots, c_n;c)$ is again an $n$-ary operation subject to the following conditions:
    \begin{itemize}
    \itemsep0em 
    \item \textbf{Composition}. There is an associative morphism $\circ$ mapping:
    \begin{equation}
        \calO(c_1, \dots, c_k; c) \otimes \calO(c_{11}, \dots, c_{1\ell_1}; c_1) \otimes \cdots \otimes \calO(c_{k1}, \dots, c_{k\ell_k}; c_k) \to \calO(c_{11}, \dots, c_{k\ell_k}; c)
    \end{equation}
    \item \textbf{Unit}. There is a unary operation $I_c \in \calO(c;c)$, called the identity on $c$\footnote{We re-iterate our warning from Footnote \ref{footnote:OperadWarning}.}
    \begin{equation}
        m \circ (I_c, \dots, I_c) = m = I_c \circ m
    \end{equation}
\end{itemize}
    As well as associativity relations like before and $S_n$ equivariance if desired.
\end{definition}
\noindent Morphisms and algebras follow like before.

\subsection{Example: Topological Operads}
Here we will introduce some topological operads: operads valued in $\calC = (\mathrm{Top},\times)$. The prototypical example of a topological operad is the little $d$-disks operad $E_d$.\footnote{This was the example that led to the definition of operads \cite{boardman2006homotopy, may2006geometry}. In \cite{may2006geometry}, the $E_d$ operad was introduced to study $E_d$ algebras. In particular, May's Recognition Principle roughly says that: if a connected topological space $Y$ is an $E_d$-algebra, then it is homotopy equivalent to the $d$-fold loop space $\Omega^d(X)$ for some pointed topological space $X$. See Appendix B of \cite{Pace:2023ccj} for an overview of this construction.}

\begin{figure}[t]
\begin{equation*}
\begin{tikzpicture}[thick, baseline={(current bounding box.center)}]
    \def\gS{2};
    \coordinate (C) at (0,0);
    \coordinate (c1) at (-0.5*\gS,-0.0*\gS);
    \coordinate (c2) at (0.4*\gS,0.5*\gS);
    \coordinate (c3) at (0.4*\gS,-0.4*\gS);
    
    \draw (C) circle (1*\gS);
    \draw (c1) circle (0.4*\gS);
    \draw (c2) circle (0.2*\gS);
    \draw (c3) circle (0.3*\gS);
    
    \draw (c1) node {$1$};
    \draw (c2) node {$2$};
    \draw (c3) node {$3$};
\end{tikzpicture}
\quad
\circ_1
\quad
\begin{tikzpicture}[thick, baseline={(current bounding box.center)}]
    \def\gS{2};
    \coordinate (C) at (0,0);
    \coordinate (c1) at (+0.2*\gS,+0.3*\gS);
    \coordinate (c2) at (-0.3*\gS,-0.5*\gS);
    
    \draw (C) circle (1*\gS);
    \draw (c1) circle (0.5*\gS);
    \draw (c2) circle (0.3*\gS);
    
    \draw (c1) node {$1$};
    \draw (c2) node {$2$};
\end{tikzpicture}
\quad
=
\quad
\begin{tikzpicture}[thick, baseline={(current bounding box.center)}]
    \def\gS{2};
    \coordinate (C) at (0,0);
    \coordinate (c1) at (-0.5*\gS,-0.0*\gS);
    \coordinate (c2) at (0.4*\gS,0.5*\gS);
    \coordinate (c3) at (0.4*\gS,-0.4*\gS);
    
    \draw (C) circle (1*\gS);
    \draw[dotted] (c1) circle (0.4*\gS);
    \draw (c2) circle (0.2*\gS);
    \draw (c3) circle (0.3*\gS);
    
    \draw (c2) node {$3$};
    \draw (c3) node {$4$};

    \coordinate (m1) at (+0.2*\gS*0.4-0.5*\gS,+0.3*\gS*0.4);
    \coordinate (m2) at (-0.3*\gS*0.4-0.5*\gS,-0.5*\gS*0.4);
    
    \draw (m1) circle (0.5*\gS*0.4);
    \draw (m2) circle (0.3*\gS*0.4);
    
    \draw (m1) node {$1$};
    \draw (m2) node {$2$};
    
\end{tikzpicture}
\end{equation*}
\caption{Partial composition $\circ_i$ in the little disks operad is given by shrinking the second disk down and inserting it into position $i$ of the first disk.}
\label{fig:littleDisks}
\end{figure}
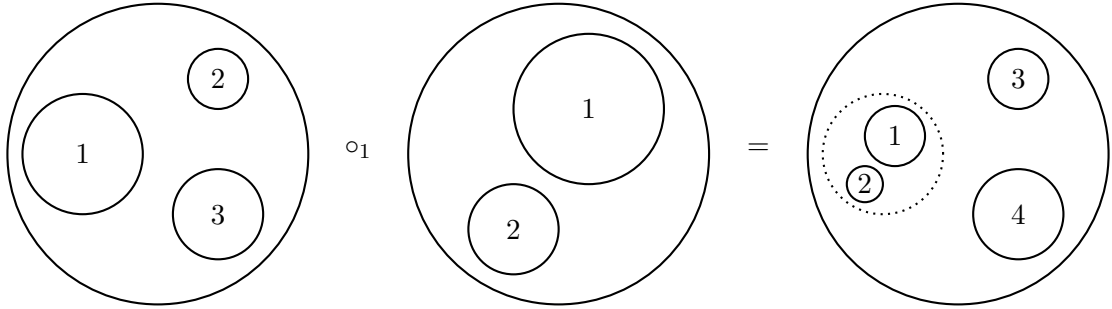

\begin{definition}[Little $d$-Disks Operad, $E_d$ Operad] 
    Let $D_d(n)$ be the topological space of $n$ disjoint subdisks in the $d$-dimensional unit disk $D^d \subset \bbR^d$.\footnote{This is a topological space because the $n$ disks can be thought of as the images of $n$ continuous maps $f_i: S^{d-1} \to D^d$.} Each of these little-disks is labelled by a number $i \in \{1,\dots, n\}$. The \textit{little $d$-disks operad} or \textit{$E_d$ operad} takes points in $D_d(n)$ as its $n$-ary operations, with $S_n$ action on labels. The (partial) composition of $m_1 \in D_d(n_1)$ with $m_2 \in D_d(n_2)$ is
    \begin{equation}
        m_{1} \circ_i m_{2} \in D_d(n_1 + n_2 - 1)\,,
    \end{equation}
    defined by shrinking $m_2$ to the size of little-disk $i$ in $m_1$, and inserting the tiny $m_2$ into $m_1$. See Figure \ref{fig:littleDisks}.
\end{definition}

This gives a distinct $n$-ary operation for each configuration of little disks in the unit disk. There are also equivalent or homotopy equivalent operads (in the sense that their spaces of $n$-ary operations are homotopic to $D_d(n)$) given by little cubes, little rectangles, and so on, all which go by the name $E_d$ operad in the literature. There is also the option to add a framing, which we will neglect for simplicity.

Since the spaces of $n$-ary operations in topological operads are topological spaces, we can apply machinery from algebraic topology to study these operads and their algebras. As a first example, consider the $E_1$ operad of disjoint little disks (segments) in the interval $[0,1]$. A priori, there is a distinct $n$-ary operation for each collection of $n$ disjoint intervals in $[0,1]$, for example there are distinct operations:\footnote{Here the colouring is just to help identify the intervals, this is not a ``coloured operad'' in any relevant nontrivial way.}
\begin{align}\label{eq:littleInterval1}
m_3(a,b,c) &= 
\begin{tikzpicture}[thick, baseline={(current bounding box.center)}]
    \def\gS{6};
    \def\tickS{0.1};
    \filldraw[red!20, opacity=0.5] (0.2*\gS,-\tickS) rectangle (0.1+0.4*\gS,+\tickS);
    \filldraw[red!20, opacity=0.5] (0.5*\gS,-\tickS) rectangle (0.1+0.7*\gS,+\tickS);
    \filldraw[red!20, opacity=0.5] (0.8*\gS,-\tickS) rectangle (0.1+0.9*\gS,+\tickS);
    \draw (-0.0*\gS,0) -- (1.0*\gS,0);
    \draw (0*\gS,-0.2) node[below] {$0$};
    \draw[arrows = {-Bracket[line cap=butt]}] (0.1*\gS,0) -- (0,0);
    \draw (1*\gS,-0.2) node[below] {$1$};
    \draw[arrows = {-Bracket[line cap=butt]}] (0.9*\gS,0) -- (1.0*\gS,0);
    \draw (0.3*\gS,+0.2) node[above] {$a$};
    \draw[arrows = {-Bracket[line cap=butt]}] (0.1*\gS+0.2*\gS,0) -- (0+0.2*\gS,0);
    \draw[arrows = {-Bracket[line cap=butt]}] (0.0*\gS+0.4*\gS,0) -- (0.1+0.4*\gS,0);
    \draw (0.6*\gS,+0.2) node[above] {$b$};
    \draw[arrows = {-Bracket[line cap=butt]}] (0.1*\gS+0.5*\gS,0) -- (0+0.5*\gS,0);
    \draw[arrows = {-Bracket[line cap=butt]}] (0.0*\gS+0.7*\gS,0) -- (0.1+0.7*\gS,0);
    \draw (0.85*\gS,+0.2) node[above] {$c$};
    \draw[arrows = {-Bracket[line cap=butt]}] (0.1*\gS+0.8*\gS,0) -- (0+0.8*\gS,0);
    \draw[arrows = {-Bracket[line cap=butt]}] (0.0*\gS+0.9*\gS,0) -- (0.1+0.9*\gS,0);
\end{tikzpicture}\,,\\
\label{eq:littleInterval2}
m_3'(a,b,c) &= 
\begin{tikzpicture}[thick, baseline={(current bounding box.center)}]
    \def\gS{6};
    \def\tickS{0.1};
    \filldraw[red!20, opacity=0.5] (0.15*\gS,-\tickS) rectangle (0.1+0.25*\gS,+\tickS);
    \filldraw[red!20, opacity=0.5] (0.5*\gS,-\tickS) rectangle (0.1+0.7*\gS,+\tickS);
    \filldraw[red!20, opacity=0.5] (0.8*\gS,-\tickS) rectangle (0.1+0.9*\gS,+\tickS);
    \draw (-0.0*\gS,0) -- (1.0*\gS,0);
    \draw (0*\gS,-0.2) node[below] {$0$};
    \draw[arrows = {-Bracket[line cap=butt]}] (0.1*\gS,0) -- (0,0);
    \draw (1*\gS,-0.2) node[below] {$1$};
    \draw[arrows = {-Bracket[line cap=butt]}] (0.9*\gS,0) -- (1.0*\gS,0);
    \draw (0.2*\gS,+0.2) node[above] {$a$};
    \draw[arrows = {-Bracket[line cap=butt]}] (0.1*\gS+0.15*\gS,0) -- (0+0.15*\gS,0);
    \draw[arrows = {-Bracket[line cap=butt]}] (0.0*\gS+0.25*\gS,0) -- (0.1+0.25*\gS,0);
    \draw (0.6*\gS,+0.2) node[above] {$b$};
    \draw[arrows = {-Bracket[line cap=butt]}] (0.1*\gS+0.5*\gS,0) -- (0+0.5*\gS,0);
    \draw[arrows = {-Bracket[line cap=butt]}] (0.0*\gS+0.7*\gS,0) -- (0.1+0.7*\gS,0);
    \draw (0.85*\gS,+0.2) node[above] {$c$};
    \draw[arrows = {-Bracket[line cap=butt]}] (0.1*\gS+0.8*\gS,0) -- (0+0.8*\gS,0);
    \draw[arrows = {-Bracket[line cap=butt]}] (0.0*\gS+0.9*\gS,0) -- (0.1+0.9*\gS,0);
\end{tikzpicture}\,.
\end{align}
However, there is an obvious sense in which the operation $m_3$ is homotopic to $m_3'$: the corresponding configurations are homotopic in the space of $3$-ary operations in $E_1$. If we work homotopically, all $n$-ary operations can be generated from the homotopically unique binary operation $m_2$ by composing, translating, and scaling intervals. 

More generally, the $E_d$ operad encodes continuous families of $n$-ary operations whose composition rules look like composition of little disks. The $E_d$ operad (or any topological operad) gives a natural operad $C_*(E_d)$ in chain complexes, or ``dg-operad,'' by applying the singular complex functor to each $D_d(n)$ (see e.g. \cite{Markl:1994by, kontsevich1999operads}). Similarly, passing from complexes to homology produces the ``homology operad'' $H_*(E_d)$ in the category of graded abelian groups.

For the $E_d$ operad, each point in $D_d(n)$ corresponds to a collection of disjoint little disks. Taking just the centers of these disks, we get a continuous map to $\mathrm{Conf}_n(\bbR^d)$ with contractible fiber. Hence there is a homotopy equivalence
\begin{equation}
    D_d(n) \simeq \mathrm{Conf}_n(\bbR^d)\,.
\end{equation}
And so, we can identify the $n$-ary chain complex
\begin{equation}
    C_*(E_d)(n) = C_*(\mathrm{Conf}_n(\bbR^d))\,.
\end{equation}

Why would one be motivated to study the dg-operad $C_*(E_d)$ or, more specifically, the homology operad $H_*(E_d)$? Given an $\calO$-algebra structure on $\Gamma$, i.e. a map $\calO \to \mathrm{End}_\Gamma$, there is an induced map $H_*(\calO) \to H_*(\mathrm{End}_\Gamma)$ which inturn induces a map to $\mathrm{End}_{H_*(\Gamma)}$ (if there is an appropriate K\"unneth-like map for $\Gamma \in \calD$) \cite{sinha2006homology}. In other words: the dg-operad (or homology operad) of $\calO$ captures essential dg (or homological) data about $\calO$-algebras. As we will see in the next section, quantum field theories are a natural source of topological operads acting on chain complexes, and often we are only interested in $n$-ary operations up to some notion of homotopy, as described in the preceding examples. Further mathematical motivation would require the introduction of $\infty$-categories, which we would like to avoid.

As an example, the $2$-ary chain complex in $C_*(E_d)$ is given by $C_*(\mathrm{Conf}_2(\bbR^d)) = C_*(S^{d-1})$, and the $2$-ary operations of the homology operad are indexed by
\begin{equation}
    H_*(\mathrm{Conf}_2(\bbR^d))
    =
    H_*(S^{d-1})
    =
    \begin{cases}
            \,\bbZ & \text{in degree $0$ and $d-1$}\\\
            0 & \text{otherwise}
    \end{cases}\,.
\end{equation}
In the $E_1$ case, this is just $\mathrm{Conf}_2(\bbR) = S^{0}$. The two disjoint points of $S^0$ correspond to the generically non-commutative 2-ary operation of an $E_1$ algebra in homotopy, reflecting our inability to phase the little intervals in \eqref{eq:littleInterval1} and \eqref{eq:littleInterval2} through each other. But note that such $E_1$ algebras are associative in homotopy. On the other hand, for $E_{d>1}$, we have a less associative but more commutative structure, e.g. an $E_2$ algebra structure in $\mathrm{Cat}$ is equivalent to a braided monoidal structure (see \cite{LurieHigherAlgebra} Example 5.1.2.4).

\begin{definition}[Disjoint Sets Operad, $\mathrm{Disj}_M$]
    Let $M$ be a topological space. Let $\mathrm{Disj}_M$ be the coloured (symmetric) operad with colours given by all open sets in $M$. For every finite collection of $n \in \mathbb{N}^0$ open sets $\{U_1, \dots, U_n\}$ in $M$, and an open set $V$, we define:
    \begin{equation}
        \mathrm{Disj}_M(U_1,\dots, U_n; V) = \begin{cases}
            * & \text{if $U_i$ pairwise disjoint and contained in $V$}\\
            \emptyset & \text{otherwise}
        \end{cases}\,.
    \end{equation}
    Here $*$ means a point.
\end{definition}
The operad of disjoint sets is dramatically larger than the little k-disks operad, allowing for a distinct $n$-ary operation for each collection of open sets in $M$. As we will see next, a general quantum field theory is essentially related to $\mathrm{Disj}_M$, but simplifications can occur that make it ``as simple'' as the $E_d$ operad.

\subsection{Factorization Algebras and the OPE}\label{app:FactAlg}
Factorization algebras axiomatically encode the observables, operator product expansion, and correlation functions of perturbative QFT, see \cite{CG1, CG2, Costello:2023knl} for rigorous definitions and reviews. Here let us record the very minimal definitions to connect to the previous sections on operads. We will assume we are working in Euclidean spacetime.

\begin{definition}[Prefactorization Algebra]
    Let $M$ be a topological space. A \textit{prefactorization algebra} $\mathrm{Obs}$ on $M$ assigns a topological vector space\footnote{Actually, $\mathrm{Obs}$ can be taken in any symmetric monoidal category $\calD$. Typically we work in a \textit{derived} setting where $\calD$ is the category of complexes; we will return to this shortly.\label{footnote:derived}} $\mathrm{Obs}(U)$ to each open $U \subset M$, subject to the following conditions:
    \begin{itemize}
    \itemsep0em
        \item For each inclusion $U \subset V$, there is a linear map $m^U_V:\mathrm{Obs}(U) \to \mathrm{Obs}(V)$
        \item For every finite collection of pairwise disjoint open sets $\{U_i\}_{i=1}^n$ contained in $V$, there is a linear map $m_V^{U_1,\dots, U_n}: \mathrm{Obs}(U_1) \otimes \cdots \otimes \mathrm{Obs}(U_n) \to \mathrm{Obs}(V)$ 
    \end{itemize}
    with compatibility morphisms on the overlapping sets.
\end{definition}
Physically, $\mathrm{Obs}(U)$ is the topological vector space of all observables supported on a region $U$. The map $m_V^U$ says that observables in a region are observables in larger regions containing them, while $m_V^{U_1,\dots, U_n}$ is essentially an operator product expansion for \textit{local observables}. It should be noted that an open set $U$ is the support of the full observable and the supports (open sets) should not overlap. Local operators are a special class of local observable supported on a point, but in any interacting theory, a genuine renormalized local operator carries a scale dependence and has a radius. A true ``operator product expansion'' is specifically an asymptotic expression about point observables, and only (provably) exists in special cases, like factorization algebras obtained from perturbation theory around a free fixed point.\footnote{We will still refer to the $n$-ary operation of the pre-factorization algebra for local observables as an ``OPE,'' where ``O'' stands for ``operator'' or ``observable'' based on context.}

By inspection, we see that prefactorization algebras are on-the-nose equivalent to algebras over the topological operad $\mathrm{Disj}_M$. This gives a physical interpretation of a large class of topological operads and their algebras as axiomatizing observables in quantum field theory from an OPE-centric viewpoint. A \textit{factorization algebra} is essentially a prefactorization algebra where observables are determined by arbitrarily small neighbourhoods of points (see \cite{CG1} for rigorous details). In other words, in a factorization algebra $\mathrm{Obs}(V)$ can be recovered from the $\mathrm{Obs}(U_i)$. 

As mentioned in footnote \ref{footnote:derived}, we can actually take prefactorization algebras to be $\mathrm{Disj}_M$-algebras for any symmetric monoidal category $\calD$. From \cite{CG1}, it is known that the observables of a Euclidean (perturbative) QFT are naturally valued in the category of cochain complexes. i.e. we work in a derived setting where local observables are naturally cochains. This is not unusual to physicists: working in a derived setting is essentially the statement that we will work in a BV-BRST formalism, where fields and observables are written as cochain complexes, and the invariant information is captured by the cohomologies -- which are unchanged by quasi-isomorphisms. From here out, all operad-algebras coming from physics are understood to be valued in complexes.

A general prefactorization algebra (roughly, observables in a quantum field theory) is described by the very general coloured operad $\mathrm{Disj}_M$. We can also ask for various equivariance conditions on the (pre)factorization algebras to encode symmetries; these can simplify or enrich the relevant underlying operad. We expect that the very simplest quantum field theories, topological quantum field theories, are described by some of the simplest topological operads. This is the case: in topological field theories $\mathrm{Obs}(U)$ is (quasi)isomorphic to $\mathrm{Obs}(V)$ whenever $U \subset V$, called a locally-constant factorization algebra. When $M=\bbR^d$, locally-constant factorization algebras are $E_d$ algebras (see Theorem 5.4.5.9 of \cite{LurieHigherAlgebra} and Theorem 4.0.2 of \cite{CG1}). Similarly, in cohomological TFTs obtained from twists, the local observables $\mathrm{Obs}^Q$ carry an action of the $E_d$ operad \cite{Elliott:2018cbx}. 

The most basic example occurs in quantum mechanics: the factorization algebra of quantum mechanics is locally constant, and hence has an $E_1$-algebra structure. By the results in the previous section, we find that the algebra of observables in quantum mechanics is necessarily associative (see \cite{CG1} page 5), but not commutative. Dimension 1 topological operators in $d=3$ are described by an $E_{3-1} = E_2$-algebra, corresponding to a braided monoidal category of anyons. Note that the fusion of anyons is commutative, owing to the extra commutativity of $E_2$ structures (see \cite{Beem:2018fng}).

As a more general example, consider a translation-invariant QFT. Let $\mathrm{Obs}(B_r(x))$ be the observables supported in a ball of radius $r$ around the point $x$.\footnote{Intuitively, we might try to identify these observables with the space states on a sphere of radius $r$ around $x$. In a CFT, the state-operator correspondence says that if we radially quantize around $x$, that just the local operators i.e. point observables (with well-defined dilitation eigenvalues) placed at $x$ can act on the (conformally invariant) vacuum to prepare (a dense subspace of) the space of states on the sphere.} Translation invariance implies that the complex of observables $\mathrm{Obs}(B_r(x))$ should be essentially independent of $x$, but not $r$. More precisely, $\mathrm{Obs}(B_r(x))$ should be quasi-isomorphic to some complex $C_r$. Let $\mathrm{Disk}_d$ be the coloured topological operad whose $n$-ary operations are the spaces\footnote{We are intentionally vague about additional structures that may be of interest, such as complex structure, framings, or other tangential or $G$-structures. E.g. formally we should consider polydisks rather than disks in the complex case.}
\begin{equation}
    \mathrm{Disk}_d(r_1,\dots, r_n; r)
\end{equation}
of $n$ disjoint balls of radius $r_1, \dots, r_n$ inside a ball of radius $r$, with composition defined as per usual. A point $p$ in the space $\mathrm{Disk}_d(r_1,\dots, r_n; r)$ is a configuration of disks (labelled by the center of the disks) and corresponds to an $n$-ary operation of the operad. The factorization algebra of the translation invariant theory is an algebra over this operad, and its OPE map can be written
\begin{equation}
    m_p: C_{r_1} \otimes \cdots \otimes C_{r_n} \to C_r\,.
\end{equation}
This map depends smoothly on $p$, so we can declare there to be just one OPE operation from $\mathrm{Disk}_d(r_1,\dots, r_n; r)$
\begin{equation}
    m: C_{r_1} \otimes \cdots \otimes C_{r_n} \to C_r \otimes C^\infty(\mathrm{Disk}_d(r_1,\dots, r_n; r))\,.
\end{equation}
So a generic $n$-ary operation depends sensitively on the locations of the local observables and their radii as we expect from the general OPE.

However, if the theory is holomorphic-topological, then some subset of translations acts homotopically trivially, and we can package them into superfields of operator-valued forms as described in Section \ref{sec:ManyFaces}:
\begin{equation}
    m: C_{r_1} \otimes \cdots \otimes C_{r_n} \to C_r \otimes \Omega_{\dd}^*(\mathrm{Disk}_d(r_1,\dots, r_n; r))\,.
\end{equation}
Here, $\Omega_{\dd}^*$ is the appropriate space of mixed holomorphic-topological forms based on context. In particular, in cohomology, we have
\begin{equation}
    m: H^*_{\dd}(C_{r_1}) \otimes \cdots \otimes H^*_{\dd}(C_{r_n}) \to H^*_{\dd}(C_r) \otimes H_{\dd}^*(\mathrm{Disk}_d(r_1,\dots, r_n; r))\,.
\end{equation}
Thus we see that the OPE axiom of the factorization algebra literally becomes the OPE of holomorphic-topological QFTs in a straightforward way (see \cite{costello2015lectures, Beem:2018fng, Budzik:2023xbr} for more detailed examples and also Section 10.5 of \cite{CG2} for a careful comparison of statements about local observables versus local operators). These $n$-ary operations $m$ are used to define our brackets $\{\,\cdot\,,\dots,\,\cdot\,\}$, see Section \ref{sec:PointSplittingRevisted}.

Consider the topological case. By the analysis above, the number of distinct OPEs of 2 observables in $Q$-cohomology should be parametrized by
\begin{equation}
    H^*_{\mathrm{dR}}(\mathrm{Disk}_d(r_1,r_2;r))
    \cong H^*_{\mathrm{dR}}(\bbR^d - 0)
    = \begin{cases}
           \,\,* & \text{in degree $0$ and $d-1$}\\\
            0 & \text{otherwise}
    \end{cases}\,.
\end{equation}
The two non-trivial cohomologies give the ``product'' and ``secondary product'' for the superfields \cite{Beem:2018fng}. An understanding of the higher configuration spaces and their homologies shows that there are no interesting new $n$-ary operations in the topological case, since they can all be generated by partial compositions of the $2$-ary operations \cite{Beem:2018fng, sinha2006homology}.

\bibliographystyle{JHEP}
\bibliography{mono}
\end{document}